\documentclass[lettersize,journal]{IEEEtran}
\usepackage{amsmath,amsfonts}
\usepackage{algorithmic}
\usepackage{algorithm}
\usepackage{array}
\usepackage{textcomp}
\usepackage{stfloats}
\usepackage{url}
\usepackage{verbatim}
\usepackage{graphicx}
\usepackage{cite}
\usepackage{blindtext}
\usepackage{graphicx} 
\usepackage{svg}
\usepackage{multirow}
\usepackage{comment}
\usepackage{subcaption}
\usepackage[hidelinks]{hyperref}
\usepackage{xurl}
\usepackage{xcolor}
\definecolor{revgreen}{RGB}{0,180,0} 
\newcommand{\rev}[1]{#1}

\makeatletter
\def\ps@IEEEtitlepagestyle{%
  \def\@oddhead{\mycopyrightnotice}%
  \def\@evenhead{}%
}
\def\mycopyrightnotice{%
{\footnotesize\parbox{\textwidth}{This manuscript has been accepted for publication in IEEE Transactions on Network and Service Management. The DOI is 10.1109/TNSM.2026.3705327. Please refer to IEEE Xplore for the version of record.}}%
  \gdef\mycopyrightnotice{}%
}
\makeatother

\begin{document}
\bstctlcite{NoDash}
\title{Hermes: A General-Purpose Proxy-Enabled Networking Architecture}
\author{
    \IEEEauthorblockN{Behrooz Farkiani, Fan Liu, Ke Yang, John DeHart, Jyoti Parwatikar, Patrick Crowley}\\
    \IEEEauthorblockA{
        Washington University in St. Louis, \\ 1 Brookings Dr., St. Louis, MO, 63130, USA\\
        Emails: \{b.farkiani, fan.liu, k.yang1, jdd, jp, pcrowley\}@wustl.edu
    }
}

\markboth {IEEE Transactions on Network and Service Management}%
{Farkiani \MakeLowercase{\textit{et al.}}: Hermes: A general-purpose proxy-enabled networking architecture}


\maketitle

\begin{abstract}
We introduce Hermes, a general-purpose networking architecture that aims to improve service delivery over the Internet. Hermes delegates networking responsibilities from applications and services to proxies and is designed as a portable, adaptable solution to four fundamental challenges of efficient service delivery over the Internet: end-to-end traffic management, backward compatibility, data-plane security and privacy {models}, and adaptable communication layers. The design centers on an overlay of reconfigurable proxies and HTTP tunneling and proxying techniques, utilizing assisting components to extend proxy functionality when needed. Through prototyping and emulation, we demonstrate that Hermes improves key performance metrics across multiple use cases: it provides backward compatibility through protocol translation and tunneling, improves reliability by delegating retry logic to proxies, enables unified policy-based Layer 3 routing across network segments, and serves as an efficient substrate for future architectures like NDN, facilitating their operation over the Internet. Beyond evaluating Hermes across various use cases, we measured the overhead of Hermes' HTTP tunneling and proxying mechanisms and found it to be modest, typically under 2 ms per proxy pair traversal \rev{in an isolated collocated setup}. \rev{Although the HTTP proxying and tunneling techniques used by Hermes increase single-connection processing overhead, we also show that, with up to 1{,}000 concurrent requests, proxies can amortize connection setup time and reduce end-to-end latency by utilizing connection pooling and multiplexing.}

\end{abstract}


\maketitle

\section{Introduction}

\IEEEPARstart{T}{he} {Internet's success as a general-purpose, packet-switched architecture rests on two complementary pillars: the universal deployment of a lean core protocol suite and the end-to-end principle, which shifts complexity to hosts while keeping intermediate nodes simple. By offering only best-effort packet delivery, the Internet can interconnect independently administered networks at low cost. Today, however, the dominant workload carried by the Internet is not best-effort host-to-host communication but service delivery. Most packets are generated in support of web services, cloud platforms, media delivery systems, and distributed applications whose performance and correctness depend on end-to-end service semantics rather than mere reachability. Inter-domain routing protocols such as BGP prioritize peering agreements over latency or reliability objectives, often yielding paths that are suboptimal from a service perspective. Addressing and routing remain fundamentally location-based, whereas modern services are defined by names, identities, and application-layer endpoints. These challenges, together with limited built-in security mechanisms, have hindered efficient service delivery over the Internet and motivated efforts to revisit and improve the Internet’s foundational design and service architectures \cite{balakrishnan_revitalizing_2021, brown_architecture_2024, clark_overlay_2006, mccauley_enabling_2019}.}

Over the years, numerous solutions have been proposed to improve service delivery over the Internet. Some advocate for complete clean-slate replacements of the global network architecture, including SCION \cite{perrig_scion_2017} and Named Data Networking (NDN) \cite{zhang_named_2014}. Despite their contributions and innovations, these clean-slate Internet replacements have not achieved widespread adoption \cite{mccauley_enabling_2019}. Furthermore, designing applications for architectures fundamentally different from the Internet requires rethinking communication models, APIs, and deployment practices, changes that may not be justifiable given the paradigmatic differences and the enormous installed base of existing Internet infrastructure.

Other approaches aim to improve specific aspects of Internet service delivery without replacing the underlying architecture. They provide important service-delivery functions through a patchwork of fragmented overlays, in-network processing, end-to-end enhancements, and application-layer workarounds that are configured and operated independently, rather than through a unified architectural substrate. These innovations include performance-enhancing proxies (PEPs) \cite{griner_performance_2001}, application-layer solutions such as Content Distribution Networks (CDNs) and Fog/Edge computing, and both public and private overlay networks like Tor \cite{dingledine_tor_2004} and Akamai \cite{nygren_akamai_2010}. PEPs represent one of the earliest efforts to improve reliability over lossy paths \rev{ \cite{ehsan_evaluation_2003, caini_pepsal_2006, abdelsalam_quic_proxy_2019, ivanovich_tcp_2008}}. They operate transparently at the transport layer by splitting end-to-end connections, buffering packets, providing enhanced congestion control, and supporting forward error correction. While PEPs transparently serve 20 to 40\% of Internet paths \cite{yuan_sidekick-network_2024}, they also function as bridges between incompatible architectures; for instance, PEP-DNA \cite{ciko_pep-dna_2021} translates between IP and NDN models.

In contrast to transport-layer PEPs, application-layer approaches such as CDNs, edge computing, and fog computing place data and computation closer to end-users and improve performance metrics. Many CDN and edge systems use overlay routing, directing traffic across dedicated networks optimized for service-specific delivery rather than relying on default Internet routing. For example, both Akamai \cite{nygren_akamai_2010} and Cloudflare \cite{cloudflare_interconnect, argo_smartrouting} operate their own overlay networks. Overlay networks vary in their traffic ingress and egress points and in the specific benefits they provide. Tor \cite{dingledine_tor_2004} is a public overlay designed primarily for anonymity. Users run Tor clients on their devices, and traffic is routed through volunteer-operated relays to an exit relay for public sites, or to the service itself in the case of onion services. Conversely, private overlays like Akamai typically route traffic from {an edge node near the source to a node} near the destination \cite{nygren_akamai_2010}.

PEPs, CDNs, edge/fog computing, and many other advances, together with end-to-end protocol innovations such as QUIC \cite{iyengar_quic_2021} and HTTP/3 \cite{bishop_http3_2022}, have materially improved service delivery over the Internet. Without these advances, today's Internet would likely not function effectively at scale. Nevertheless, significant challenges remain.

\subsection{Motivation}
In this paper, we identify four structural limitations that affect efficient service delivery over the Internet and motivate our work. The \textbf{first} is end-to-end traffic management. \rev{Most existing solutions focus on upstream, provider-controlled segments, where performance can be more easily optimized due to lower heterogeneity and greater operational control. As a result, they often bypass the end-user side, particularly the last mile, the network segment that reaches end-user devices. However, the last mile experiences much of the performance degradation observed in practice \cite{fontugne_persistent_2020, bajpai_dissecting_2017, brown_architecture_2024}.} The \textbf{second} limitation is backward compatibility. While services evolve rapidly, customer clients may not be able to evolve at the same pace. This mismatch leads to fragmentation in the customer base and slow adoption of new services. The \textbf{third} limitation concerns security and privacy models. Various and often conflicting models exist across the path that handles end-user traffic, both across network segments and across network layers: IP reachability rules, middlebox filtering, and application-layer authorization are each defined and enforced in isolation, with no shared notion of identity or policy continuity. The \textbf{fourth} limitation is the lack of an adaptable communication layer. Systems that are built on top of fixed protocols and assumptions inherit their limitations, and these limitations propagate into any solution layered above them. When those limitations conflict with service goals, service providers cannot easily change the communication layer itself and are instead forced to add workarounds in the application. A communication layer that can adapt to network conditions and service requirements without modifying application logic would directly address this gap.

\subsection{Goal}
Our goal is to provide a unified end-to-end solution to these challenges to improve service delivery over the Internet. We design our solution as a general-purpose architecture so service providers can apply it across different use cases. As a general-purpose solution, it must handle any type of traffic at the IP layer and above, and it must be extensible to support new functionality. It must also be adaptable, meaning that after deployment, data-plane configuration can be updated to address changes in network conditions and service requirements over time. Finally, it must be portable, meaning it can be used to address various use cases with minimal architectural change.

Service providers can use our solution to gain end-to-end control of \rev{traffic} routing, policy \rev{enforcement}, and observability from user devices to services. This improves service performance and simplifies cross-layer policy management without requiring changes to application logic. Our solution can be deployed selectively for a subset of users or services that need special traffic handling, and it coexists with and complements other mechanisms that providers already use, such as CDNs and service meshes \cite{farkiani_service_2022}.

\subsection{Approach}
\begin{figure}[htbp]
    \centering
    \includegraphics[width=\columnwidth]{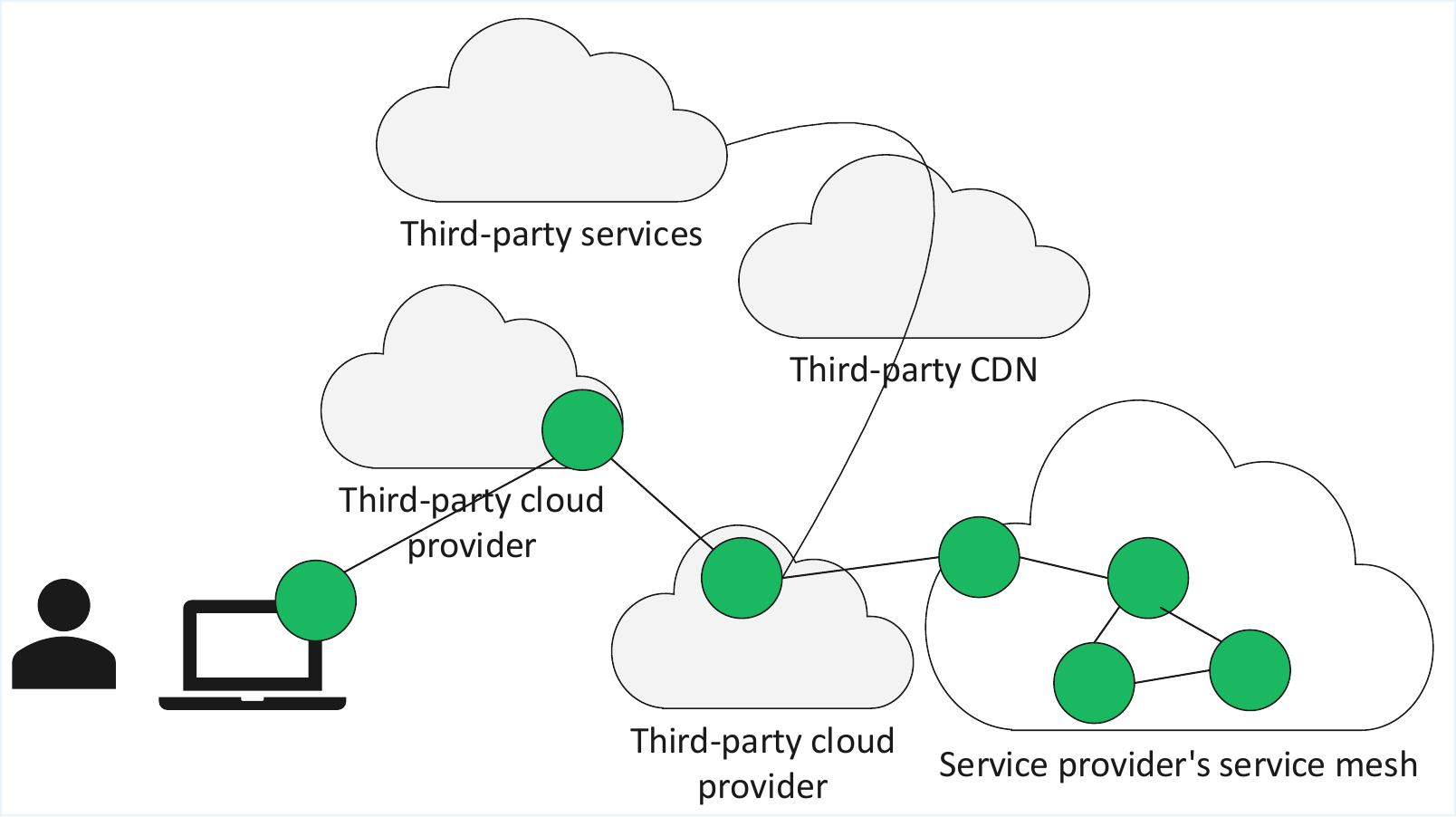}
    \caption{A deployment example. Hermes proxy nodes that are controlled by the service provider are shown in green.}
    \label{fig:vision}
\end{figure}

We build a general-purpose architecture named \textbf{Hermes} using the following elements. First, we delegate networking tasks from application logic to centralized components, namely proxies. Proxies can be updated independently to react to network and service changes without disrupting applications. Next, we deploy an overlay of proxies from end-user devices to wherever the service requires. For example, an end-to-end overlay can extend from end-user devices to a service mesh that handles service-to-service communication. This overlay controls traffic at the source and enables fine-grained routing, resource allocation, and security and privacy management across the Internet and network segments. Because Hermes needs to be general-purpose, it must handle IP traffic. We therefore extend standard proxy functionality with assisting components such as tunnel devices to capture and process IP traffic. We support multiple proxy deployment options, enabling Hermes to operate in a wide range of settings.

An architecture must specify its data-plane protocols. Hermes aims to be portable, so it relies only on UDP, TCP, and HTTP proxying and tunneling. These protocols are ubiquitous, and HTTP tunneling allows encapsulation of other protocols using the MASQUE framework \cite{schinazi_masque_2024}. We select HTTP as the backbone of the architecture. Using HTTP makes Hermes easily extensible; new functionality can be introduced by defining custom HTTP headers, provided the logic exists in proxies and assisting components.

A logically centralized control plane, operated by the overlay operator (either the service owner or a third party), manages the Hermes overlay. A vision of the architecture is shown in Figure \ref{fig:vision}. Here, the service provider operates its own Hermes overlay over the Internet, and the overlay is integrated with the service provider's service mesh. This creates an end-to-end overlay that carries traffic from the user device to the target service. The service provider uses different cloud providers to host Hermes proxies across the Internet in various regions. When user traffic requires third-party services, Hermes proxies can forward requests to external CDNs and receive traffic back from them.

\subsection{Contributions}
This paper makes two main contributions. First, it introduces the design and components of Hermes. We developed a prototype using commercially relevant open-source projects and custom software, including an overlay controller and Android and iOS applications.\footnote{The code for this paper is available at: \url{https://github.com/Bfarkiani/Hermes}} Although we describe control-plane components and their features, our focus is on the data plane: what the components are, which features they require, and how they operate to address challenges.

Second, by evaluating the prototype developed in this work, we show that Hermes:
\begin{itemize}
\item can provide backward compatibility and reliable delivery in noisy environments (Section \ref{video});
\item can securely transfer data in highly unstable networks (Section \ref{intermittent});
\item can provide end-to-end control for general IP packets (Section \ref{IP});
\item can dynamically adapt to network condition changes according to business policies (Section \ref{IP});
\item can act as a reliable underlying network for future networking architectures (Section \ref{NDN}).
\end{itemize}
Additionally, because Hermes relies on HTTP tunneling and HTTP-layer semantic processing, we evaluate multiple tunneling and proxying approaches and find that the \rev{per proxy pair traversal overhead in an isolated collocated setup is at most 2 ms.} We also evaluate using workloads of up to 1000 concurrent connections and show that proxies can even reduce end-to-end latency compared with direct no-proxy approaches by utilizing HTTP/2 multiplexing and connection pooling.

The rest of this paper investigates the details of the Hermes architecture. Section \ref{related} explores the challenges and related work in greater detail, and Section \ref{hermes} describes Hermes components, features, and how Hermes addresses challenges. Section \ref{imp} describes prototype implementations of Hermes overlays, and we evaluate their performance in solving various problems. In Section \ref{perf}, we investigate the performance of HTTP proxying and tunneling approaches that Hermes uses. Finally, Section \ref{conclusion} concludes this paper.

\section{Challenges and related work} \label{related}

We first explore the challenges Hermes addresses in detail, and then discuss related architectures and solutions.

\textbf{End-to-end traffic management:}
In practice, providers rely on DNS and BGP to route end-user traffic to their infrastructure. DNS caching and long TTLs, along with BGP's policy-based path selection, limit fast per-flow steering and inhibit fine-grained per-customer control. Therefore, providers usually have little leverage over traffic before it reaches their ingress, and they defer control decisions until traffic reaches ingress. A prominent example is rate limiting. HTTP APIs, the dominant channel for Internet service delivery, are usually protected by provider-side rate limiters that define and enforce quotas. Because providers have little or no control over when end-users send requests, clients typically resort to exponential backoff when they receive HTTP 429 responses or, in some services, 503 responses (for example, Amazon \cite{amazon_ratelimit} and OpenAI \cite{openai_exponential}) after exceeding service quotas. When an endpoint is shared by uncoordinated clients, independent exponential backoff wastes retry attempts because users lack visibility into aggregate load, as we showed in \cite{farkiani_rethinking_2025}.

Generally, having end-to-end control over traffic can improve routing, resource management, and service delivery. For example, PAINTER \cite{koch_painter_2023} showed that end-to-end traffic management can help in enterprise networks: an enterprise-managed edge proxy tunnels outbound flows to selected cloud ingress points based on live measurements and policy, reducing latency and improving resilience. However, PAINTER targeted enterprise cloud deployments with controllable edges such as SD-WAN stacks, rather than the general-purpose endpoints addressed in this paper.

\textbf{Backward compatibility:}
Services continuously evolve, including updates to both application logic and client-facing interfaces. Interface evolution, in particular, requires updating how users access service endpoints, often across a wide range of dedicated clients and even web browsers. Given the diversity of user devices and software stacks, developers must ensure consistent behavior across all environments. However, this is challenging for two reasons: (1) the environments in which end-users run their clients are often heterogeneous, making it difficult to achieve consistent behavior across all configurations; and (2) users may simply refuse to update their hardware or software. As a result, service providers often face fragmented customer bases, where some users can access newer services while others remain dependent on legacy paths, requiring continued support for both.

A notable example is Server Name Indication (SNI) for TLS-encrypted communication. During secure connection setup, the server must return the correct certificate. Without SNI, virtual hosting requires mapping each hostname to a separate IP address. SNI enables multiple destinations to share the same IP address and allows the server to select the appropriate certificate based on the unencrypted SNI field, thereby reducing cost and improving scalability. Although SNI was introduced in 2003 as part of RFC 3546 \cite{blake-wilson_transport_2003}, deployment lagged for years because many clients, especially Internet Explorer 6 on Windows XP, lacked SNI support. As XP's share declined in the mid-2010s, SNI became safe to rely on broadly \cite{sni_story, sni_story2}.

\textbf{Data-plane security and privacy models:}
Modern deployments implement a patchwork of security and privacy mechanisms that operate under fundamentally different models across network layers and network segments. For example, service mesh implementations like Istio \cite{istio_l4} offer Layers 4-7 zero-trust policy enforcement independent of the underlying infrastructure \cite{farkiani_service_2022, li_service_2019}, whereas Layer 3 policies depend entirely on platform support. Layer 3 policies control what addresses can communicate with each other, irrespective of any application logic at upper layers. Layer 3 policy enforcement is a standard feature in many infrastructures. For example, in Kubernetes it can be enforced using network plugins \cite{kubernetes_plugin}. Kubernetes by default allows entities to communicate freely, even across tenant boundaries, making Layer 3 policy enforcement necessary. However, 91\% of clusters do not implement any network policies \cite{wiz_report, pizzato_intent-based_2024}, leaving Layer 3 unprotected. Therefore, authorization must be fully implemented for data-plane endpoints according to application logic, which is a complicated process considering the size of service components. This lack of unified policy enforcement across network layers makes the application layer vulnerable, since application developers might incorrectly presume a safe Layer 3 and leave application layer authorization incomplete.

Furthermore, treatment of traffic varies significantly before and after the ingress point. User traffic may be unencrypted or protected only by simple TLS (i.e., server certificate plus application-layer credentials) whereas service-to-service traffic typically uses mutual TLS. Privacy guarantees also vary, as data may be decrypted, re-encrypted, and stored multiple times along the path before it reaches the target service, increasing the risk of exposure.

\textbf{Adaptable communication layer:} Services often rely on fixed communication protocols and assumptions that have inherent design limits. For example, the Internet is primarily location-addressed, so any service built on top of it is forced to run over location-based addressing. If a service prefers content-based addressing, it must implement content naming and lookup in the application layer. If these features were already present in the communication layer, service complexity would be greatly reduced. Reliability shows a similar constraint. Transports such as TCP and QUIC provide congestion control and loss recovery, but services that rely only on these standard designs cannot easily and dynamically customize core behaviors beyond what the specifications expose. As another example, if providers rely only on QUIC and HTTP/3, for example, to improve client mobility and avoid HTTP/2 head-of-line blocking, users may experience up to 45\% worse performance than HTTP/2 over TCP on high-speed links \cite{zhang_quic_2024}. All protocol designs have inherent limitations, so a communication layer with predetermined protocol choices cannot perform well in all conditions, especially on the Internet where conditions fluctuate rapidly and unpredictably.

Services therefore benefit from a communication layer that adapts to service needs and network conditions and can update its protocols and configurations to avoid limitations that hurt service delivery. Such a layer should be extensible based on service needs, reconfigurable and policy-driven, and able to steer traffic according to application and network requirements.

To the best of our knowledge, there is no general-purpose solution that addresses these challenges in a unified way. In our prior work \cite{farkiani_rethinking_2025}, we addressed HTTP API rate limiting by introducing proxies that manage client retries. These proxies can be implemented as service workers \cite{service_workers} in web browsers, and we showed that using proxies reduced HTTP errors by up to 97\%. In addition, in \cite{farkiani_l3mesh} we demonstrated that assisting components can enable portable service mesh implementations such as Istio \cite{istio} to handle Layer 3 network policies in a portable, infrastructure-agnostic way. More importantly, by building a lightweight Layer 3 overlay that added less than 1 ms to end-to-end latency, we showed that service meshes can handle Layer 3 through Layer 7 traffic management and policy enforcement in an integrated way over diverse infrastructures. This paper generalizes our earlier work to address the fundamental challenges of efficient service delivery over the Internet. 
%{Hermes is not intended to replace the Internet or legacy protocols. Unlike clean-slate designs such as NDN and SCION, Hermes can be deployed incrementally and leverages widely deployed Internet protocols to remain portable while providing a range of benefits on top of today's Internet.}

Hermes targets service providers as its primary users. {From the perspective of end-to-end traffic management, Hermes can be compared to Tailscale \cite{tailscale}, Cloudflare Zero Trust \cite{cloudflare_zerotrust}, and OpenZITI \cite{openziti}. These systems operate from end-user devices using proxying and tunneling techniques and provide a secure access layer, but they generally do not integrate with upper-layer service semantics. Hermes also operates from end-user devices, but it is intended as a general-purpose, service-oriented communication substrate that is configurable to satisfy service requirements. When a service requires secure access, Hermes can provide an end-to-end secure access layer that is integrated with service semantics. More broadly, Hermes also provides other capabilities through the same substrate, including operating as an adaptable communication layer. Integrating these capabilities with the secure access layer and service semantics differentiates Hermes from the systems above and targets improving service delivery.}

\begin{table}[hb]
\caption{Comparison with service meshes}
\label{tab:servicemesh}
\renewcommand{\arraystretch}{1.1}
\begin{tabular}{|l|ll|}
\hline
\multirow{2}{*}{\textbf{Aspect}} & \multicolumn{2}{c|}{\textbf{Architecture}} \\ \cline{2-3}
& \multicolumn{1}{c|}{Service Mesh} & \multicolumn{1}{c|}{Hermes} \\ \hline
Goal & \multicolumn{1}{l|}{\begin{tabular}[c]{@{}l@{}}Improve service to service \\communication \cite{howard_introducing_nodate}\end{tabular}} & \begin{tabular}
[c]{@{}l@{}}Improve service delivery\\ over the Internet\end{tabular} \\ \hline
\begin{tabular}[c]{@{}l@{}}Target \\ environments\end{tabular} & \multicolumn{1}{l|}{\begin{tabular}[c]{@{}l@{}}Managed service\\ environments\end{tabular}} & \begin{tabular}[c]{@{}l@{}}End-user devices,\\ across Internet, and \\ managed service \\ environments \end{tabular} \\ \hline
\begin{tabular}[c]{@{}l@{}}Proxy \\ locations\end{tabular} & \multicolumn{1}{l|}{Between services} & \begin{tabular}[c]{@{}l@{}}Mandatory at end-user \\ devices, optionally \\ everywhere\end{tabular} \\ \hline
\begin{tabular}[c]{@{}l@{}}Traffic \\ type\end{tabular} & \multicolumn{1}{l|}{\begin{tabular}[c]{@{}l@{}}Mostly TCP and \\HTTP \cite{istiotcp} \cite{linkerdtcp}\end{tabular}} & IP packets \\ \hline
\end{tabular}
\end{table}

From a design viewpoint, as Hermes is composed of an overlay of proxies, it is comparable to architectures that use proxy overlays. The most closely related are service mesh architectures, which specialize in managing service-to-service communication. Although various service mesh implementations exist with different features, including Istio \cite{istio}, Linkerd \cite{linkerd_site}, and Cilium mesh \cite{cilium_mesh}, the features listed in Table~\ref{tab:servicemesh} are common among them. As shown, Hermes and service mesh architectures are inherently different. Although some research has explored using service meshes in edge systems \cite{elkhatib_evaluation_2023} and as a networking layer \cite{ashok_leveraging_2021} in cloud-native environments, service meshes are not designed for general-purpose networking and operate within managed and controlled environments. In contrast, Hermes creates an overlay that extends from end-user devices and may terminate at service components, the public Internet, or other networking architectures. Hermes does not assume the service provider has control over the end-user devices where proxies run, except for proxy configuration. For this reason, we designed Hermes to be portable with various deployment options so it can operate in diverse settings. Additionally, Hermes operates over the Internet, where there is often no control over the entire end-to-end path due to the nature of Internet routing. More importantly, service meshes are typically designed to handle TCP and HTTP traffic, whereas Hermes is built to carry general IP packets and encapsulates traffic from various architectures using HTTP tunneling. These fundamental differences result in distinct concepts, features, and use cases for each architecture. Notably, one application of Hermes is extending service meshes to end-user devices, enabling end-to-end service provisioning and improving resource allocation by incorporating real-time end-user statistics.

Hermes can also be compared with other overlay architectures like Akamai. CDN platforms like Akamai are not end-to-end; they steer end-user traffic using DNS-based resolution techniques to one of their edge points. In contrast, with Hermes, end-users do not need to resolve a name to an IP address, and their traffic is captured directly at the end-user device. As we show in Section~\ref{hermes}, clients can submit a name to a proxy, and proxies forward it until it reaches a proxy that is configured to resolve it to a destination address. This reduces the burden of globally consistent name resolution while preserving the provider's control over policy and routing.

\rev{Hermes is also related to PEPs. Similar to PEPs, Hermes proxies can improve communication by interposing proxy functionality on the data path and can help enable communication across different networking architectures. However, unlike PEP-based approaches that rely on protocol translation to support communication between architectures (e.g., PEP-DNA~\cite{ciko_pep-dna_2021}, COIN~\cite{jahanian_managing_2020}), Hermes primarily uses HTTP-based encapsulation, including the MASQUE framework, to carry different types of traffic over the overlay. Supporting multiple architectures through translation requires deploying and maintaining multiple translation gateways, which is operationally burdensome and can lead to incomplete or inconsistent translations.} 

%Another relevant design comparison is between Hermes and performance-enhancing proxies (PEPs). PEPs can improve the performance of Internet protocols at the transport and application layers. The effectiveness of PEPs in reducing end-to-end delay and packet loss in satellite communication and cellular environments has been demonstrated in studies such as \cite{ehsan_evaluation_2003, caini_pepsal_2006, abdelsalam_quic_proxy_2019, ivanovich_tcp_2008}. Additionally, the authors of \cite{ciko_pep-dna_2021} showed how PEPs can be used to connect the Internet to other clean-slate architectures, {such as Content-Centric Networking \cite{jacobson_networking_2009}}. Their approach involves breaking the end-to-end TCP logic and translating TCP packets into new architecture formats using a proxy. Hermes proxies indeed act as PEPs. Additionally, they use encapsulation and decapsulation of IP packets in HTTP via the MASQUE framework to carry any type of traffic without requiring protocol translation.

\begin{table*}[htbp]
\caption{Comparison with related work}
\label{tab:related}
\renewcommand{\arraystretch}{1.1}
\begin{tabular}{|l|l|l|l|}
\hline
\multicolumn{1}{|c|}{\textbf{Solution}} & \multicolumn{1}{c|}{\textbf{Type}} & \multicolumn{1}{c|}{\textbf{Goal}} & \multicolumn{1}{c|}{\textbf{Technique}} \\ \hline
\begin{tabular}[c]{@{}l@{}}\rev{PEP-DNA}, \\ \rev{COIN} \end{tabular} & PEP & Enabling TCP/IP to communicate with other architectures & Using a proxy \rev{gateway} that translates \rev{between architectures} \\ \hline
Akamai & Overlay & Private overlay that improves service delivery & Using DNS to direct traffic to the nearest edge server \\ \hline
Tailscale & Overlay & Secure end-to-end access layer & An overlay of Wireguard \cite{wireguard} endpoints and relay nodes \\ \hline
\textbf{Hermes} & \begin{tabular}[l]{@{}l@{}}PEP and \\ Overlay\end{tabular} & \begin{tabular}[l]{@{}l@{}}Improving end-user experience \\ by enhancing service delivery over the Internet \end{tabular} & \begin{tabular}[l]{@{}l@{}}Using an overlay of proxies\\ Using proxying and tunneling \\ Decoupling communication knowledge from application logic\end{tabular} \\ \hline
\end{tabular}
\end{table*}

\rev{We summarize the comparison of Hermes with PEPs and related approaches in Table~\ref{tab:related}. Compared to Tailscale and other zero-trust approaches that provide a secure access layer, Hermes targets a broader, service-oriented, adaptable communication layer, while still supporting end-to-end secure access when required by the service.} Hermes also does not replace current CDN and edge service providers such as Akamai and Cloudflare, or architectures that unify access to edge services such as InterEdge~\cite{brown_architecture_2024}. CDNs place points of presence at well-connected managed locations, whereas Hermes is explicitly end-to-end and can handle traffic from unmanaged end-user devices. Instead, Hermes complements these building blocks by providing end-to-end control for the service provider across the public Internet. For example, a provider may use CDNs to reduce latency while using Hermes to control how end-user traffic reaches CDN edge endpoints under service-specific policy; if reliability issues arise, Hermes can enforce remedies at the source or steer traffic onto alternate overlay paths.

Hermes specifies the capabilities that proxies should provide and how they operate to address concrete challenges. Each service provider instantiates its own proxy configurations to meet specific needs, which in turn define how traffic is processed. Hermes targets service-delivery problems under the assumption that providers do not control end-user devices and that traffic traverses the public Internet with limited control. Once traffic enters the provider's controlled infrastructure, Hermes is neutral: the provider decides whether proxies are still needed, how to implement proxies, and where traffic should be delivered. Hermes integrates with existing infrastructure and is agnostic to deployment models inside controlled environments (e.g., containers, VNFs, physical appliances) and to lower-layer routing choices (e.g., SD-WAN, VNF-FG, VPNs). 

\rev{Taken together, two key characteristics of Hermes design are its unifying approach and its flexible integration with existing tools: it provides a unified, end-to-end overlay substrate that enforces service policy from the end-user device, through an overlay of proxies, to services across network layers and segments. In addition, Hermes does not try to replace existing tools and technologies; rather, it can integrate with and complement them to improve service delivery. These two aspects position Hermes as a distinct approach to addressing fundamental service-delivery issues.}

\section{Hermes Design}\label{hermes}
We first introduce the core concepts and solution elements we use in the design of Hermes. Then we discuss how Hermes addresses the challenges.

\begin{figure}[htbp]
\centering
\includegraphics[width=\columnwidth]{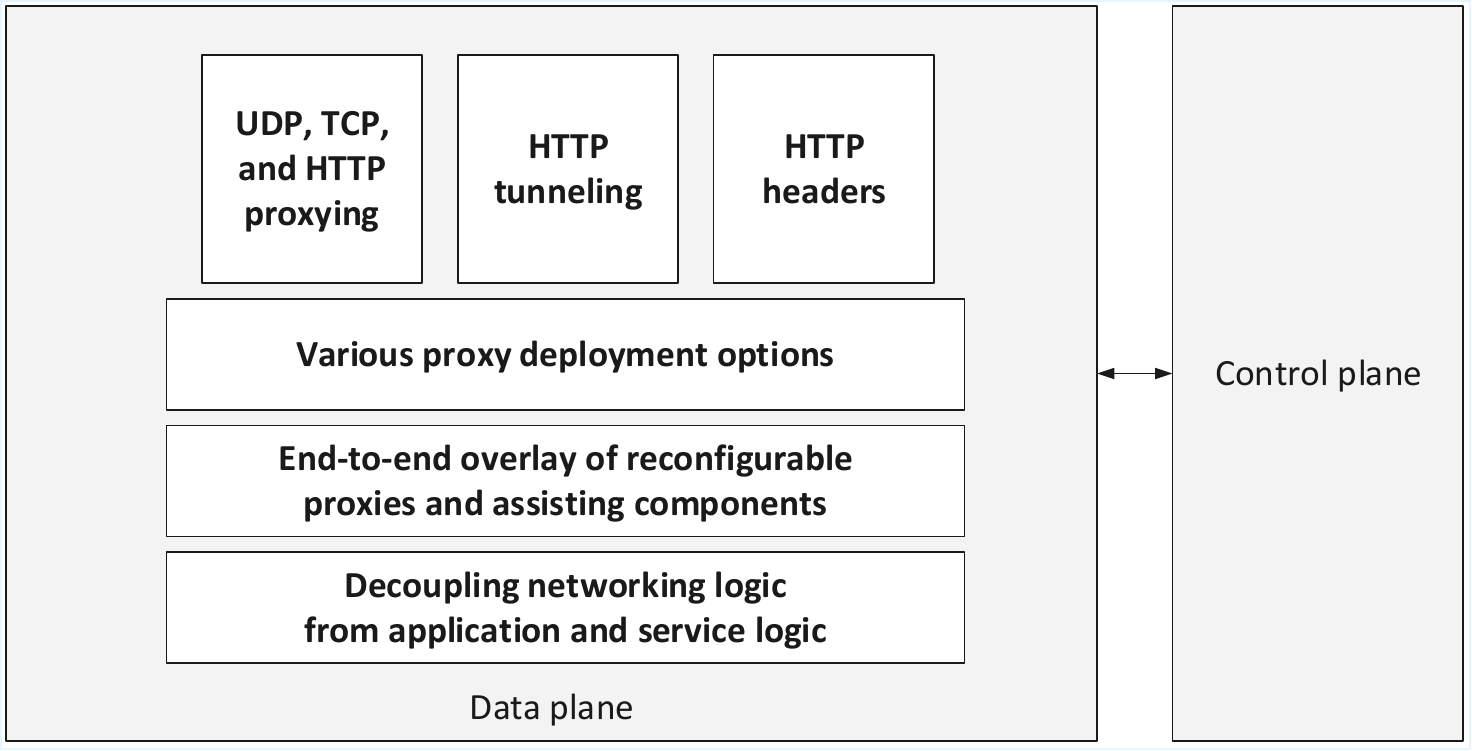}
\caption{{The solution elements used in the design of Hermes.}}
\label{fig:elements}
\end{figure}

\begin{table}[htb]
\caption{Mapping goals to solution elements}
\label{tab:goal_map}
\renewcommand{\arraystretch}{1.1}
\begin{tabular}{|l|l|}
\hline
\multicolumn{1}{|c|}{\textbf{Goal}} & \multicolumn{1}{c|}{\textbf{Solution element}} \\ \hline
Portability & \begin{tabular}[c]{@{}l@{}}Relying on UDP, TCP, and HTTP proxying \\and tunneling\\ Relying on user-space solutions\\ Various proxy deployment options\end{tabular} \\ \hline
Adaptability & \begin{tabular}[c]{@{}l@{}}Reconfigurable proxies and assisting \\ components that handle Layers 3-7 traffic\end{tabular} \\ \hline
General-purpose & \begin{tabular}[c]{@{}l@{}}Handling IP traffic using tunnel devices\\ Relying on HTTP headers for new semantics\end{tabular} \\ \hline
End-to-end & \begin{tabular}[c]{@{}l@{}}An overlay of proxies extended from \\end-user devices\end{tabular} \\ \hline
\end{tabular}
\end{table}

\begin{figure*}[h!]
\centering
\begin{subfigure}[b]{0.45\textwidth}
\centering
\includegraphics[height=3cm]{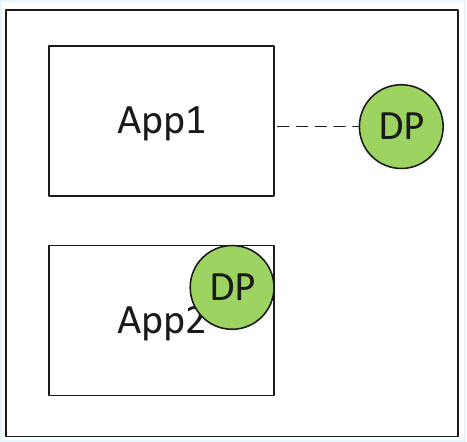}
\caption{Separate dependent proxies}
\label{fig:proxy1}
\end{subfigure}
\hfill
\begin{subfigure}[b]{0.45\textwidth}
\centering
\includegraphics[height=3cm]{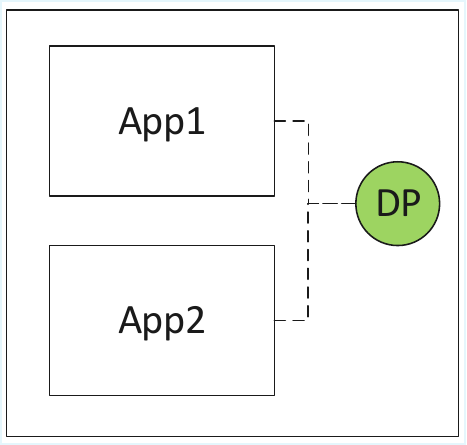}
\caption{A shared dependent proxy}
\label{fig:proxy2}
\end{subfigure}

\vspace{0.5cm}
\begin{subfigure}[b]{0.45\textwidth}
\centering
\hspace*{22mm}\includegraphics[height=3cm]{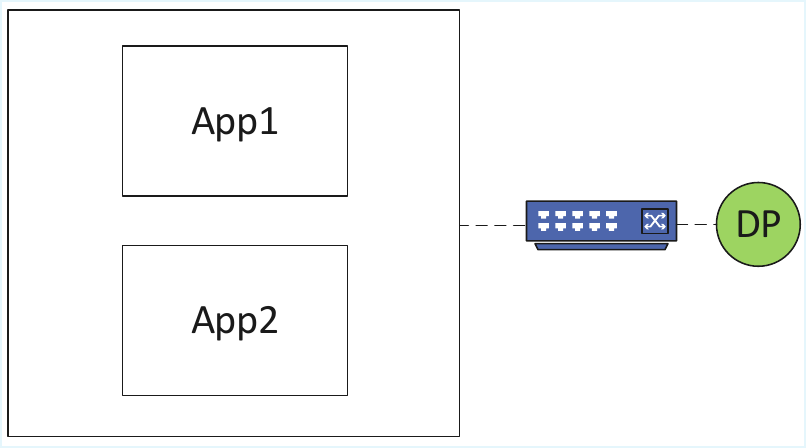}
\caption{A shared dependent proxy running on another machine}
\label{fig:proxy3}
\end{subfigure}
\hfill
\begin{subfigure}[b]{0.45\textwidth}
\centering
\raisebox{2mm}{\includegraphics[height=3cm]{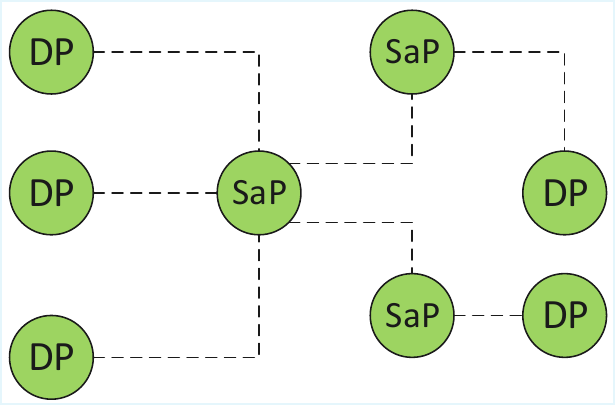}}
\caption{An overlay of standalone and dependent proxies}
\label{fig:SaProxy}
\end{subfigure}

\caption{Different proxy configurations}
\label{fig:proxy_configs}
\end{figure*}
\subsection{Core concepts}\label{concepts}
Hermes is a general-purpose networking architecture that is composed of an overlay network of dynamically reconfigurable proxies and managed by an overlay control plane. The elements we used in the design of Hermes and their relations are shown in Figure \ref{fig:elements}. Table \ref{tab:goal_map} shows how each element is related to the Hermes goals.

Hermes builds on the idea that the separation of networking logic from application logic, which has been valuable in service mesh solutions \cite{farkiani_service_2022, li_service_2019}, can also benefit general-purpose networking. This enables applications to communicate with fixed endpoints (i.e., proxies) without worrying about networking aspects, as the endpoint handles all networking complexities. Additionally, this separation enables fast adaptation to network and service changes without modifying application logic. However, a notable difference between service meshes and Hermes is that traffic redirection in service meshes is usually done transparently through iptables \cite{istio_redirect} or \rev{other} kernel-based \cite{cilium_redirect} approaches. Hermes targets end-user devices, where we assume the service provider does not have control over the software and hardware except for proxy configuration through the control plane. Therefore, traffic must be explicitly redirected to proxies either by sending traffic directly to the proxy endpoint, configuring applications to use proxies, or using tunnel devices that capture related traffic. This highlights the second element, which is the \rev{end-to-end overlay of} proxies and assisting components.

Proxies are versatile tools that can be configured for various benefits, such as privacy, performance, and security enhancements. For example, Tor \cite{dingledine_tor_2004} uses proxies to create an anonymous overlay network that preserves privacy, while Cloudflare \cite{cloudflare_work} employs reverse proxies to enhance Internet security, performance, and reliability. Proxies designed for the Internet architecture process traffic at the transport or application layer and serve as forwarding agents \cite{fielding_hypertext_1997}. In Hermes, we extend proxy capabilities with assisting components like caching mechanisms, advanced traffic manipulation algorithms, policy enforcement components, and most importantly, tunnel devices. Tunnel devices capture traffic at Layer 3 and deliver it to proxies for service-specific processing. In addition, Hermes requires proxies to be reconfigurable, i.e., their configuration, including the protocols and ports they use, can be updated at any time through the control plane. This enables the overlay of proxies to become adaptable. 

{As Hermes aims to enable end-to-end solutions, proxies are mandatory components on the end-user side of the communication. The overlay thus starts from end-user devices and can be extended as far as the service requires. For example, a service provider might deploy proxies across the Internet and terminate the overlay at the ingress point of the service environment, or another provider might extend the overlay and integrate it with the service mesh it operates. All proxies along the path may modify headers and process traffic as required by the service to prepare the flow for the next hop.}

{We define two roles for proxies: dependent proxies (\textbf{DP}) and standalone proxies (\textbf{SaP}). Dependent proxies receive traffic directly from the source or destination endpoint, while SaPs act as routers that connect DPs together. We examine their details in Section~\ref{components}. To support diverse environments, we consider multiple proxy deployment options, as shown in Figure~\ref{fig:proxy_configs}. The outer box represents a physical machine. Proxies can be part of the application, for example, service workers in web browsers~\cite{service_workers}, or can be deployed as a separate component outside the application. Proxies can also be shared among applications. Even if a device cannot run a proxy, one can be deployed inside the local network, and devices can communicate with that proxy endpoint directly. We also use user-space solutions on end-user devices, as they provide wider compatibility and portability.}

Hermes DPs listen to UDP and TCP ports and process packets at UDP, TCP, and HTTP layers, ensuring that all exiting traffic is either HTTP or encapsulated in HTTP. In this way, traffic along its path to the destination can be processed by SaPs solely based on its HTTP headers. Hermes SaPs can also process traffic based on UDP or TCP attributes if they are configured to, but the traffic remains HTTP. There are four main reasons why we chose HTTP as the carrier for every kind of traffic. First, we need to associate users with their traffic. This association either requires new protocols and designs, or it should be done using current protocols. HTTP headers have long been used for associating web traffic with user identities, and the Internet itself is engineered to provide best support for HTTP. Over 50\% of Internet traffic is HTTP(S) \cite{schumann_impact_2022, tsareva_decade_2023} and HTTP acts as the narrow waist of the Internet \cite{popa_http_2010}. For instance, if we chose a custom UDP-based protocol to encapsulate data, it is possible that \rev{such traffic might face different network treatment and manageability challenges} in some access ISPs \cite{kuhlewind_manageability_2022}.

Second, HTTP is highly extensible and adaptable to carry virtually any type of traffic. Indeed, HTTP can act as a unifying protocol for different traffic types: HTTP can be used to encapsulate UDP, IP, and Ethernet (RFC 9298 \cite{schinazi_proxying_2022}, RFC 9484 \cite{pauly_proxying_2023}, and \cite{sedeno_proxying_2024}). These are all part of the MASQUE framework \cite{schinazi_masque_2024}, which extends HTTP's functionality and positions it as a versatile, general-purpose communication protocol well-suited for unified processing of heterogeneous traffic types.  In addition, HTTP/3 maps HTTP semantics onto QUIC, which natively supports multiplexed streams, and can also carry unreliable {datagrams} using the QUIC DATAGRAM extension \cite{rfc_9297, pauly_unreliable_2022}.

Third, by using HTTP headers, Hermes provides native support for custom namespaces and operates as a name-based communication layer rather than relying solely on location-based Internet mechanisms. Namespaces can follow any structure, for example, a concatenation of strings and numbers or even an IP-like token. Consider a request to \texttt{http://DEV}. Alice's web browser is configured to forward HTTP requests to her dependent proxy. The DP forwards the request upstream without rewriting the host information (HTTP/1.1 \texttt{Host} header or HTTP/2 and HTTP/3 \texttt{:authority}). Overlay proxies along the path inspect the host information and route to the next destination accordingly. The request eventually reaches a proxy that can resolve \texttt{DEV} for Alice and forward it to the correct endpoint. Two points are important here. First, overlay proxies can resolve \texttt{DEV} uniquely for each user, since identity is present in headers and can be used in routing decisions. Second, Alice's device does not rely on DNS to map \texttt{DEV} to an IP address. To compare with similar solutions, in Tailscale's MagicDNS, users can have custom namespaces that are converted to IP addresses using DNS resolution on the user's device \cite{magicdns}. \rev{Instead}, Hermes \rev{can} keep names in application-layer headers and relies on overlay proxies to resolve names without \rev{necessarily} performing DNS resolution on the end-user's device. Headers can also carry new semantics. For example, to broadcast a request to multiple overlay destinations, we can define a header such as \texttt{X-Hermes-Broadcast: true}. SaPs that {process} this header duplicate the request to all configured upstream destinations. This approach lets Hermes support a variety of use cases that require custom behavior without modifying applications.

Fourth, using HTTP CONNECT \cite{thomson_http2_2022} and CONNECT-UDP \cite{schinazi_proxying_2022}, a client can tunnel an end-to-end secure connection {(e.g., TLS)} through proxies without allowing those proxies to undetectably alter the protected payload. With CONNECT, the client sends the authority ({\texttt{host}:\texttt{port}}) to its configured proxy, the proxy opens a TCP connection to the target, returns 200 OK, and switches to blind byte forwarding. TLS over TCP inside this tunnel provides end-to-end confidentiality and integrity between client and destination. With CONNECT-UDP, the client is configured with a MASQUE URI template that carries \texttt{target\_host} and \texttt{target\_port}; the proxy extracts these values, opens a UDP socket to the target, and relays UDP payloads. End-to-end security then comes from the tunneled UDP protocol, for example QUIC or DTLS. These properties also enable shared third-party overlays, since a proxy cannot read or modify the inner TLS or QUIC  payload without detection. {We discuss applications of this approach in Hermes in Appendix~\ref{app:security}.}
Please also note that Hermes is not restricted to a specific version of HTTP, and choosing which version depends directly on service requirements. As long as proxies are reconfigurable, even the HTTP version in use can be updated to match changes in service and network conditions.

Finally, the overlay requires a control plane to manage the lifecycle of proxy components and their configuration, monitoring, observations, and other aspects of management. This paper is mostly focused on the data-plane components and their specification, but we will review the control-plane components in the next section.

\subsection{Components and roles}\label{components}
\begin{figure*}[htbp]
\centering
\includegraphics[width=0.8\textwidth]{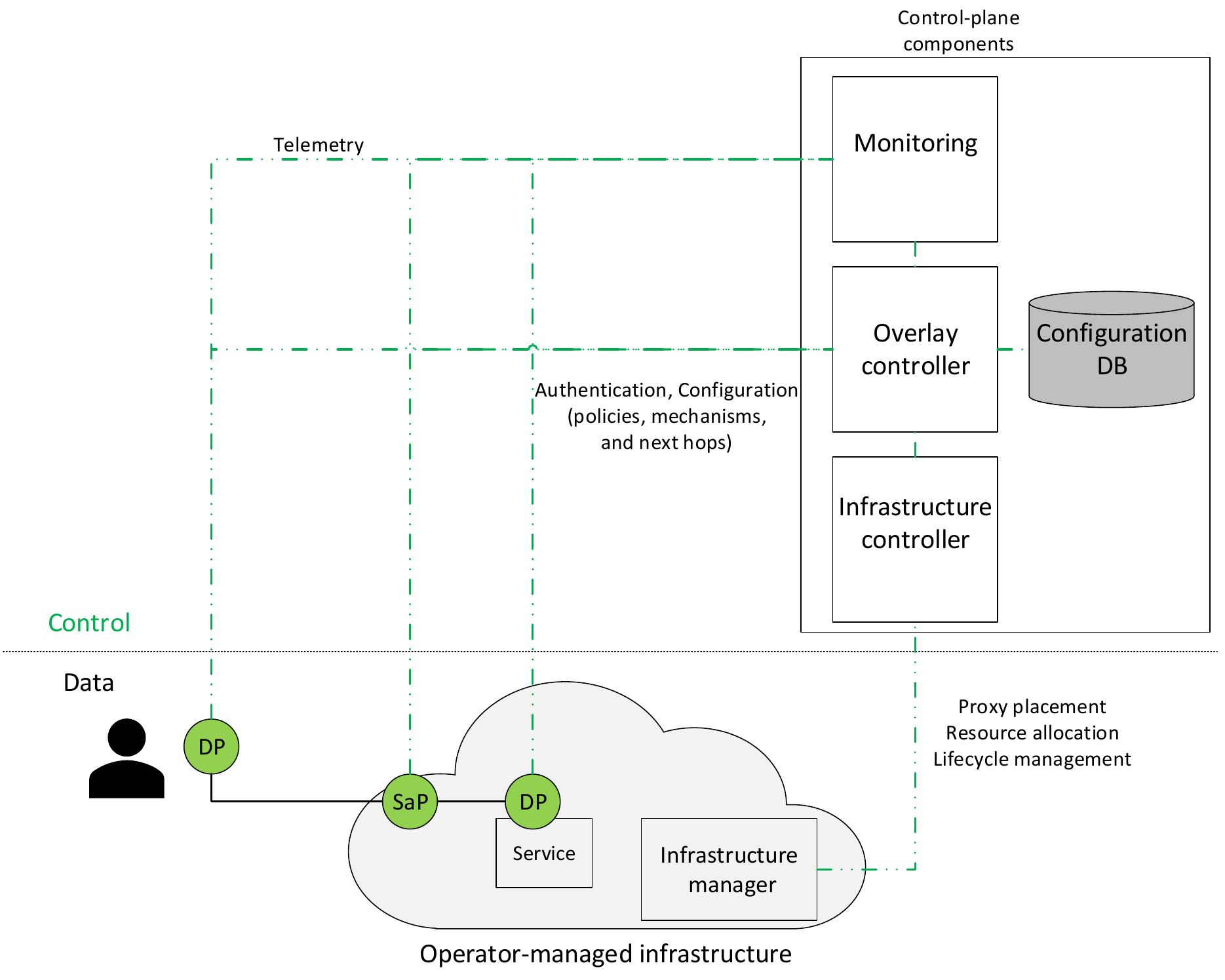}
\caption{\rev{Data-plane and control-plane components and their interactions. Assisting components are omitted for brevity.}}
\label{fig:controlPlane}
\end{figure*}

\rev{This section examines the data-plane and control-plane components of Hermes in more detail. These components and their interactions are shown in Figure \ref{fig:controlPlane}.}

\subsubsection{\textbf{Data-plane components}}

Hermes overlay proxies are software components and IP endpoints with IPv4 or IPv6 addresses and associated TCP and UDP ports. They process traffic using UDP, TCP, and HTTP header information. Hermes extends proxy capabilities with assisting components that enable IP traffic capture and processing, policy-based routing, advanced traffic management, and other tasks. As already mentioned, Hermes assumes two roles for proxies: dependent (DP) and standalone (SaP). As a core element of the architecture, end-user devices must connect to dependent proxies, but the inclusion of SaPs is optional. Please note that both types of proxies can create new tunnels, perform protocol translation, or have separate configuration for users based on their identities, and the architecture does not restrict them.

\paragraph{{\textit{Dependent proxies}}} Dependent proxies accept traffic from end hosts, including end-user \rev{devices}, and act as communication gateways \rev{for} one or more applications or services, as shown in Figures \ref{fig:proxy1} to \ref{fig:proxy3}. All relevant traffic entering the overlay or exiting to applications or services must pass through the dependent {proxies}. While these proxies may have little or no knowledge of the applications or services, they play a crucial role in ensuring smooth communication. The architecture mandates running at least one dependent proxy on the end-user side of communication to be able to manage the traffic from the source and also augment end-user traffic with overlay-related information.

As discussed above, dependent proxies utilize HTTP headers to augment incoming traffic with overlay-specific information, which the overlay uses to determine subsequent actions. This information can include user identity, custom name-space information, or tunneling details, among other examples. {If the incoming traffic is not HTTP, dependent proxies can tunnel it using HTTP CONNECT, CONNECT-UDP, or CONNECT-IP and include the relevant headers on the encapsulating request.} However, proxies still need to associate incoming traffic with the appropriate headers. One approach is to link headers to the ports on which the proxy is listening. For instance, if only database-related traffic arrives on port 1000, database-specific headers will be appended to the proxied traffic.

\paragraph{{\textit{Standalone proxies}}} Standalone proxies act as software routers that route traffic between different sections of the overlay, similar to Figure \ref{fig:SaProxy}. These proxies can act as policy enforcers and enforce access control across Layers 3-7 between different overlay sections. Additionally, standalone proxies can rely solely on TCP or UDP forwarding to enhance overlay performance or when HTTP processing is not necessary. For example, as shown in Section \ref{intermittent}, proxies can act as TCP forwarders to ensure an end-to-end secure TLS connection between the end-user and the destination, eliminating the need to trust the overlay proxies.

Please note that a proxy can have both roles of dependent and standalone proxies simultaneously. It may accept traffic on one of its ports from other proxies, acting as a standalone proxy, while also communicating with a local application on another port, acting as a dependent proxy.

\paragraph{{\textit{Assisting components}}}
{Assisting components extend proxy capabilities to match service requirements and can be built into the proxy or deployed as separate processes. In some deployments, assisting components may implement functionality that is impractical to integrate into the proxy itself. If deployed separately, an assisting component must communicate directly with the proxy it assists, typically over a local interface. Assisting components may or may not be co-located with the proxy on the same host, depending on their functionality and privilege requirements. In all cases, assisting components must be configurable by the control plane and should communicate only through the proxy.}

{An important assisting component that enables Hermes to capture L3 traffic is a packet tunnel interface, which can either be integrated into the proxy (so the proxy creates the tunnel interface and reads captured packets directly, similar to \cite{singbox}) or deployed as a separate assisting process that forwards captured packets to the proxy (as in our IP use case, Section~\ref{IP}). We utilize WireGuard-style \cite{wireguard} user-space packet tunnel interfaces (Linux TUN/TAP \cite{linux_tuntap}, Windows Wintun \cite{WinTun}, Apple NetworkExtension packetFlow \cite{apple_packetflow}, and Android VpnService \cite{android_vpn}), where selected traffic is routed by the host OS to a virtual L3 interface and delivered to user space for classification and forwarding. Which traffic is captured is determined by tunnel and routing configuration, and the captured packets can be classified and filtered according to the overlay policies using IP and transport metadata prior to encapsulating them for carriage across the overlay.}

When assisting components are external to the proxy, the proxy must be able to recognize their incoming traffic and associate it with the appropriate headers. For example, if an assisting component enforces policies, the local proxy must be able to detect whether traffic is allowed or denied. If the proxy can create a tunnel device, it can directly encapsulate traffic in HTTP, append the relevant headers, and forward it. However, if the proxy uses a separate assisting component, that component must encapsulate IP packets in TCP or UDP and deliver them to a specific port on the proxy. The port can be chosen based on the IP destination. For instance, traffic destined for 10.10.0.0/16 can be encapsulated in UDP and directed to UDP port 10000 on the proxy. The control plane then configures the assisting component to send such traffic to UDP port 10000 and configures the proxy to encapsulate data arriving on UDP port 10000 in HTTP while appending headers that indicate the destination network 10.10.0.0/16.

\subsubsection{\textbf{\rev{Control-plane components}}}

The focus of this paper is the Hermes data plane. Here we briefly discuss the control-plane components. We also implemented various control-plane components, as we elaborate in Section \ref{imp}, but the discussion about exact control plane functionalities and its interfaces is out of the scope of this paper.

The control plane, as shown in Figure \ref{fig:controlPlane}, consists of four key components: the overlay controller, infrastructure controller, configuration database, and observability and monitoring tools.

\paragraph{The overlay controller} The overlay controller is responsible for the placement, configuration, and lifecycle management of overlay proxies and their supporting components. It ensures that the overlay remains healthy and efficient by implementing various network management solutions as needed.

\paragraph{{The infrastructure controller}} The infrastructure controller serves as the interface between the overlay controller and the underlying infrastructure manager. When proxies are deployed on managed infrastructure, this component assists the control plane in deploying and managing proxies effectively. It communicates directly with the overlay controller to manage resources that are assigned to the overlay.

\paragraph{{Configuration database}} The configuration database stores the configurations for all overlay components, \rev{specifying} how each component processes traffic. The overlay controller writes to the configuration database to update the overlay configuration and, when necessary, pushes the latest configurations to the overlay components to ensure they stay up to date.

\paragraph{{The observability and monitoring component}} The observability and monitoring component collects and stores data on overlay status, resource usage, and end-user telemetry, among other metrics. The overlay controller relies on this data to detect network and service issues and optimize the overlay's efficiency.

\paragraph{\rev{Control-plane operations}}
\rev{As discussed, Hermes defines the capabilities and operations that proxies should support, but it does not enforce a specific proxy implementation or deployment environment. In addition, Hermes only requires IP connectivity between overlay components and does not mandate how connectivity between them is provided (e.g., public Internet, private links, or cloud networking). This design keeps Hermes neutral to service providers' choices. End-user proxies are typically constrained by the user's device, operating system, and access network. In contrast, service providers have more flexibility in deploying the remaining overlay proxies. Depending on the overlay topology, scale, service requirements, and business requirements, they may run proxies on owned infrastructure, rented cloud resources, VMs, containers, serverless functions, or other deployment environments.}

\rev{Hermes distinguishes between end-user-hosted data-plane components and operator-controlled data-plane components. End users have full control over their own devices and may start or stop end-user proxies and assisting components at any time. When these components are running, they must authenticate with the overlay controller and receive service-provider policy and configuration through the control-plane configuration channel. In contrast, non-end-user data-plane components operate within an operator-controlled administrative domain and can be fully managed by the control plane.}

\rev{Accordingly, proxy discovery, trust establishment, and authentication differ across these two domains. End-user components become active overlay participants only when users run them and authenticate with the control plane, whereas operator-controlled components can be discovered and managed through automated control-plane mechanisms. Because end-user devices are outside the operator's administrative control, compromise of end-user components is treated as a higher risk than compromise of operator-controlled overlay components and is mitigated through overlay mechanisms. Operator-controlled infrastructure can also use a wider set of credentials and attestation mechanisms than end-user devices. In addition, while Hermes does not enforce a specific end-user deployment model, overlay operators can choose secure allocation and multi-tenant isolation mechanisms for operator-controlled proxies according to their infrastructure and business objectives. We discuss security assumptions and threat models in more detail in Appendix~\ref{app:security}.}

\subsubsection{\textbf{Overlay users}}
Having discussed the overlay components, we now define the users within the overlay.

\paragraph{{End-user}} An end-user device is the source of traffic passing through the overlay. Traffic may originate from IoT devices, mobile devices, or other systems. End-users have full control over their devices and can start or stop proxies and supporting components like any other user-space application. {End-user proxy configuration is provided by the service provider through the control-plane interfaces.} 

\paragraph{{Service providers}} Service providers are organizations or individuals that use the overlay to improve service delivery over the Internet and enhance end-user experience.

\paragraph{{Overlay operator}} {The overlay operator is the organization or individual that sets up the control plane and manages and deploys data-plane components and proxy configuration. The overlay operator is not necessarily the same as the service provider and may be a third party. In such cases, a third-party operator can manage an overlay used by multiple service providers. These third-party shared overlays are similar in business model to offerings such as Tailscale and Cloudflare Zero Trust.}

An overlay may also span multiple administrative domains, where independent operators collaborate to maintain a global-scale overlay while each controls the proxies within its own domain. In such cases, in the data plane, operators need to standardize the subset of HTTP headers they exchange, and in the control plane, they need cross-operator policy distribution and enforcement. Although multi-operator overlays are feasible, this paper focuses on the single-operator model, and we treat multi-operator overlays as out of scope.

%The components of the control plane are logically centralized. Each domain requires its own separate control plane, and the connection between controllers in autonomous domains can be established through custom interfaces or out-of-band communication. The next section illustrates an example of a multi-domain Hermes overlay deployment.

\subsection{How Hermes features address \rev{the service-delivery} challenges}

\rev{Section~\ref{related} described the service-delivery challenges that motivate Hermes. This section explains how the Hermes building blocks introduced above address these challenges. Specifically, we map Hermes features to end-to-end traffic management, backward compatibility, data-plane security and privacy, and adaptability, and we point to the use cases in Section~\ref{imp} that demonstrate these capabilities.}

\textbf{End-to-end traffic management:}
Hermes provides fine-grained, end-to-end IP layer traffic control and observation through an overlay of proxies that extends {from} end-user devices. Deploying a proxy on the end-user device allows the organization to extend its control and processing capabilities directly to the user's device. This enables early and end-to-end control over user traffic from the moment it is generated, offering several significant advantages. These include fine-grained and customer-specific traffic management, better load balancing and endpoint selection based on early insight into last-mile path characteristics, and the ability to perform local algorithms on user devices for improved traffic control.

Additionally, if the overlay is integrated with the service environment, such as a service mesh, end-user traffic can be traced from the moment it is generated, across the Internet, until it is consumed by the target services. This simplifies distributed tracing and root cause identification for poor service delivery over the Internet. Since Hermes handles IP traffic and is not limited to TCP and HTTP traffic, this end-to-end control enables Hermes to replace legacy VPNs in remote access use cases, as we show in Section \ref{IP}. 

\textbf{Backward compatibility:}
Proxies capture traffic and can process its header and payload if they are configured to. Hermes provides a diverse range of techniques to provide backward compatibility. Hermes proxies can provide automated service interface translation between different interface versions. In addition, Hermes proxies can provide various protocol translation and tunneling techniques. For example, a proxy can translate between HTTP/1.1 and HTTP/3, or it can tunnel UDP traffic over TCP and HTTP, although the tunneling techniques require proxies at each end to be configured in the same way. In this way, the application logic remains simple; application traffic uses fixed proxy endpoints and does not need to update the protocols it uses or modify its communication logic. Instead, the proxy component provides compatibility between application traffic and service {requirements}. This simplifies application development, as whenever network and service requirements change, proxy components can be automatically updated to reflect changes without modifying the application logic. In Section \ref{video}, we demonstrate how an RTP-based video streaming service can achieve reliable, lossless streaming in a noisy network by tunneling UDP over HTTP, without requiring any modifications to the client or server implementations.
 
\textbf{Data-plane security and privacy models:}
Hermes provides an end-to-end overlay that can span all network segments and controls traffic at the IP layer and above. Using assisting components that capture IP packets, tunnel them over HTTP, and carry IP metadata in HTTP headers, overlay proxies can enforce Layer 3 policies by evaluating the required information directly from headers. Layer 3 policies specify which entities may communicate at the IP layer. Proxies can also enforce Layer 4 policies, restricting which overlay entities may access specific port and protocol combinations, and Layer 7 policies, determining which application-level functions are accessible to an overlay entity. Hermes therefore enables unified security policy enforcement across network layers and segments. Please refer to our prior work \cite{farkiani_l3mesh} for an example. 

Because traffic processing begins on the end-user device rather than only at the service edge, Hermes can enforce a unified privacy model. By instantiating a virtual Layer 3 network from the end-user device to the service, Hermes can create a single, homogeneous Layer 3 policy domain across network segments; payloads can be encrypted on the device and decrypted only inside the target service process, enabling true end-to-end encryption for confidential and privacy-preserving computing. The data remains routable because it is tunneled inside HTTP with headers that carry identity and destination metadata, without exposing payloads to overlay nodes. In effect, traffic \rev{can be} treated under one unified security and privacy policy \rev{model} from its origin at the device, through ingress into the service provider domain, and onward to the target service. Please refer to Section~\ref{IP} for an example.

\textbf{Adaptable communication layer:}
Proxies have end-to-end control over traffic, and the overlay controller can update their configuration to respond to changes in network and service conditions. In this model, UDP, TCP, and HTTP proxying and tunneling are tools rather than constraints, since proxy behavior is policy-driven and reconfigurable. For example, proxies and assisting components can be configured with custom algorithms for severe network disruption that outperform the default reliability features of standard transport protocols. In Section \ref{intermittent} we show that while direct TCP communication is infeasible in highly unstable networks \cite{farkiani_mitigating_2023}, a Hermes overlay with tuned proxy configuration achieved a 100\% delivery rate using TCP as the underlying transport.

Additionally, Hermes can serve as an efficient communication layer for experimental networking architectures by providing capabilities that are often missing from basic Internet delivery. Many new architectures require features such as custom name resolution and reliable delivery to operate efficiently over the Internet. Hermes can offer these as common services, letting architectures leverage them without implementing their own native mechanisms. For example, an IP-based architecture can tunnel its traffic through Hermes, using HTTP headers for custom naming schemes and relying on Hermes for reliability. In Section~\ref{NDN}, we show that the Hermes can carry traffic between NDN nodes by mapping namespaces to HTTP headers while providing reliability and load balancing beneath NDN, reducing the need for NDN to rely on its native mechanisms for these functions.

\subsection{Practical considerations}
There are several additional aspects that must be considered before deploying the Hermes overlay. The challenges highlighted here represent the most significant obstacles to the wide-scale deployment of Hermes. As we continue to implement and use the architecture more extensively, new challenges will undoubtedly emerge.

\subsubsection{Motivations} 
A central question is: why would a service provider deliver its service via the Hermes overlay and assume the associated costs?

Service providers can use Hermes to deliver better service quality to end-users. \rev{Rather than being limited by underlying network conditions, especially last-mile path characteristics, Hermes gives providers additional control points at the end-user device and along the path to the provider ingress network. These control points allow providers to mitigate impairments, steer traffic, and reconfigure delivery behavior according to service-provider objectives.} In addition, end-to-end control over incoming traffic also enables global load balancing and end-to-end distributed tracing. \rev{Moreover}, because the overlay can handle traffic at Layers 3-7 in an integrated way across network segments and network layers, it lets providers implement complex business policies efficiently in a unified and integrated way without stitching together multiple tools for each layer and segment. Together, these capabilities improve service delivery and customer satisfaction while enhancing resource management, increasing profit, and reducing operational costs.

\subsubsection{Feasibility} Each device can support a limited number of ports, and users may simultaneously utilize multiple services. Two key questions arise for end-users: 1) How many proxies are needed on their devices? 2) Are service providers required to migrate all their users and services to Hermes?

To address the first question, it is important to note that a dependent proxy can be deployed on a separate machine and can simultaneously serve multiple applications (Figures \ref{fig:proxy2} and \ref{fig:proxy3}). However, managing different proxies may become a tedious task for end-users, potentially hindering the adoption of the architecture. {A potential solution, which we have already discussed in the paper, is to use a shared overlay offered by a third-party overlay provider, similar to how Tailscale~\cite{tailscale} or Cloudflare Zero Trust~\cite{cloudflare_zerotrust} operates today. In this model, an end-user runs a single dependent proxy instance provided by the third-party operator and configurable through the operator's control plane. Each service provider contracts separately with the overlay operator and is granted an interface to configure the proxy according to its needs, while the third-party operator manages conflicts across providers' configurations.} Therefore, instead of working with many proxy instances, end-user will only deal with one proxy {instance}.

As for the second question, it is important to highlight that Hermes acts as a service \textit{facilitator}. The answer to this question depends on the problem the service provider is facing, how many customers are affected, and what benefits Hermes can provide. Therefore, it is not necessary to migrate the entire user base to the Hermes architecture. A service provider might even selectively route the traffic of certain services through Hermes while allowing other services to bypass it. Consequently, Hermes can be deployed solely for users and services requiring special traffic handling, and it can coexist with any other already deployed solutions.
 
\subsubsection{Performance}\label{perfReview}
Does using Hermes improve or degrade performance? It depends. Performance measures must be defined in alignment with the system's goals. Using an overlay of proxies inevitably adds per-connection processing overhead compared to operating without a proxy. However, proxies can provide other benefits such as connection pooling and HTTP/2 or HTTP/3 stream multiplexing, which can reduce average end-to-end latency at scale, as we show in Section \ref{perf}.

Hermes requires HTTP-layer processing at dependent proxies to prepare end-user traffic for the overlay. Hermes is designed to be portable and to run on end-user devices, where we assume the service provider controls proxy configuration but not the device itself; therefore, client-side deployment options are limited. Once traffic leaves the end-user device, it is under the service provider's control and can be processed as desired. Because Hermes is HTTP-based, at least one other entity along the path must perform HTTP-layer processing to interpret and act on HTTP headers; other overlay hops, if they exist, can rely on transport-layer forwarding to reduce per-hop overhead. In Section~\ref{perf}, we measure tunneling overhead and show that HTTP encapsulation and decapsulation add at most 2 ms of latency per proxy pair traversal.

\rev{As demonstrated in Section~\ref{imp}, Hermes can also} improve other performance metrics. For example, in Section~\ref{video}, the overlay prevented packet loss and improved delivered video quality. In Section~\ref{intermittent}, where performance is defined as successful delivery ratio, the overlay increased this metric from 0\% to 100\%. We also showed that proxies can amortize connection setup overhead and, through connection pooling and multiplexing over HTTP/2, reduce end-to-end latency compared with no-proxy baselines. Hermes' ability to handle traffic end-to-end also creates opportunities for further improvements. For example, a Hermes overlay can act as a global load balancer that steers end-user traffic across the Internet to a specific service instance within the provider's environment, enabling global-scale resource allocation.

In general, the performance of a Hermes overlay depends on the overlay configuration, the processing performed at each proxy, and the proxy implementations. As with any new technology, adopters should evaluate performance gains, capital and operating expenditures, and potential disadvantages before deployment. Hermes is no exception.

\subsubsection{Security and Privacy}
Another set of questions arises around security and privacy. Notably, how a Hermes overlay is deployed and who operates it changes the security assumptions, threats, and mitigation strategies. For example, when the service provider is the overlay operator, end-users can trust overlay proxies, and the primary concerns shift to preventing overlay misuse, mitigating compromised clients, and preventing unauthorized Hermes-related header modifications. In contrast, when the overlay operator is a third party, additional concerns arise, including preventing confidentiality violations of Hermes traffic and mitigating multitenancy risks. We address these questions, along with others, and discuss the threat models and mitigation strategies for both deployment modes in Appendix~\ref{app:security}. \rev{Our discussion identifies assumptions, threats, and mitigation strategies for these deployment modes from an architectural perspective. We also elaborate that endpoint compromise remains a residual risk, although it can be mitigated through overlay mechanisms.}

\subsubsection{Management}
{Will using proxies improve network management tasks? In many deployments, yes. Proxies function as IP endpoints and tunnel terminators, and they control traffic from source to destination. They can emit fine-grained telemetry and participate in distributed tracing. A control plane integrated with observability can reconfigure proxies on demand to change routes and push new policies, allowing the network to respond quickly to service and network changes. Proxies can also run local fallback behavior when the control plane is unreachable to keep traffic flowing until control is restored.}

{Hermes makes management a first-class concern by introducing explicit observation and enforcement points (proxies and assisting components) along the end-to-end path, all configurable through the control plane. We highlight how Hermes improves service management below.
\begin{itemize}
\item {Telemetry:} Proxies and assisting components can export per-hop and end-to-end measurements such as latency, loss and retransmission indicators, throughput, and error rates across the layers the service targets (e.g., the telemetry exposed by Envoy~\cite{envoy_stats}).
\item {Actuators:} In Hermes, proxies use Hermes metadata carried in HTTP headers to perform header-based routing and request processing across the overlay. Other actuators include security mode selection (e.g., hop-by-hop encryption between proxies or end-to-end secure tunneling), and reliability and protection controls such as timeouts, retries, backoff policies, and rate limits. Because Hermes overlay \rev{proxies} are reconfigurable, the actions that a proxy performs for a given class of traffic can be updated over time.
\item {Policy enforcement:} Since Hermes can provide end-to-end control over Layers 3-7 traffic, it can enable unified policy enforcement across layers and network segments. Policy granularity in Hermes is service-defined and can range from per-identity and per-destination intent to coarser scopes such as geographic region. When an external policy engine is used (e.g., Open Policy Agent~\cite{opa_web}) as an assisting component, proxies provide traffic metadata as input, and the resulting decision can affect proxy processing (see the Video use case, Section~\ref{video}, for an example).
\item {Configuration:} Because Hermes supports reconfiguration, the configuration database should support versioning, and the control plane should verify that updates are applied successfully (for example, by checking proxy configuration state via configuration dumps or acknowledgments). If the control plane detects that an update fails or is only partially applied, it can revert the affected proxy to the last-known-good configuration. If a proxy becomes unavailable or cannot be recovered, the control plane can isolate it by updating the overlay path and instantiating a replacement. To reduce the risk of correlated misconfiguration, configuration and policy updates can be deployed gradually.
\end{itemize}}

\subsubsection{Deployment and \rev{operational} considerations}
\rev{As described in Section~\ref{components}, Hermes is flexible in the selection of deployment environments. In addition, a Hermes overlay can be deployed incrementally. An operator can enable capture and policy enforcement for a limited traffic class (e.g., a subset of users, destinations, or requests) while continuing to use existing mechanisms for the remaining traffic.}

Hermes has two distinct data-plane scaling domains. End-user dependent proxies run on unmanaged end-user devices, and thus scale naturally with the user population. In contrast, operator-managed overlay components are conventional software services that can be scaled horizontally. In particular, standalone proxies and service-side dependent proxies can be replicated to handle higher traffic volumes using standard mechanisms, for example DNS-based load balancing across replica pools, and Layer 4 or Layer 7 load balancing.

\rev{In addition to scaling, effective overlay management requires the overlay controller to make proxy-placement and routing decisions efficiently. A large body of work studies related problems in container placement and migration, VM/VNF placement, Kubernetes scheduling, and service-mesh optimization~\cite{kaur_container_2022, laghrissi_survey_2019, carrion_kubernetes_2022, saxena_copper_2025}. Prior work in these areas shows that placement and routing decisions can affect many metrics, including resource utilization, QoS, energy consumption, resilience, and cost. However, the metric to optimize depends on the infrastructure, service requirements, business requirements, and optimization objectives of the deployment. For example, a video-delivery service may prioritize latency, throughput, and continuity under failures, whereas a compute service may prioritize resource utilization, cost, and job-completion time. Therefore, Hermes, as a general-purpose architecture, does not prescribe a single universal optimization objective. Instead, once a service provider selects its deployment environment and optimization target, existing placement and routing techniques can be applied through Hermes' control plane. At the same time, Hermes introduces new optimization opportunities and tradeoffs because it coordinates proxy behavior, routing decisions, and policy enforcement in a unified way across network layers and segments. Quantifying these benefits and tradeoffs requires a large-scale deployment and evaluation, which is beyond the scope of this paper and left for future work.}

Control-plane scalability and operational complexity are driven primarily by the number of enrolled endpoints and proxies, the rate of policy updates, and telemetry volume. These parameters directly impact configuration distribution, rollout management, and monitoring overhead. Hermes can therefore reuse established control-plane scaling techniques from service meshes and managed proxies, such as tenant or region-scoped sharding, incremental configuration distribution, and telemetry aggregation or sampling. 

A Hermes deployment should also define explicit failure behavior that is service and traffic-class dependent. Proxy failures can be detected through health checks, and the control plane can instantiate replacement instances or reroute traffic. Transient control-plane unavailability can be handled via local fallback mechanisms and last-known-good configuration. Hermes implementations should treat overlay-path failures (e.g., an upstream proxy becoming unavailable) as a residual availability risk and mitigate them through redundancy, such as maintaining replica pools of upstream proxies (e.g., per region or destination prefix). By default, proxies should detect upstream unavailability via health checks (and/or control-plane signals) and shift traffic to other healthy upstream proxies in the pool. If no upstream proxy is available, behavior depends on policy: for traffic classes that require Hermes treatment, fail closed; for traffic classes where bypass is permitted, fall back to direct (non-overlay) delivery.

\section {Implementation and Evaluation}\label{imp}
\begin{table}[htbp]
\caption{Mapping of evaluation sections to challenges they address}
\label{tab:evalmap}
\renewcommand{\arraystretch}{1.1}
\begin{tabular}{|l|l|}
\hline
\multicolumn{1}{|c|}{\textbf{Use case}} & \multicolumn{1}{c|}{\textbf{Challenges}} \\ \hline
Video (Section \ref{video}) & \begin{tabular}[c]{@{}l@{}}End-to-end traffic management\\ Backward compatibility\\ Adaptable communication layer \end{tabular}  \\\hline
Intermittent (Section \ref{intermittent}) & \begin{tabular}[c]{@{}l@{}}End-to-end traffic management\\ Adaptable communication layer\end{tabular} \\ \hline
IP (Section \ref{IP}) & \begin{tabular}[c]{@{}l@{}}End-to-end traffic management\\ Data-plane security and privacy models \\ Adaptable communication layer \end{tabular} \\ \hline
NDN (Section \ref{NDN}) & Adaptable communication layer \\ \hline
\end{tabular}
\end{table}

\rev{This section demonstrates the capabilities of the Hermes architecture by prototyping and evaluating its performance across several use cases. We implemented the necessary control-plane components and built data-plane prototypes using available open-source tools. This approach shows that the Hermes data plane is feasible and can be assembled easily from existing tools. Although this implementation choice limits some optimization opportunities and affects the absolute performance results, it supports our goal of demonstrating Hermes' data-plane feasibility. Therefore, the measurements reported here should be viewed as baseline results that can be improved with implementations more directly tailored to the Hermes architecture. In addition, the evaluations in this section primarily focus on data-plane features and mechanisms, consistent with this paper's goal of specifying the capabilities that proxies should provide and how they operate.}

In this section, we handpicked four introductory use cases to show Hermes can address challenges listed in Section \ref{related}. For each use case, we deploy an overlay topology over an underlying physical network. Dependent proxies and standalone proxies are abbreviated as \textbf{\textit{DP}} and \textbf{\textit{SaP}} as before, respectively. Configuration details are briefly discussed here, with further information provided in the appendix. Mapping of each use case to challenges is shown in Table \ref{tab:evalmap}.

\paragraph{Infrastructure setup.}
The hardware network testbed used in the experiments, except for the IP use case (Section \ref{IP}), is a remotely accessible testbed for networking researchers and educators \cite{onl}. We utilized software routers and switches running on separate Ubuntu 20.04.6 machines with 1 Gbps Ethernet interfaces as the physical network components. Link bandwidth and packet drop rates were controlled using the Ubuntu tc utility \cite{ubuntu_tc}, with bandwidth limited to 10 Mbps to facilitate easy comparison across different implementations. The end-user was a machine within the remote topology, also running Ubuntu 20.04.6 and Docker. For the IP use case, we used servers located in Warsaw and Chicago, with all systems running Ubuntu 22.04.4, {and the experiment was performed over the public Internet without enforcing any additional controls.}

\paragraph{Proxies}
There are many proxy implementations available, but we required a proxy that can process TCP, UDP, and HTTP headers, supports reconfiguration, supports creating tunnels and capturing IP packets, and is capable of tunneling UDP, TCP, and IP packets over HTTP. Unfortunately, no single proxy implementation meets all of these requirements. Even advanced, production-grade proxies like Linkerd \cite{linkerd} and Envoy \cite{envoy_web} lack some of the necessary features. Ultimately, we selected Envoy, as it supports UDP-over-HTTP tunneling, although this feature is still in development. Envoy is a high-performance proxy that provides both layer 4 and layer 7 filtering, load balancing, and comprehensive statistics, among other features \cite{farkiani_service_2022}. Envoy can be configured to use a set of filters to process traffic and control routing. These filters can process HTTP headers, interact with external entities for authorization, create additional tunnels, and perform protocol translation. Envoy also supports custom plugins, which can act as assisting components in the architecture. If an assisting component is complex, such as tunneling device or an advanced cache, it can be deployed as a separate process that communicates through the proxy.

For Ubuntu systems, we use the envoy-dev container image with digest 10d0784b9cdb \cite{envoy_dev}. Additionally, we developed Android and iOS proxy clients, screenshots shown in Figures \ref{fig:mobileAndroid} and \ref{fig:mobileIOS}, using Envoy's mobile library \cite{envoy_mobile}. These clients, with limited HTTP filtering capabilities, accept traffic from other applications on the phone and route it over the overlay. All proxy instances, including the mobile proxies, use an initial configuration to boot up and then receive their full configuration from the control plane.

\paragraph{Control plane}
We developed an overlay controller \rev{in Go using Envoy's xDS APIs to dynamically configure Envoy proxies \cite{envoy_control_plane}}. The controller connects to an Apache CouchDB v3.3.3 \cite{couchdb_web} database, which stores the configuration of overlay proxies. \rev{Configurations can be updated directly through the database interface. The controller polls the database} every five seconds and updates the overlay when configuration changes occur. The controller also supports on-demand update requests to \rev{update} specific nodes. Initial proxy configurations contain unique IDs, which are sent to the controller over gRPC \cite{xds}, and the controller responds with the corresponding configuration. The connection between overlay proxies and the controller is TLS-encrypted, and proxies can be authenticated using passphrases. The controller can push new configurations to initiate a hot restart \cite{hotrestart} in response to changes in network or service conditions. Although we designed an infrastructure controller using Ansible \cite{ansible} and other relevant packages, it was not used in the following use cases because this paper focuses on demonstrating the data-plane benefits of Hermes in addressing various challenges.

\paragraph{Assisting components}
For policy-based routing and authorization, we use Open Policy Agent (OPA) \cite{opa_web} and its Envoy plugin as assisting components. OPA \rev{uses the declarative policy language Rego} to define policies based on input data provided by a collocated Envoy instance. In our implementation, the OPA instance makes an allow/deny decision, returning it along with an HTTP header. We append this header to allowed traffic and use it in policy-based routing decisions. As the Envoy proxy versions we used could not create tunnel devices and capture IP packets, unlike other proxies such as \cite{singbox}, we developed a custom assisting component. This component creates tunnel devices, performs UDP encapsulation and decapsulation, and supports split tunneling {by directing traffic for each destination to a controller-selected UDP port on the local proxy.}

\subsection{Providing \rev{compatibility with legacy applications and protocols}}\label{video}
In this section, we show how Hermes can address backward compatibility challenges without requiring modifications to existing client or server applications. Because proxies act as IP endpoints, they can perform protocol translation (e.g., HTTP/1.1 to HTTP/3) or API translation (e.g., converting old request formats into new ones) transparently to clients. Hermes can also enhance functionality without traffic translation. For example, if end-users are limited to unreliable protocols, Hermes can provide reliability \textit{without} translating the protocols into a different reliable protocol. In this way, end-users benefit from reliability while continuing to use their unmodified applications. The only change required is directing traffic to a local Hermes proxy endpoint instead of the original destination.

\subsubsection{Problem (Video Use Case)}
As a concrete example, consider video streaming over the RTP protocol \cite{schulzrinne_rtp_2003}. RTP over UDP is widely used on the Internet, for instance, by Zoom \cite{michel_enabling_2022}, Vimeo \cite{vimeo_rtp}, and others \cite{rtp_ffmpeg, blair_how_2024}. Although many streaming services use reliable protocols, RTP does not guarantee delivery or enforce packet ordering \cite{schulzrinne_rtp_2003}. Most implementations, therefore, add their own application-layer recovery mechanisms. For example, Zoom retransmits lost packets up to two times \cite{michel_enabling_2022}, whereas FFmpeg provides no recovery mechanisms out of the box.

A customer using an RTP-based service over UDP reports degraded video quality. Implementing a recovery layer would normally require modifying either the client or the server, an impractical solution if customers resist updates or if the streaming infrastructure is closed source. Additionally, customers in different regions may have different reliability requirements. For example, Zoom's two retries might suffice for one customer, but another may need more retries depending on network conditions. We therefore ask: can Hermes enable reliable video delivery over RTP \textit{without} any modifications to existing client or server implementations?

\begin{figure}[htbp]
\centering
\includegraphics[width=0.8\columnwidth]{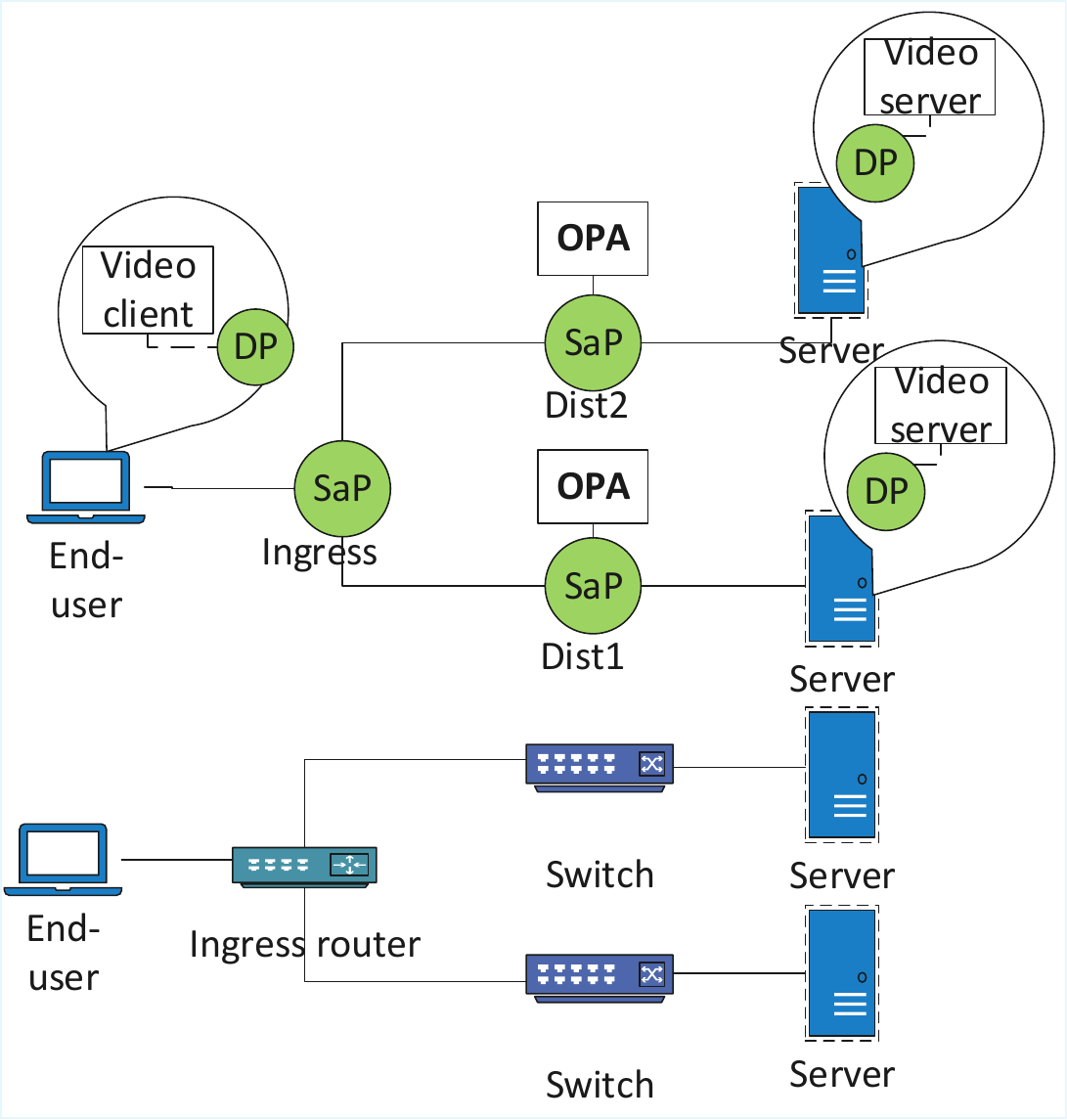}
\caption{Video use case topology. Overlay is above.}
\label{fig:topo_video}
\end{figure}

\begin{table}[ht] 
    \centering 
    \begin{minipage}{0.40\textwidth}
        \centering
        \caption{Video use case results. \textit{Num}: average number of, \textit{S.D.}: standard deviation.}
    \renewcommand{\arraystretch}{1.1}
		\begin{tabular}{|l|ll|ll|}
		\hline
		\multicolumn{1}{|c|}{\multirow{2}{*}{\textbf{Solutions}}} & \multicolumn{2}{l|}{\textbf{\begin{tabular}[c]{@{}l@{}}Macroblock\\ errors\end{tabular}}} & \multicolumn{2}{l|}{\textbf{\begin{tabular}[c]{@{}l@{}}Processed\\ frames\end{tabular}}} \\ \cline{2-5} 
		\multicolumn{1}{|c|}{} & \multicolumn{1}{c|}{Num} & \multicolumn{1}{c|}{S.D.} & \multicolumn{1}{c|}{Num} & \multicolumn{1}{c|}{S.D.} \\ \hline
		Overlay -1 min & \multicolumn{1}{r|}{0.00} & \multicolumn{1}{r|}{0.00} & \multicolumn{1}{r|}{1800.00} & \multicolumn{1}{r|}{0.00} \\ \hline
		Overlay -2 min & \multicolumn{1}{r|}{0.00} & \multicolumn{1}{r|}{0.00} & \multicolumn{1}{r|}{3600.00} & \multicolumn{1}{r|}{0.00} \\ \hline
		Direct -1 min & \multicolumn{1}{r|}{26.74} & \multicolumn{1}{r|}{5.26} & \multicolumn{1}{r|}{1781} & \multicolumn{1}{r|}{10.28} \\ \hline
		Direct -2 min & \multicolumn{1}{r|}{55.88} & \multicolumn{1}{r|}{7.02} & \multicolumn{1}{r|}{3561.47} & \multicolumn{1}{r|}{12.57} \\ \hline
		\end{tabular}
        \label{tab:video}
    \end{minipage}\hfill 
\end{table}

\subsubsection{Solution}
We deploy a dependent proxy on both the customer's devices and servers. When the server starts streaming RTP, it sends the UDP packets to the dependent proxy's configured port. The dependent proxy then tunnels the UDP packets over an HTTP/2 tunnel, sending them to the customer's dependent proxy. The customer's dependent proxy receives the HTTP packets, terminates the HTTP tunnel, decapsulates the UDP packets, and delivers them to the client. This process requires no changes to the client or server implementations. Since HTTP/2 runs over TCP, it can recover from small transmission errors and packet losses. Although this approach adds tunneling overhead and TCP-related latencies, the end-user receives video without lost frames.
Additionally, Hermes proxies can be configured dynamically for retry options. Since these proxies sit in the network path, their reliability settings can be easily tuned to match specific network conditions and end-user requirements. This flexibility avoids one-size-fits-all solutions and instead provides customized reliability and improves performance across diverse environments without requiring any application changes.
\subsubsection{Implementation}
We used ffmpeg \cite{ffmpeg} as both client and server applications. We used the topology shown in Figure~\ref{fig:topo_video}, and software details are shown in Tables~\ref{tab:Videoconfig} and~\ref{tab:ffmpeg}. We introduced 1\% loss along the path with Ubuntu tc to investigate the effect of packet loss on streaming. We developed a video server that accepts HTTP connections. It passes requests to its local ffmpeg application to stream the requested file back to the IP address of the client that requested it over RTP.

The video client sends requests in the format \textit{http://VIDEO/filepath} with an authorization token to its local dependent proxy endpoint. The "VIDEO" portion is part of the overlay custom names, and proxies handle "VIDEO" to route traffic to the appropriate destination based on user identity and access level. The dependent proxy tunnels the requests over an HTTP/2 TLS-encrypted link to the ingress proxy of the video provider's network. Once received, the ingress proxy decrypts the requests, processes the HTTP headers, and forwards them to one of the distributor proxies. The distributor proxies then forward the request to the local OPA instance to verify the user's access level.

The local OPA instance authorizes the user based on the token and the configured policy. If authorized, the request is forwarded to the dependent proxy at the video server. The server's dependent proxy delivers the request to the local video server running ffmpeg. The video server only sees the IP address of its local dependent proxy and instructs ffmpeg to stream back to that proxy. The dependent proxy at the video server tunnels the RTP-over-UDP stream using CONNECT-UDP \cite{CONNECT_UDP} and sends it along the same path. When the tunneled datagrams reach the end-user's dependent proxy, it decapsulates and delivers the UDP packets to the client ffmpeg. This ffmpeg instance is configured to receive UDP on its local proxy port and parse the payload as RTP.

We compare the overlay performance with direct requests to the video servers using RTP. The direct requests, shown in Figure \ref{fig:topo_video}, followed the format \textit{http://IP:Port/filepath}, along with a token that the video server ignored. The video server instructed ffmpeg to stream the video directly to the end-user using RTP over UDP. We used a video from NASA \cite{nasa} with a duration of three minutes and extracted two MP4 videos of one and two minutes. We conducted 30 iterations of the experiment, where both files were requested in each run. After retrieving the files, we analyzed them using ffmpeg and counted the number of processed frames and macroblock errors as performance and quality measures. The results are shown in Table \ref{tab:video}.

The Hermes overlay uses HTTP tunneling, which benefits from the reliability of TCP. In contrast, the direct connection uses an unreliable RTP-over-UDP connection. As shown in Table \ref{tab:video}, with the overlay approach, the user received the video without any errors. However, with the direct connection, we observed frame losses and macroblock errors. Specifically, we recorded an average loss of 19 and 38.53 frames for the 1- and 2-minute video files, respectively. We also observed 26.74 and 55.88 macroblocks in error for the same files.

Compared to the direct approach, the tunneling approach increases processing overhead, and TCP can introduce head-of-line blocking and added jitter. However, with the buffering and timeout configurations that are equal for both direct and overlay approaches, packets arrive within the configured playout deadline at the receiver. 

\subsubsection{Discussion} Using HTTP tunneling within the Hermes overlay mitigated loss and improved user experience under noisy network conditions. This lets users who cannot or do not wish to upgrade their clients obtain reliable delivery while continuing to use RTP over UDP. Proxies need not translate RTP into another streaming protocol, they simply carry the UDP datagrams inside a reliable HTTP tunnel, so no changes to client or server applications are required. The overlay also adds practical features such as global load balancing, authorization, encryption, and service discovery. Users address "VIDEO" rather than a specific server name, and proxies route requests to the correct destination based on identity and policy.

Tunneling RTP over HTTP for reliability has limitations. It adds processing overhead, and TCP-based tunnels suffer from head-of-line blocking. UDP-based tunneling (e.g., HTTP/3 over QUIC) mitigates head-of-line blocking, but severe packet loss or repeated disconnections along the path can still disrupt any end-to-end transport, preventing delivery \cite{farkiani_mitigating_2023}. This limitation applies to both QUIC and TCP. In the next section, we demonstrate how Hermes enables successful delivery under extreme disruption and allows TCP to achieve a 100\% delivery ratio.

\subsection {Supporting communication in unstable networks}\label{intermittent}
This section examines how Hermes can act as an adaptable communication layer. Reliable transport protocols react to communication failures according to their designs. However, when multiple links fail along the path, and in extreme cases no end-to-end path exists at a given time, end-to-end transports fail simply because no continuous route is available to deliver traffic. Such conditions arise in Delay Tolerant Networks (DTNs), and there is a substantial literature on protocols and approaches for these settings \cite{ginzboorg_message_2014, fall_dtn_2008}. Adopting a DTN stack, however, typically requires changes to clients, servers, or network nodes.

Our goal is to provide DTN-like resilience without modifying applications. Hermes enables TCP to deliver data by segmenting the path with proxies. When proxies fail to transfer data to the upstream destination, they retry based on configured retry policies. The service provider configures retry policies, including retry limits and backoff schedules, through the control plane based on service requirements and network conditions. When the control plane is unreachable, proxies can fall back to local retry policies. This ensures that retry behavior continues autonomously until coordination is restored.

\begin{table}[htbp]
    \centering
    \caption{Download time (s). Data is {reported as average download time} (standard deviation). Base average values are shown {in the column headers.}}
    \renewcommand{\arraystretch}{1.1}
    \begin{tabular}{|l|ll|ll|}
    \hline
    \multicolumn{1}{|c|}{\multirow{2}{*}{\textbf{Solutions}}} & \multicolumn{2}{c|}{\textbf{\begin{tabular}[c]{@{}c@{}}1MB\\ (Base=1.02)\end{tabular}}} & \multicolumn{2}{c|}{\textbf{\begin{tabular}[c]{@{}c@{}}5MB\\ (Base=4.85)\end{tabular}}} \\ \cline{2-5} 
    \multicolumn{1}{|c|}{} & \multicolumn{1}{c|}{1st} & \multicolumn{1}{c|}{2nd} & \multicolumn{1}{c|}{1st} & \multicolumn{1}{c|}{2nd} \\ \hline
    \multirow{2}{*}{Overlay} & \multicolumn{1}{l|}{\multirow{2}{*}{\begin{tabular}[c]{@{}l@{}}115.82\\ (36.18)\end{tabular}}} & \multirow{2}{*}{\begin{tabular}[c]{@{}l@{}}133.69\\ (57.42)\end{tabular}} & \multicolumn{1}{l|}{\multirow{2}{*}{\begin{tabular}[c]{@{}l@{}}601.49\\ (110.21)\end{tabular}}} & \multirow{2}{*}{\begin{tabular}[c]{@{}l@{}}618.75\\ (101.46)\end{tabular}} \\
     & \multicolumn{1}{l|}{} &  & \multicolumn{1}{l|}{} &  \\ \hline
    \begin{tabular}[c]{@{}l@{}}Overlay w/\\ CONN\end{tabular} & \multicolumn{1}{l|}{\begin{tabular}[c]{@{}l@{}}101.67\\ (44.51)\end{tabular}} & \begin{tabular}[c]{@{}l@{}}86.96\\ (46.68)\end{tabular} & \multicolumn{1}{l|}{\begin{tabular}[c]{@{}l@{}}567.3\\ (134.73)\end{tabular}} & \begin{tabular}[c]{@{}l@{}}498.28\\ (178.32)\end{tabular} \\ \hline
    \end{tabular}
    \label{tab:intermittentDownload}
\end{table}

\begin{figure}[htbp]
    \centering
    \includegraphics[width=0.95\columnwidth]{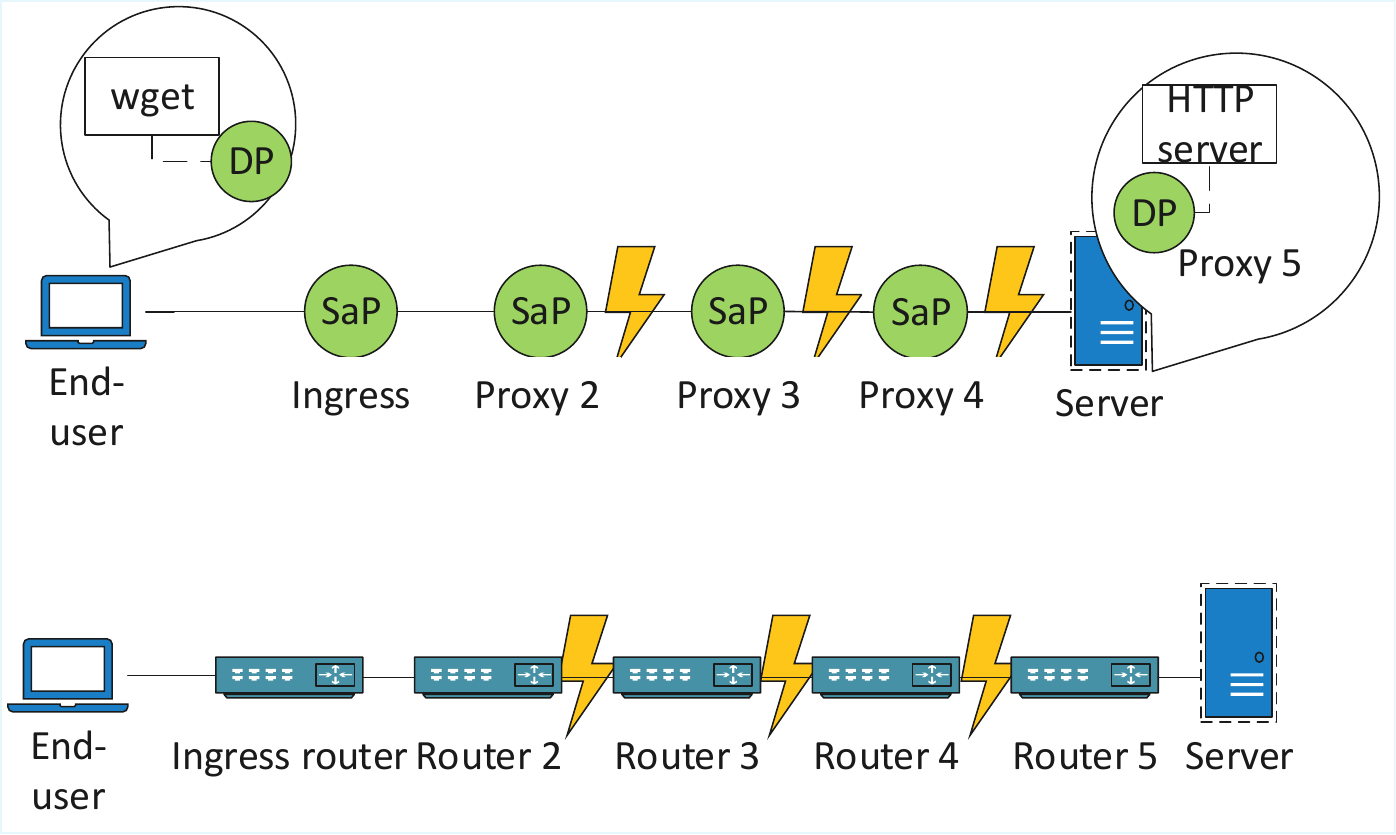}
    \caption{Intermittent use case topology. Overlay is above.}
    \label{fig:topo_intermittent}
\end{figure}

\subsubsection{Problem (Intermittent use case)}
A user wants to securely download a file from an HTTP web server over TCP. The network between the user and the server is severely unstable, for example when nodes are mobile and there is no stable wireless coverage for all nodes. We consider a path with three lossy links where, in each time slot, exactly one link is up and the others are down. Because no continuous end-to-end path exists at any instant, a direct TCP connection between client and server cannot make progress, as shown in \cite{farkiani_mitigating_2023}.
\subsubsection{Solution}
Proxies are IP endpoints that can initiate and terminate TCP connections on each segment and control them according to local conditions. We deploy an overlay along the path between the client and the web server, with a proxy on each side of each unstable link. Each proxy retries until it successfully delivers data to the next proxy. For the client, we use wget \cite{wget} to take advantage of its retry options. For end-to-end security, there are two options:
\begin{itemize}
\item \textbf{Hop-by-hop TLS:} Each adjacent proxy pair establishes a TLS connection. This allows caching at each node because encryption terminates at the proxies. It requires the user to trust the overlay and fits deployments where the service provider also operates the overlay.
\item \textbf{End-to-end HTTPS via CONNECT:} A proxy processes the HTTP CONNECT request and creates a TCP tunnel to the target website, then blindly forwards encrypted bytes so that security remains end-to-end between the user and the web server. This is appropriate when the service provider and the overlay operator are different.
\end{itemize}

\subsubsection{Implementation}
The topology is shown in Figure~\ref{fig:topo_intermittent}. Software configurations are shown in Table~\ref{tab:Intermittentconfig}, and retry parameters are shown in Table~\ref{tab:intermittentRetry}. The links between Proxies 2-5 were intermittent; each link was only connected for one second, and no two links were connected simultaneously. This ensures that no end-to-end path is available at any given time. We used Ubuntu tc and set packet drop to 100\% to simulate link disconnection.

We used wget with a timeout of 60s to download 1MB and 5MB files. Wget tried 10 times and continued from whatever was downloaded before. The web server configuration is shown in Table \ref{tab:nginxConfig}. We compare four scenarios: 1) direct connection to servers with no intermittent link as the base measurement (Base), 2) direct connection to the servers in the presence of three intermittent links, shown in the bottom of Figure \ref{fig:topo_intermittent}, (Base-Intermittent), 3) hop-by-hop TLS connection through overlay proxies in the presence of intermittent links with caching at all overlay nodes except the ingress (Overlay), and 4) No caching and using CONNECT at client that was processed at Proxy 2, after that in Proxies 3-5 we only used TCP listeners (Overlay w/CONN).

In the "Overlay" scenarios, client requests were in the form of \textit{http://UNSTABLE/filepath}, and they were sent to the dependent proxy. The dependent proxy then encrypted and forwarded the requests to the ingress proxies. Proxies 2-5 were configured with an HTTP retry policy \cite{http_route}, as shown in Table \ref{tab:intermittentRetry}. In the "Overlay w/CONN" scenario, the client sent requests in the form of \textit{https://UNSTABLE/filepath}. While Proxy 2 followed the same retry policy as in the Overlay scenario, Proxies 3-5 applied a retry policy with max\_connect\_attempts=20 \cite{tcp_retry}, as they were TCP forwarders. The ingress proxy processed "UNSTABLE" and then forwarded the data.
For the "Base" and "Base-Intermittent" scenarios, the client used wget with the same retry parameters and sent requests in the format \textit{https://IP:PORT/unstable/filepath}, which were routed through Routers 2-5. The experiment consisted of 20 batches, each with two iterations. Each batch started with the intermittent pattern, and random sleep was often used to minimize the influence of the pattern stage on the experimental results. Performance was measured by the download time, the duration it took for wget to complete, measured using the \textit{time} \cite{time} command, and the success ratio, calculated as the number of successful downloads divided by the total number of iterations.

The average success ratios for Base and Base-Intermittent are 100\% and 0\% for both files, respectively. The direct TCP connection (Base-Intermittent) always fails in the presence of three intermittent links. In contrast, for the Overlay scenario, we observed a success ratio of 100\% for both files. The Overlay w/CONN success ratios for the 1MB and 5MB files are 80\% and 32.5\%, respectively. Although the success ratio is not 100\%, this shows w/CONN can also successfully deliver traffic. The lower success ratio occurred because wget exited with the error "Unable to establish SSL connection" before completing its configured number of retries, which is a known issue \cite{wget_bug}. Below we discuss why we did not wrap wget in a script to retry after receiving this {early termination}.

Table \ref{tab:intermittentDownload} shows the average download time for all scenarios except Base-Intermittent, as it always fails. For the Overlay and Overlay w/CONNECT scenarios, we averaged the first and second iterations separately. As shown, both overlay implementations show a significant increase in download time. In addition, caches were rarely used in the Overlay scenario. This was mainly due to partial data being downloaded over multiple iterations.

\subsubsection{Discussion} In our prototype, wget exited before exercising its configured retries due to a known {issue}. We did not wrap it in a supervisor script to re-launch retries. The observed 32.5\% success rate therefore reflects premature termination by the client rather than a limit of the overlay. If a lightweight wrapper had restarted wget when it exited early, the success rate would have {likely} reached 100\% under the same conditions. 

We did not wrap wget in a script to auto-retry because wget may not be used in practical deployments. We chose it for its convenient retry parameters. The key point is that proxies can be coupled with assisting components. One such component can be a small retrier, similar to wget, that issues idempotent requests until success, with parameters such as timeouts and retry counts configured directly by the control plane. Another component can provide advanced caching to augment Envoy's basic HTTP cache. With these assisting components, users can run any client application, even if it lacks native retry support, while the overlay supplies the required resilience. This highlights the role of assisting components, and the next section investigates their applications.

\subsection{Overcoming network degradation through adjusting overlay configuration}\label{IP}
This section examines how Hermes can extend an end-to-end Layer 3 overlay across network segments. The overlay can provide policy-based routing, and we show how quickly the overlay can be reconfigured to mitigate increases in latency. These updates occur without disrupting underlying IP routing or requiring user actions, and they reduce end-to-end latency by updating network paths.

\subsubsection{Problem (IP use case)}
An organization allows employees to connect remotely to internal servers to perform daily tasks and access services. The organization operates in two regions, one in Poland and one in the United States. Each remote developer has access to three IP prefixes: 10.0.0.1/32 for a personal server, 10.0.0.2/32 for a shared service such as internal mail, and 128.252.0.0/16 for general internal servers. Each user is assigned a dedicated virtual network slice, so two users can target the same IP address, for example 10.0.0.1, while the physical destination might differ and is resolved based on the user's identity. The user to destination mapping is unique.

The organization needs a solution that enforces end-to-end policies and supports seamless updates to endpoints and address spaces. In our scenario, a user who lives in Poland moves to the United States but continues to connect to an ingress and a personal server located in Poland. The overlay controller and the user detect a latency increase. The policy requires end-to-end latency to remain below a threshold. If this threshold is violated, the control plane should adjust overlay configuration to improve performance. In this case, the control plane chooses to relocate both the user's ingress and the user's personal server (10.0.0.1) to the United States.

A traditional setup can be built with WireGuard as the VPN client and server. On the server, traffic from each user can be steered to a specific routing table to resolve destinations according to the user's assigned IP space, which provides per-user isolation. Additional configuration, such as source NAT, may be required for a complete solution. This approach has several limitations:
\begin{itemize}
\item WireGuard and other standard VPN protocol traffic can be readily identified by middleboxes and may be blocked by ISPs along the path. 
\item Updating the binding between a private IP and the physical host is operationally heavy. It typically requires changes to routes, NAT, and firewall rules on the VPN gateway. Time-based or context-based steering, for example moving {10.0.0.1} after working hours, becomes brittle and slow to roll out. 
\item With traditional VPNs, authentication, authorization, policy enforcement, and cryptography are limited to what the protocol and its implementation provide. The organization may need finer control: some users or destinations may not require IP-level encryption because applications already use TLS, others may prefer TLS between intermediary nodes, another group may need obfuscation to avoid ISP blocking. These choices are constrained in flexibility under a fixed VPN protocol.
\end{itemize}

\subsubsection{Solution}
In this use case, we capture traffic at Layer 3 to create a virtual Layer 3 service over the Internet. The approach is to capture IP traffic using a tunnel device, encapsulate it in HTTP, attach HTTP headers that encode destination and user identity, and route traffic according to those headers. This way, we treat the destination as an HTTP header and rely on the overlay's Layer 7 forwarding rules to determine packet destination. Additionally, changing the binding between a private IP and a physical host only requires updating HTTP forwarding rules inside overlay proxies. Therefore, IP traffic encapsulated in HTTP is routed across network segments by processing its HTTP headers until it reaches the destination. After decapsulation, it is passed to the tunnel device at the destination to deliver it to the target process.

Because Envoy {did not support creating tunnel devices at the time of our evaluation}, we {implemented} an assisting component that interfaces with the proxy over UDP. Proxies listen on UDP endpoints. Tunnel devices on end hosts capture IP packets, encapsulate them in UDP, and deliver them to a proxy port. In our prototype, this results in two nested tunnels: IP-in-UDP and UDP-in-HTTP. The optimal design would use a single IP-in-HTTP tunnel from end-user to destination when proxy software supports tunnel interfaces.
Therefore, two features are required:
\begin{itemize}
\item Encapsulation of IP traffic in UDP.
\item Split tunneling that maps each destination network to a specific port on the user's dependent proxy that is configured to accept traffic to that destination network.
\end{itemize}

We implemented an assisting component that provides IP-in-UDP encapsulation and split tunneling. Traffic arriving on each DP port is tagged with headers that identify the destination network. The DP then encapsulates the UDP packets inside HTTP and appends user-related headers to bind destination addresses to the user identity. It forwards the request to the next proxy, which consults HTTP headers to select the next hop. This continues until the chain reaches a proxy that decapsulates UDP and hands it to an assisting component. The assisting component then decapsulates IP and delivers the packets to the final destination.

If endpoint applications already use encryption, for example HTTPS or SSH, the overlay does not need to add link encryption between proxies. Otherwise, the overlay can selectively enable TLS between specific proxy pairs. The overlay also performs policy-based routing and authorization using HTTP headers. When the control plane detects that user latency exceeds a configured threshold, it updates proxy configuration to steer traffic to a closer server. The user continues to use the same IP address, and no user action is required. The destination address remains fixed while routing decisions are handled by the overlay proxies.
 
\subsubsection{Implementation}
\begin{table}[t]
\caption{Initial latency results from {User 2}. RTT is in ms and S.D. is standard deviation}
\label{tab:IPinitial}
\renewcommand{\arraystretch}{1.1}
\begin{tabular}{|llllll|}
\hline
\multicolumn{6}{|c|}{\textbf{Poland}} \\ \hline
\multicolumn{2}{|l|}{Latency} & \multicolumn{2}{l|}{\begin{tabular}[c]{@{}l@{}}Latency w/ UDP tunnel \\ \& w/o HTTP tunnel\end{tabular}} & \multicolumn{2}{l|}{\begin{tabular}[c]{@{}l@{}}Latency w/ UDP tunnel \\ \& w/ HTTP tunnel\end{tabular}} \\ \hline
\multicolumn{1}{|l|}{RTT} & \multicolumn{1}{l|}{S.D.} & \multicolumn{1}{l|}{RTT} & \multicolumn{1}{l|}{S.D.} & \multicolumn{1}{l|}{RTT} & S.D. \\ \hline
\multicolumn{1}{|l|}{129.73} & \multicolumn{1}{l|}{5.8} & \multicolumn{1}{l|}{131.23} & \multicolumn{1}{l|}{6.31} & \multicolumn{1}{l|}{134.27} & 22.92 \\ \hline
\multicolumn{6}{|c|}{\textbf{US}} \\ \hline
\multicolumn{2}{|l|}{Latency} & \multicolumn{2}{l|}{\begin{tabular}[c]{@{}l@{}}Latency w/ UDP tunnel\\ \& w/o HTTP tunnel\end{tabular}} & \multicolumn{2}{l|}{\begin{tabular}[c]{@{}l@{}}Latency w/ UDP tunnel \\ \& w/ HTTP tunnel\end{tabular}} \\ \hline
\multicolumn{1}{|l|}{RTT} & \multicolumn{1}{l|}{S.D.} & \multicolumn{1}{l|}{RTT} & \multicolumn{1}{l|}{S.D.} & \multicolumn{1}{l|}{RTT} & S.D. \\ \hline
\multicolumn{1}{|l|}{21.71} & \multicolumn{1}{l|}{37.85} & \multicolumn{1}{l|}{21.45} & \multicolumn{1}{l|}{7.55} & \multicolumn{1}{l|}{24.05} & 23.9 \\ \hline
\end{tabular}
\end{table} 

\begin{figure*}[htbp]
    \centering
    % First row: two figures side by side
    \begin{subfigure}[b]{0.48\textwidth}
        \centering
        \includegraphics[height=0.25\textheight]{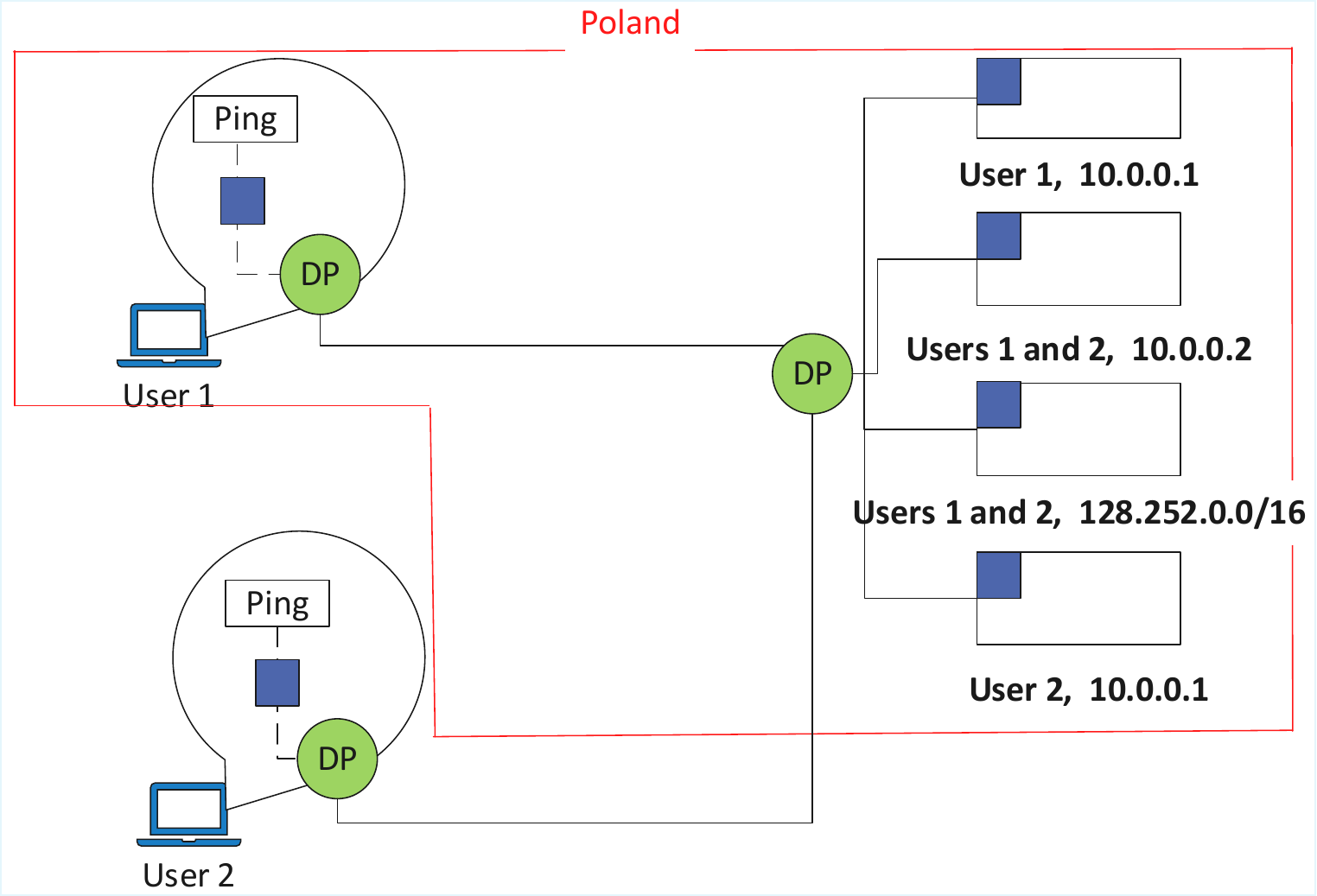} 
        \caption{{The overlay before the update: User 2 is in the US, while others are in Poland.}}
	\label{fig:IP1}
    \end{subfigure}
    \hfill
    \begin{subfigure}[b]{0.48\textwidth}
        \centering
        \includegraphics[height=0.25\textheight]{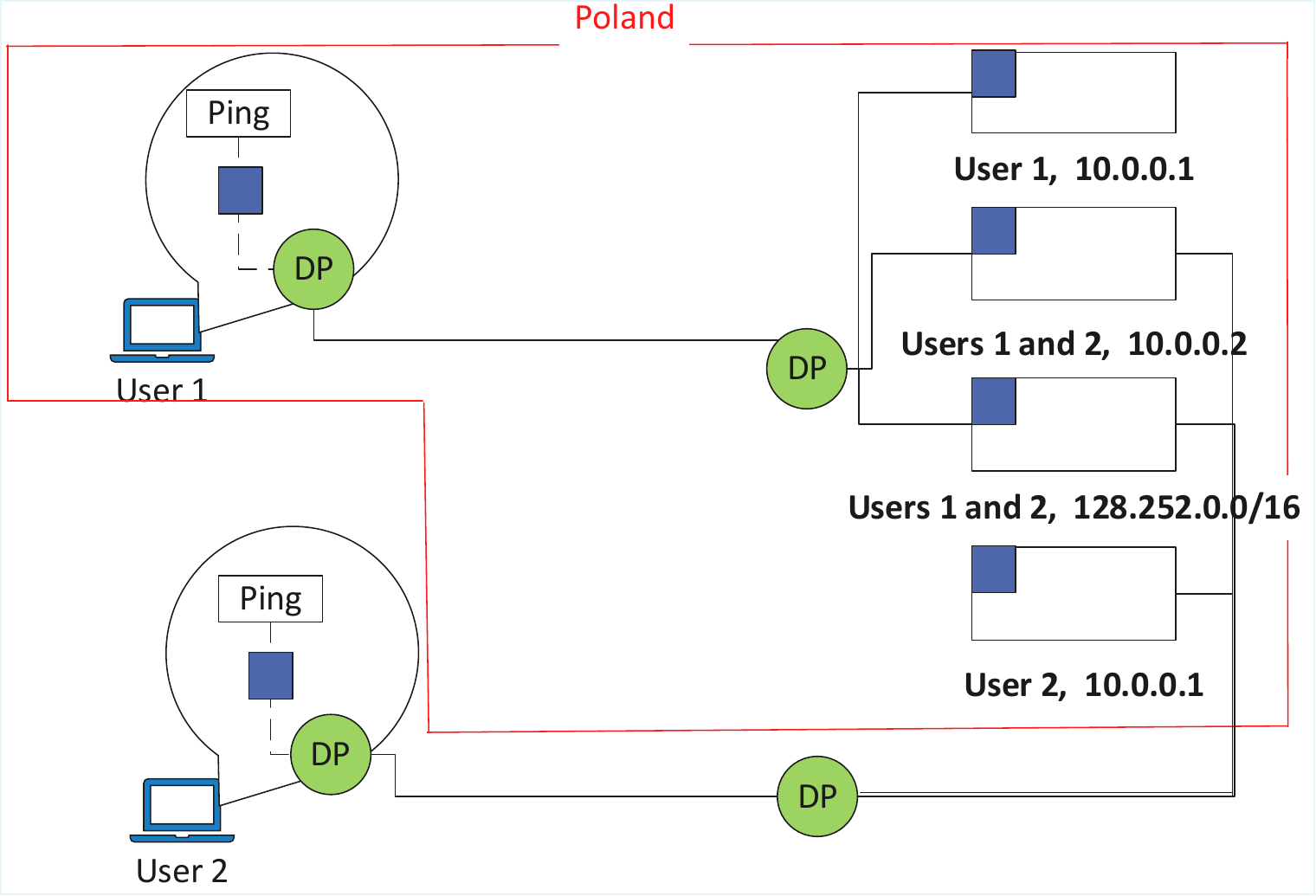}
        \caption{{The overlay after moving User 2's ingress proxy to the US (step 1)}}
	\label{fig:IP2}
    \end{subfigure}
    
    % Second row: one figure on the left and another on the right
    \vskip\baselineskip
    \begin{subfigure}[b]{0.48\textwidth}
        \centering
        \includegraphics[height=0.25\textheight]{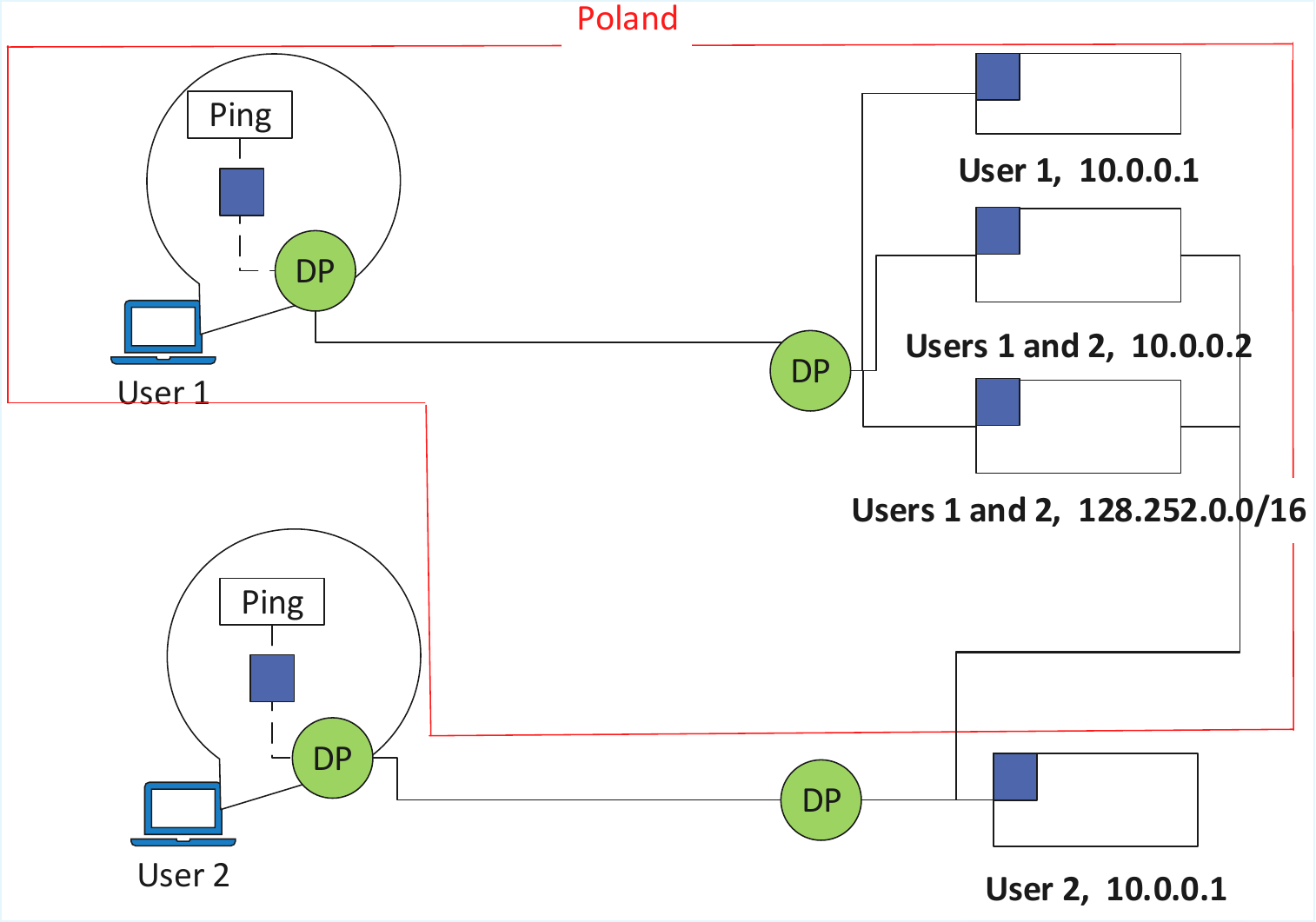}
        \caption{{The overlay after moving User 2's personal server (10.0.0.1/32) to the US (step 2)}}
	\label{fig:IP3}
    \end{subfigure}
    \hfill
    \begin{subfigure}[b]{0.48\textwidth}
        \centering
        \includegraphics[width=\textwidth]{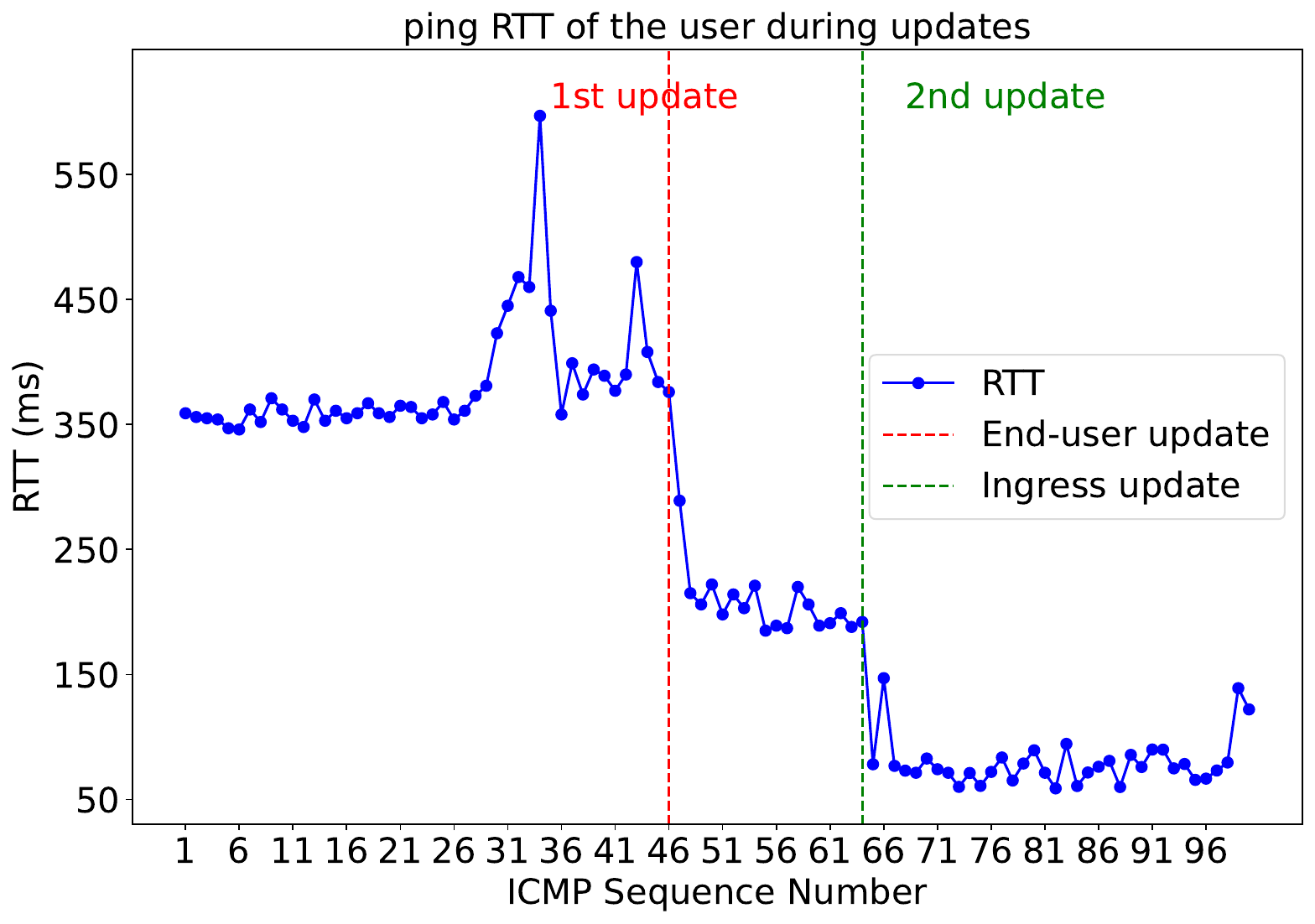}
        \caption{{Reported ping RTTs from User 2's device to 10.0.0.1 over the course of the evaluation}}
	\label{fig:IPresult}
    \end{subfigure}

\caption{{Overall network topology for the overlay update process and performance evaluation over the public Internet. Blue boxes represent assisting components (tunnel endpoints) on end-user devices and target servers. The server responsible for the 128.252.0.0/16 address range acts as a router.}}
    \label{fig:overall}
\end{figure*}
The network configuration is shown in Figure~\ref{fig:overall}. All systems ran Ubuntu 22.04. There are two users, each running an assisting component that creates a tunnel device and captures IP traffic. User 2 is in the United States, while User 1, all ingress points, and all servers are in Poland. Proxy configurations are shown in Table~\ref{IPconfig}. The experiment was performed over the public Internet, not in a controlled lab network. We did not enable TLS links between overlay nodes because latency was measured with \texttt{ping}, so link encryption was unnecessary.

The assisting components on end-user devices were configured to send traffic for 10.0.0.1/32, 10.0.0.2/32, and 128.252.0.0/16 to three distinct UDP ports on each user's local proxy. The local proxy listens on these ports and encapsulates incoming UDP packets in HTTP/2 using CONNECT-UDP (MASQUE). It also appends two HTTP headers, \texttt{User} and \texttt{IPaddress}, which carry the user identifier and the destination network prefix. For example, \texttt{User: user2} and \texttt{IPaddress: 128.252.0.0/16}. The overlay uses these headers to route traffic to the correct next hop.

Because assisting components are independent processes, we can deploy them on nodes other than the proxy nodes they assist. Hermes requires only that (i) assisting components communicate through their paired proxies and (ii) the control plane can configure them. In Figure \ref{fig:overall}, the ingress proxy uses an assisting component (i.e., the tunnel endpoints) on each destination node. This extends its reach to the destinations without requiring another proxy instance on those nodes and reduces processing overhead. It also makes the ingress proxy a dependent proxy, because it {directly communicates with the destinations} through its assisting components. The HTTP tunnel terminates at the ingress proxy, which decapsulates UDP and forwards traffic to the appropriate assisting component on the destination node based on header information. The assisting component then decapsulates IP and delivers packets to the local target process. In this way, the overlay maintains end-to-end control over IP traffic.

We evaluated the end-to-end latencies between users and servers by measuring ICMP echo round-trip time (RTT). Initial latencies are shown in Table \ref{tab:IPinitial} with ping sending one packet per second. The "latency" column reflects latency from User 2 in the United States. The "latency with UDP tunnel" columns show RTT when only IP-in-UDP encapsulation was used and the traffic was sent directly to the destination UDP tunnel endpoint. IP-in-UDP added less than 3 ms to the latency. When we also encapsulated UDP in HTTP, last column, up to 5 ms was added. This is because the proxy must establish the HTTP connection before the first ICMP packet is forwarded.

The update process has three steps. The initial state, shown in Figure \ref{fig:IP1}, has User 2 in the United States connecting to an ingress and a personal server that are both in Poland. We started with one ingress, and all users connected to it. The controller first updated the end-user proxy configuration to point to a new ingress in the United States with the aim of reducing its end-to-end latency below the threshold, as shown in Figure \ref{fig:IP2}. Since the latency was still above the threshold, in the next step, the controller performed the second update, shown in Figure \ref{fig:IP3}, and updated the ingress proxy in the United States to route User 2's traffic to a new server in the United States.

To measure update procedure impact, we sent ICMP packets every 10 seconds and used a 3-second timeout. We also updated User 2's Envoy to version 1.32.0 \cite{envoy_132} and set max\_stream\_duration to 3 seconds to avoid Envoy crashes. The results are shown in Figure \ref{fig:IPresult}. Due to the 3-second timeout, listed in Table \ref{IPconfig}, each ping required a fresh HTTP connection, which increased RTT relative to Table \ref{tab:IPinitial}. Note that Table \ref{tab:IPinitial} reports results with 1 packet per second, while Figure \ref{fig:IPresult} uses 1 packet every 10 seconds. Initially, the average RTT was 382.33 {ms}. After the first update, it dropped to 206.33 {ms} as the ingress moved closer to the user. After the second update, it dropped to 79.68 {ms} because both the ingress and the personal server were now near the user. There were no dropped packets, and each update step completed in less than 10 seconds from initiation, including the controller loop to fetch new configuration and communication time between the controller and proxies.

\subsubsection{Discussion}
We demonstrated that Hermes can provide IP routing by attaching the destination address and user identity as HTTP headers and processing them along the path. Together with the prior use case, this shows that Hermes can control traffic at Layers 3-7. A notable point in the IP use case is that the user continues to use the same destination IP address. By capturing traffic at IP via a tunnel device and encapsulating it in HTTP, Hermes can route based on high-level policies rather than being limited to Layer-3 IP-based routing.

Although HTTP tunneling added up to 5 ms of latency in our measurements, updates were smooth because we changed a Layer 7 routing rule rather than applying complex Layer 3 modifications. From updating the configuration database to completing proxy hot restarts, each update completed in under 10 seconds, including the time for pushing configuration to proxies. If proxy software could create tunnel devices and natively support IP-in-HTTP, such as CONNECT-IP from RFC 9484 \cite{pauly_proxying_2023}, configuration would be simpler and latency would likely be lower. However, we are not aware of any current implementation that provides both capabilities.

\subsection{Supporting experimental network architectures} \label{NDN}
Hermes can serve as an adaptable communication layer by providing built-in support for global load balancing, name-based routing, reliable delivery, and policy-based routing. Other architectures can leverage these features with minimal configuration to operate more effectively over the Internet. In this use case, we map NDN \cite{zhang_named_2014} namespaces to HTTP headers so that Hermes can perform name-based load balancing among replicas serving a given NDN name prefix.

\begin{table}[htbp]
    \centering
    \caption{NDN use case results. D: average transfer time (s), S.D.: standard deviation of transfer time.}
\begin{tabular}{|l|ll|ll|}
\hline
\multicolumn{1}{|c|}{\multirow{2}{*}{File}} & \multicolumn{2}{c|}{Direct} & \multicolumn{2}{c|}{Overlay} \\ \cline{2-5} 
\multicolumn{1}{|c|}{} & \multicolumn{1}{c|}{D} & \multicolumn{1}{c|}{S.D.} & \multicolumn{1}{c|}{D} & \multicolumn{1}{c|}{S.D.} \\ \hline
\multirow{2}{*}{0KB} & \multicolumn{1}{l|}{\multirow{2}{*}{0.0019}} & \multirow{2}{*}{0.0001} & \multicolumn{1}{l|}{\multirow{2}{*}{0.0041}} & \multirow{2}{*}{0.0001} \\
 & \multicolumn{1}{l|}{} &  & \multicolumn{1}{l|}{} &  \\ \hline
\multirow{2}{*}{1MB} & \multicolumn{1}{l|}{\multirow{2}{*}{0.9046}} & \multirow{2}{*}{0.0002} & \multicolumn{1}{l|}{\multirow{2}{*}{1.0321}} & \multirow{2}{*}{0.0006} \\
 & \multicolumn{1}{l|}{} &  & \multicolumn{1}{l|}{} &  \\ \hline
\multirow{2}{*}{26MB} & \multicolumn{1}{l|}{\multirow{2}{*}{21.7869}} & \multirow{2}{*}{0.0002} & \multicolumn{1}{l|}{\multirow{2}{*}{24.1847}} & \multirow{2}{*}{0.0334} \\
 & \multicolumn{1}{l|}{} &  & \multicolumn{1}{l|}{} &  \\ \hline
\multirow{2}{*}{49MB} & \multicolumn{1}{l|}{\multirow{2}{*}{41.5233}} & \multirow{2}{*}{0.0003} & \multicolumn{1}{l|}{\multirow{2}{*}{46.1002}} & \multirow{2}{*}{0.0687} \\
 & \multicolumn{1}{l|}{} &  & \multicolumn{1}{l|}{} &  \\ \hline
\end{tabular}
    \label{tab:ndn}
\end{table}

\begin{figure}[htbp]
    \centering
    \includegraphics[width=\columnwidth]{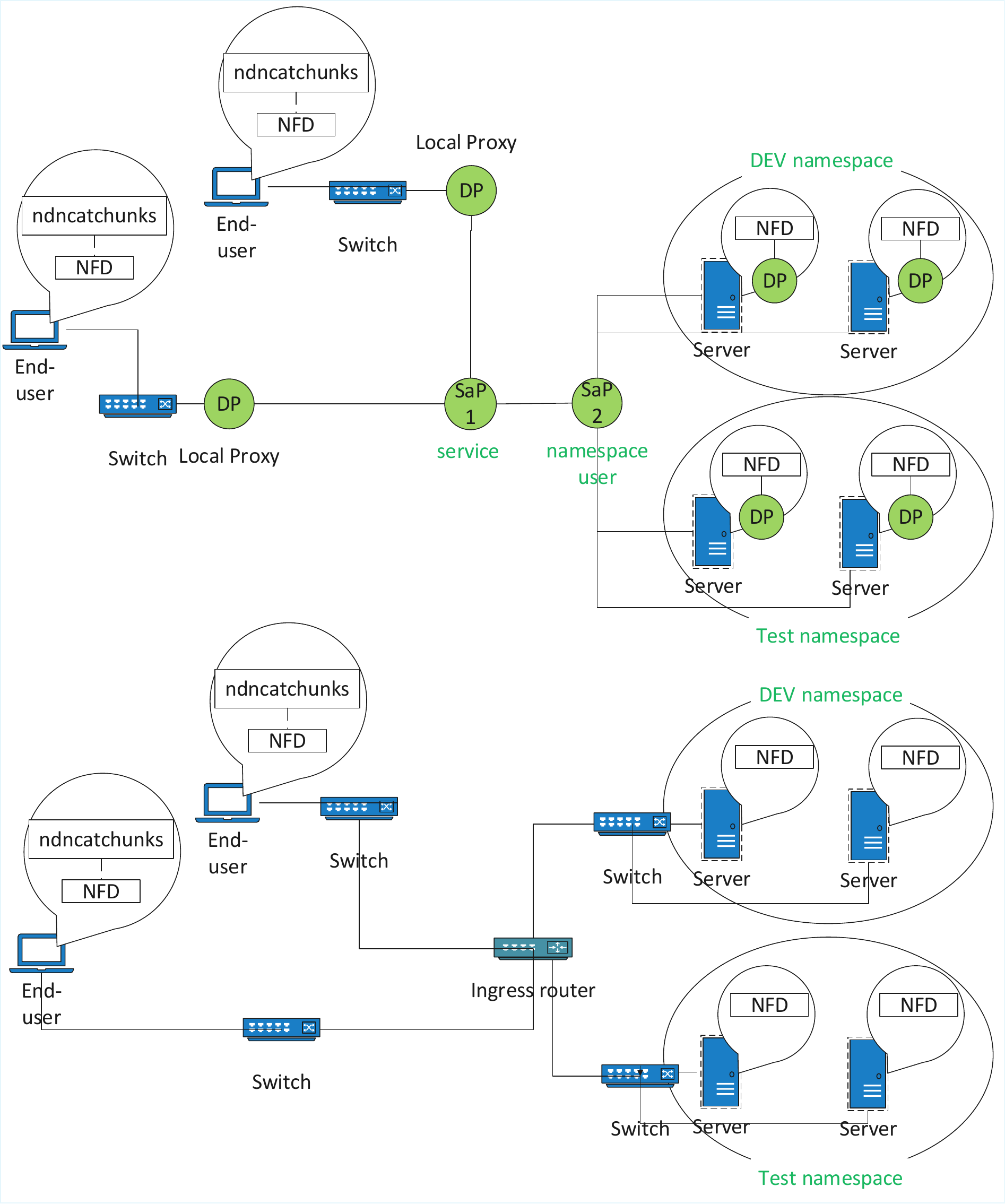} 
    \caption{NDN use case topology. Overlay is above.}
    \label{fig:ndn}
\end{figure}

\subsubsection{Problem (NDN Use Case)}
A small organization has adopted the NDN architecture for several branches worldwide. NDN nodes use Faces \cite{shi_what_2021} to connect over the Internet by encapsulating and decapsulating NDN packets over TCP or UDP so they can traverse IP networks. The organization chose UDP Faces, so NDN payload is encapsulated in UDP.

The organization exposes two namespaces, \texttt{DEV} and \texttt{TEST}, each served by multiple replicas. It needs load balancing within each namespace and, specifically, it wants identity-aware load balancing, where requests are routed to replicas based on the requester's identity as well as the name prefix. For example, requests for \texttt{/TEST/res1} from user X should go to server1, while the same name from user Y should go to server2. Prefix-based load balancing is achievable in NDN, but per-user steering is not available by default because forwarding is name-centric rather than user-centric. In addition, exposing UDP ports on the public Internet without protection is risky, and an effective firewall would require NDN-aware filtering.

\subsubsection{Solution}
A practical solution is to use a Hermes overlay as a communication layer between NDN nodes. The overlay encapsulates NDN over HTTP and maps the NDN namespace into HTTP headers. It also adds other information, such as user identity, as headers. Overlay proxies then route traffic and perform load balancing based on the namespace and user identity headers. Because traffic is carried over HTTP, standard commercial firewalls and security tools can filter, rate limit, and protect the exposed endpoints using familiar approaches. Since the organization is small, it uses a third-party Hermes overlay that allows it to configure dependent proxies. The organization also configures its NDN clients to communicate with the dependent proxies and publishes its load balancing policy and endpoint addresses to the third-party operator.

Each NDN Forwarding Daemon (NFD) instance configures one UDP Face per namespace that targets a dedicated port on the local dependent proxy. The DP encapsulates incoming NDN over UDP datagrams using CONNECT-UDP over HTTP/2 and attaches headers that carry the destination namespace and the user identity. The overlay is operated by a third party, so the organization only needs to configure its NDN clients to send each namespace to the designated DP ports, and configure the DP, via the overlay control plane, to map these ports to the corresponding HTTP headers. Standalone proxies in the third-party overlay read these headers and route traffic toward the organization's NDN servers. NDN nodes remain unmodified.

\subsubsection{Implementation}
The network configuration is shown in Figure~\ref{fig:ndn}. The end-user used ndncatchunks \cite{ndncatput} to request content from the DEV and TEST namespaces, and servers published content with ndnputchunks \cite{ndncatput}. Only the end-user and the servers ran NFDs. The end-user's NFD was configured to send Interests for each namespace to a distinct port on its dependent proxy. We observed that when using UDP Faces and targeting a UDP port of the proxy on the same host, NFD could not regulate its outgoing rate. To address this, we deployed the DP on a separate machine, allowing NFD to regulate its outgoing rate. Running the DP off-host is one of the deployment modes Hermes supports and allows the architecture to fit a wide range of use cases and operational constraints.

When the local proxy received the UDP packets, it tunneled them over HTTP and appended two headers: \texttt{service: NDN} and a namespace header, either \texttt{namespace: DEV} or \texttt{namespace: TEST}, depending on the ingress port. In this way, NDN namespaces were mapped to HTTP headers. The local proxy also added user identity as an HTTP header. The proxy then forwarded the data to SaP1 closer to the user. This SaP was configured to partially process headers, so it only processed the \texttt{service: NDN} header and ignored the rest. It then forwarded the traffic related to the NDN architecture to SaP2, which is responsible for parsing the \texttt{namespace} and \texttt{user} headers. SaP2 distributed the traffic among the HTTP proxies collocated with the NFDs at the destination based on namespace and user identity. The DP proxies then terminated the HTTP tunnel, decapsulated the UDP traffic, and delivered it to the local NFDs.

In the direct NDN connection, shown in the bottom of Figure \ref{fig:ndn}, the client knew the addresses of the servers, and its NFD was configured with those addresses. We published three files with approximate sizes of 1MB, 26MB, and 49MB, and measured the Transfer Time, which is the total time it takes for ndncatchunks to receive the requested file. Additionally, we published a 0KB file to measure latency. The ndncatchunks tool was configured to always request fresh content and used a fixed interest pipeline size of 20. For the 0KB file, the pipeline size was set to 1.

We repeated the experiments at least 20 times, requesting both namespaces twice in each iteration. The average results for both namespaces are reported in Table \ref{tab:ndn}. Although the Hermes approach is up to 12\% slower than the direct approach when downloading large file sizes and its latency is slightly higher, it provides various load-balancing features. This facilitates the development and usage of the NDN architecture even when NDN uses UDP faces.
\subsubsection{Discussion}
As demonstrated, Hermes can serve as the underlying communication network for other experimental protocols and architectures. These systems need only to deliver their data units to their local proxy, after which Hermes manages reliability, load balancing, namespace routing, and other concerns. This approach greatly simplifies deployment and management of NDN and other networking architectures over the Internet.

\section{Performance Measurement} \label{perf}
In the previous section, we used open-source tools to show how the Hermes data plane can be easily built. We also used Envoy because it is a feature-rich proxy, though it may incur higher overhead than lighter alternatives such as Linkerd \cite{linkerd_bench}. Generally, the performance of a Hermes overlay depends on the number of proxies on the end-to-end path, the processing performed at each proxy, and the proxy implementation, as discussed in Sections\rev{~\ref{hermes}} and~\ref{imp}. Note that, as discussed earlier, {Hermes requires HTTP processing on the end-user device because traffic needs to be associated with appropriate Hermes headers for overlay routing and policy enforcement. As a result, an overlay chain of proxies begins with HTTP processing at the end-user device, and then continues with Layer~4 or Layer~7 processing along the overlay path, depending on service requirements.} Additionally, the service provider needs to decide on the extent of the overlay. {Once} traffic reaches the managed infrastructure, the service provider can terminate the overlay and use kernel-based Layer~4 processing with eBPF to improve performance instead of user-space proxies, similar to Cilium~\cite{cilium-mesh}.

{Overlays of Layer 7 proxies are frequently used in service environments because they provide various benefits. For example, Meta's ServiceRouter architecture uses millions of Layer 7 proxies to route billions of requests per second across thousands of services~\cite{saokar_servicerouter_2023}. Since our prototype builds an overlay of Envoy proxies, and because Envoy is widely used in service mesh implementations such as Istio~\cite{istioEnvoy} and Consul~\cite{consul}, studies that quantify Envoy overhead in these deployments are directly relevant to Hermes performance. Zhu et al.~\cite{zhu_dissecting_2023} evaluated Envoy proxying overhead in sidecar-based service meshes and reported that, for their benchmark applications, sidecars increase end-to-end latency by up to 61\% in TCP mode, with CPU increases up to 92\%. They also show that protocol parsing dominates Layer 7 proxying overhead, accounting for about 74\% of the per-proxy overhead in HTTP mode. Saxena et al.~\cite{saxena_copper_2025} observed similar amplification across multi-hop service chains, reporting that 99th-percentile end-to-end latency increased from 9.2~ms to 27.5~ms across a four-service chain when sidecars were added. Because of the noticeable overhead in sidecar-based service meshes, newer designs such as Ambient Mode~\cite{sidecar_ambient, maturing-ambient}, Canal Mesh~\cite{canal}, and Cilium Mesh~\cite{cilium-mesh} have been proposed. Finally, for other components used in our prototype implementation, published evaluations are available for Envoy Mobile~\cite{envoy_mobile_perf}, Envoy filter overhead~\cite{zhu_dissecting_2023}, and external policy engines such as OPA~\cite{opa_perf}. In addition, in our prior work we measured the overhead of our L3 tunneling prototype to be under 1~ms~\cite{farkiani_l3mesh}.}

Our design differs from conventional service meshes in three key respects.
\begin{enumerate}
\item \textbf{Origin point:} It begins on end-user devices and may optionally merge with a downstream service mesh, depending on the application.
\item \textbf{Redirection method:} Because we deploy across heterogeneous end-user devices, we cannot rely on redirection techniques typical of service meshes on the end-user device.
\item \textbf{Protocol coverage:} We handle the full TCP/IP stack via tunneling mechanisms and assisting components, whereas most service meshes focus primarily on TCP and HTTP.
\end{enumerate}

{In service mesh environments, measured overhead depends on the traffic redirection strategy and on whether optional security mechanisms such as HBONE are enabled~\cite{hbone}. Despite these differences, the overheads reported above remain relevant to Hermes and align with the end-to-end results in the previous section.}

%In the previous section, we evaluated Hermes on representative use cases, each with a small number of users to highlight its functionality and discuss how it can address various challenges. At larger scales, Hermes follows standard service mesh practice for scalability: operators can scale resources horizontally or select a better proxy placement \cite{saxena_copper_2025}. Hermes itself does not prescribe an implementation, so detailed scaling mechanics are beyond the scope of this paper.

This section examines an important aspect of the architecture that, to the best of our knowledge, has not been studied in depth: the overhead of various HTTP proxying and tunneling techniques. These mechanisms are core to Hermes because they enable heterogeneous traffic to be carried over the overlay. To isolate the impact of tunneling and proxying, we configure proxies to perform only the minimal functions required for those mechanisms and exclude application-specific processing.

\begin{figure}[ht]
    \centering
    \includegraphics[width=\columnwidth]{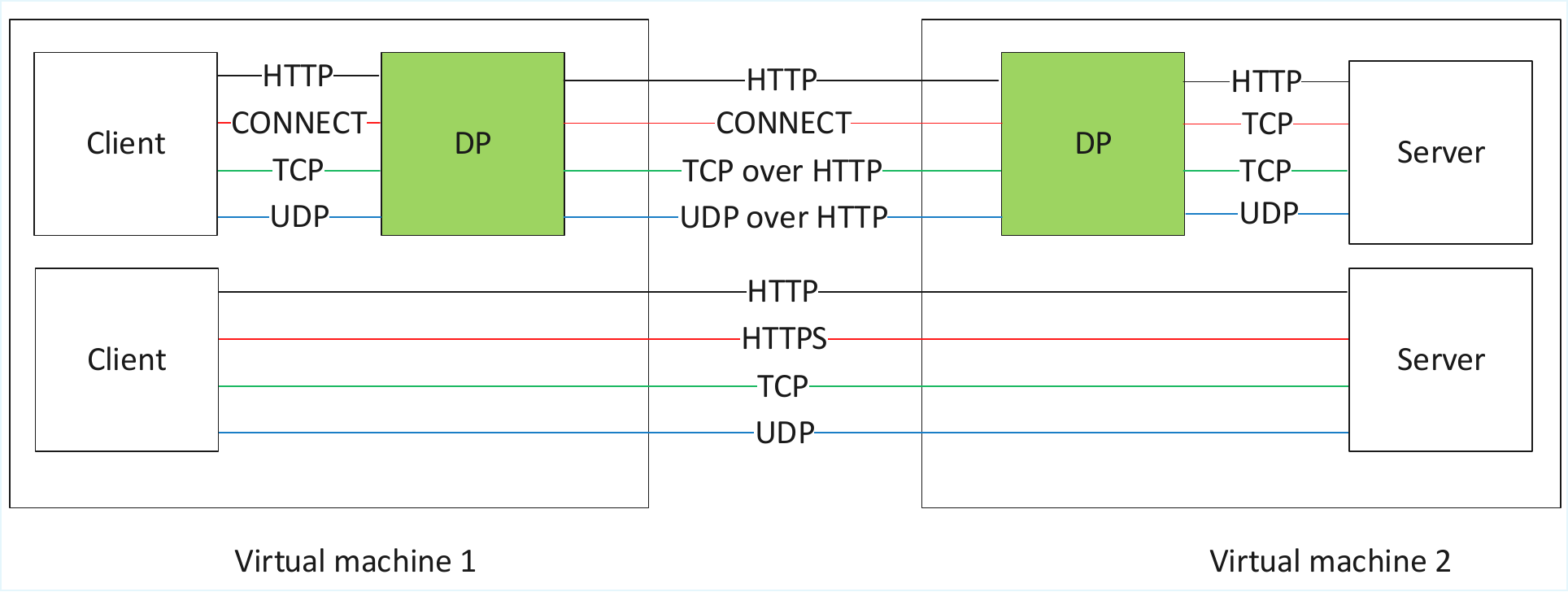} 
    \caption{Performance evaluation scenarios. Baseline {configuration is} below.}
    \label{fig:perf}
\end{figure}

\begin{table}[htbp]
\caption{Proxy overhead measurement. D: average request completion time (ms), S.D.: standard deviation}
\centering
\label{tab:perf}
\renewcommand{\arraystretch}{1.1}
\begin{tabular}{l|ll|ll|}
\cline{2-5}
 & \multicolumn{2}{c|}{Baseline} & \multicolumn{2}{c|}{Proxied} \\ \cline{2-5} 
 & \multicolumn{1}{c|}{D.} & \multicolumn{1}{c|}{S.D.} & \multicolumn{1}{c|}{D.} & \multicolumn{1}{c|}{S.D.} \\ \hline
\multicolumn{1}{|l|}{HTTP proxying} & \multicolumn{1}{l|}{2.81} & 0.72 & \multicolumn{1}{l|}{3.48} & 0.89 \\ \hline
\multicolumn{1}{|l|}{CONNECT} & \multicolumn{1}{l|}{19.45} & 1.38 & \multicolumn{1}{l|}{20.22} & 1.86 \\ \hline
\multicolumn{1}{|l|}{TCP over HTTP} & \multicolumn{1}{l|}{1.87} & 0.96 & \multicolumn{1}{l|}{3.04} & 0.93 \\ \hline
\multicolumn{1}{|l|}{UDP over HTTP} & \multicolumn{1}{l|}{0.88} & 0.2 & \multicolumn{1}{l|}{2.26} & 0.51 \\ \hline
\end{tabular}
\end{table}

Our setup (Figure \ref{fig:perf}) uses two virtual machines: one contains a client application and a dependent proxy and the other contains the server application and a dependent proxy. The server is a simple Python echo server. All experiments ran collocated on the same physical host to eliminate variability from network latency, hardware heterogeneity, and protocol-stack differences. We used Envoy commit \texttt{10d0784b9cdb}, consistent with our prior use cases. We measure the average request completion time with and without proxies to compare proxying overhead. In all experiments, the client issues a request and the server responds with a fixed payload. We compare four cases against their direct-connection baselines.

\begin{itemize}
\item \textbf{CONNECT proxying:} The client sends an HTTP/1.1 CONNECT request to its dependent proxy, which forwards it to the server proxy. The server proxy upgrades the connection, establishes a TCP link to the server, and blindly forwards TCP packets, thus preserving an end-to-end encrypted HTTP/1.1 channel. We use direct HTTPS connection to the server as baseline.
\item \textbf{TCP tunneling over HTTP/2:} The client delivers raw TCP packets to its dependent proxy, which tunnels them in an HTTP/2 POST to the server proxy. The server proxy decapsulates and forwards the TCP stream. We compare it with direct TCP connection to server.
\item \textbf{UDP tunneling over HTTP/2:} The client sends UDP datagrams to its dependent proxy, which tunnels them via HTTP/2 using the CONNECT-UDP method. The server proxy decapsulates and delivers UDP to the server. We compare it with direct UDP connection to server.
\item \textbf{Simple HTTP proxying:} The client issues HTTP/1.1 GET requests to its dependent proxy, which forwards them to server proxy without extra processing. Baseline is direct HTTP request to server.
\end{itemize}

Table \ref{tab:perf} summarizes the results. Although completion times increase, ranging from 4\% for CONNECT proxying to 63\% for TCP tunneling, the absolute delays remain small: measured proxy overhead spans from 0.67 ms for HTTP proxying to 1.38 ms for UDP tunneling. These findings indicate that the tunneling and proxying techniques at the core of our architecture add only modest latency. Note that we measured tunneling and proxying with minimal proxy configuration. More complex HTTP processing tailored to specific use cases would increase overhead, consistent with prior studies we reviewed.

\begin{figure*}[htbp]
    \centering
    \begin{subfigure}[b]{0.4\textwidth}
        \centering
        \includegraphics[width=0.9\textwidth]{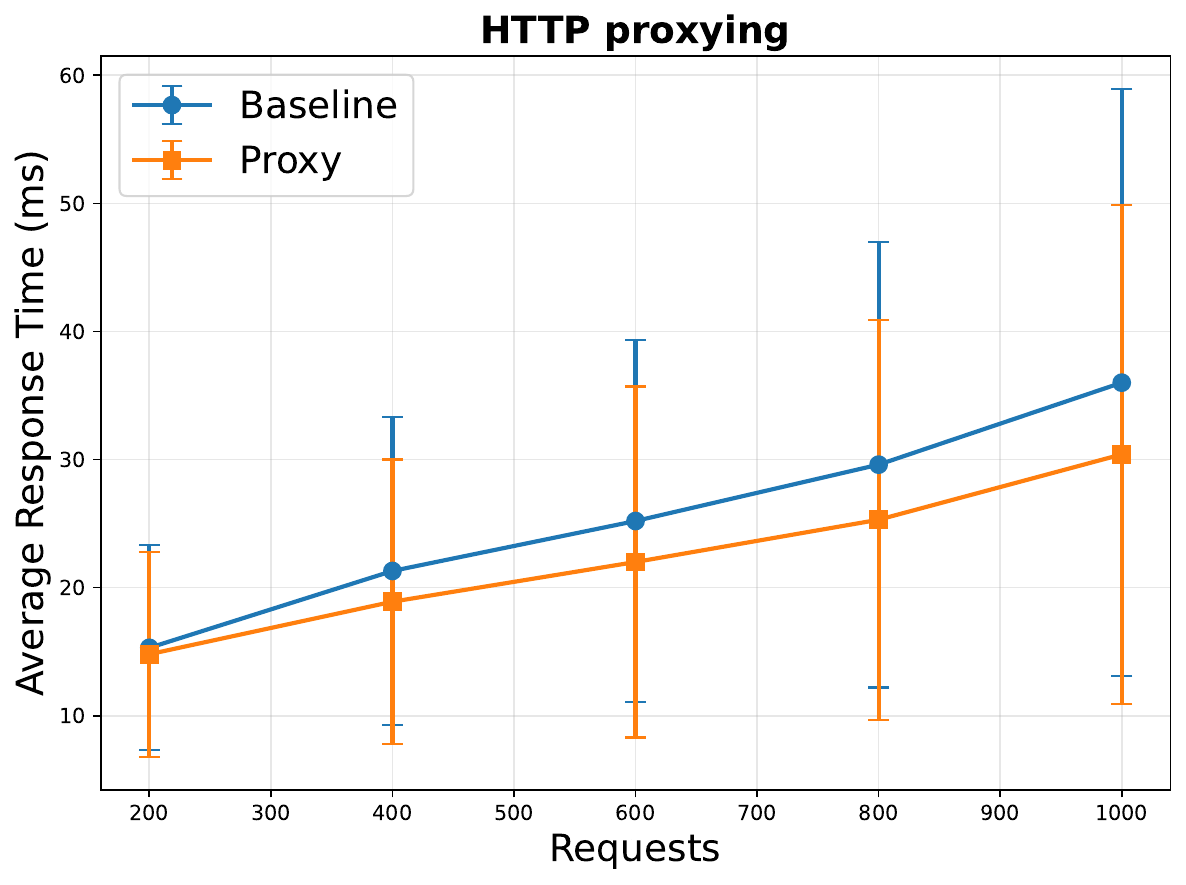}
        \caption{HTTP proxying}
        \label{fig:http}
    \end{subfigure}
    \hfill
    \begin{subfigure}[b]{0.4\textwidth}
        \centering
        \includegraphics[width=0.9\textwidth]{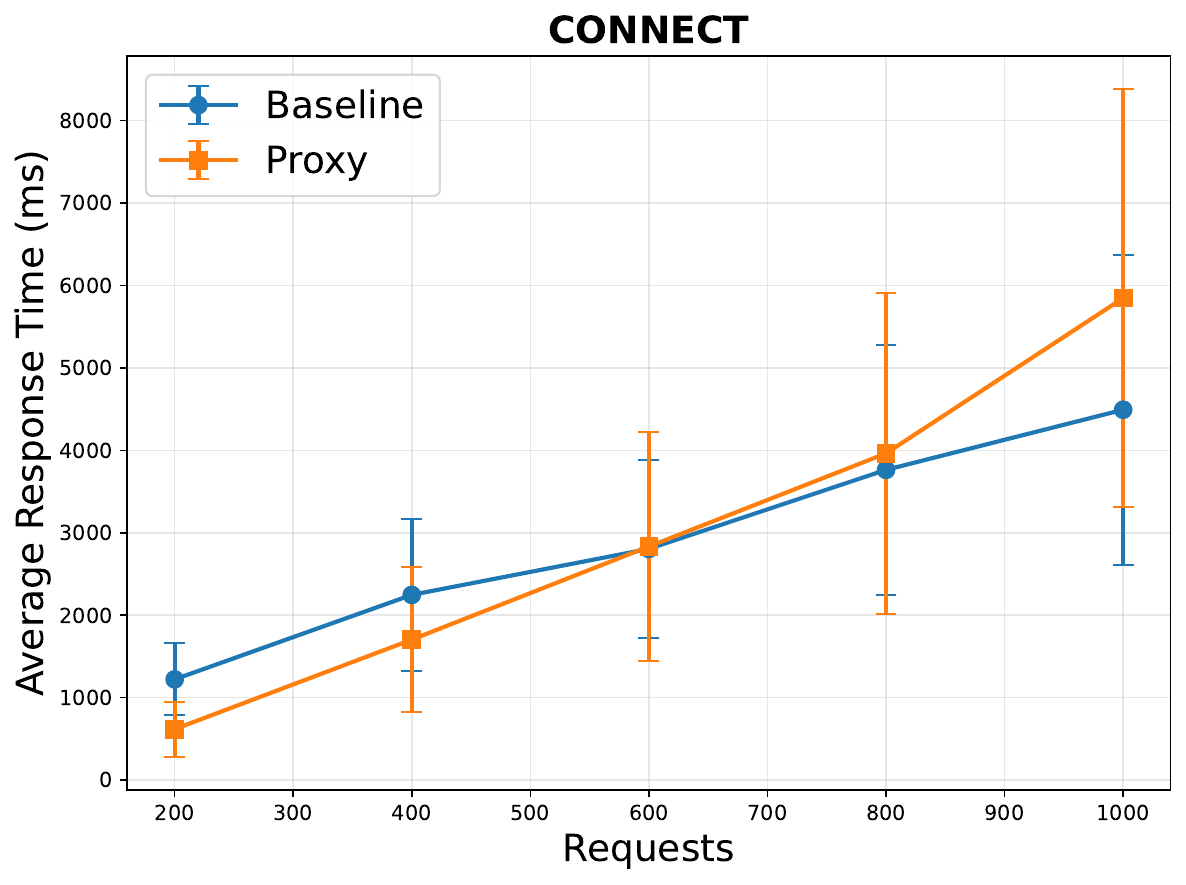}
        \caption{CONNECT}
        \label{fig:connect}
    \end{subfigure}
    
    \vspace{0.5cm}
    
    \begin{subfigure}[b]{0.4\textwidth}
        \centering
        \includegraphics[width=0.9\textwidth]{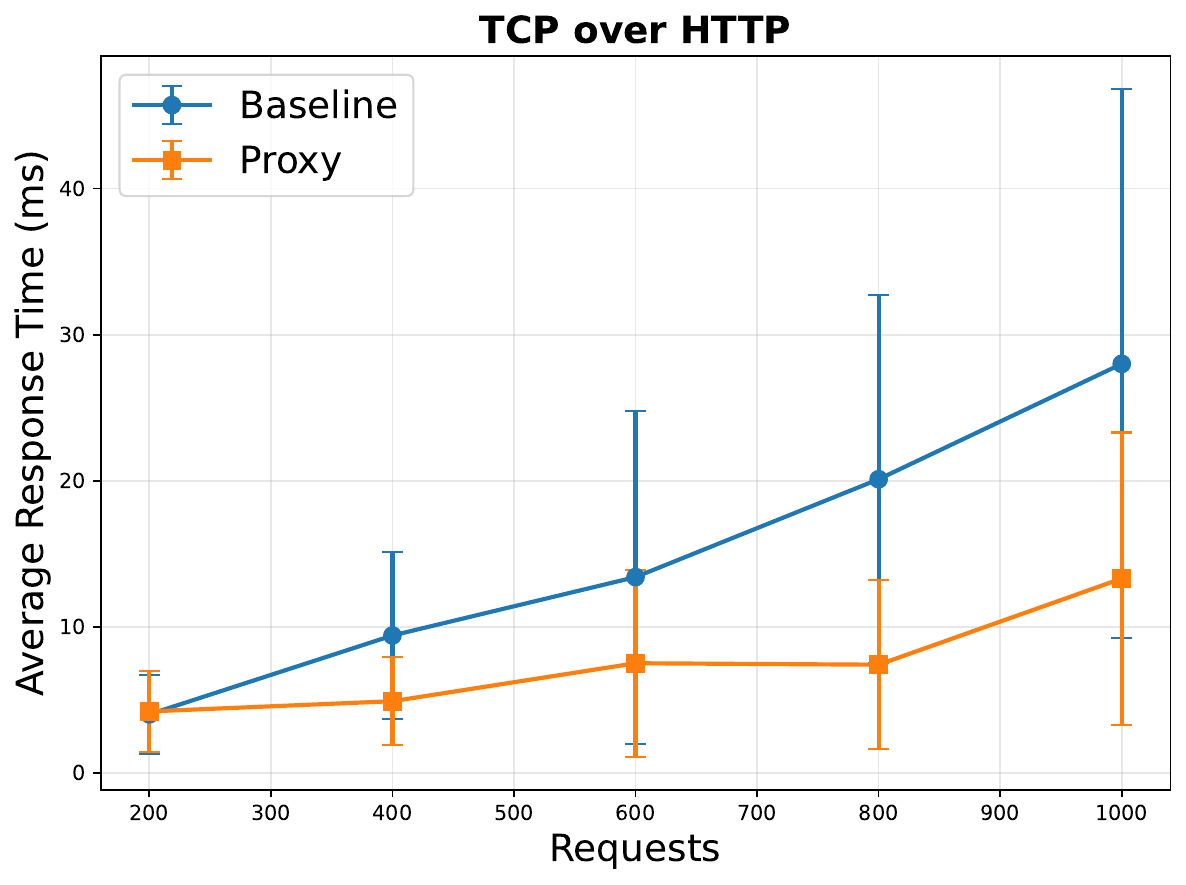}
        \caption{TCP over HTTP}
        \label{fig:tcp}
    \end{subfigure}
    \hfill
    \begin{subfigure}[b]{0.4\textwidth}
        \centering
        \includegraphics[width=0.9\textwidth]{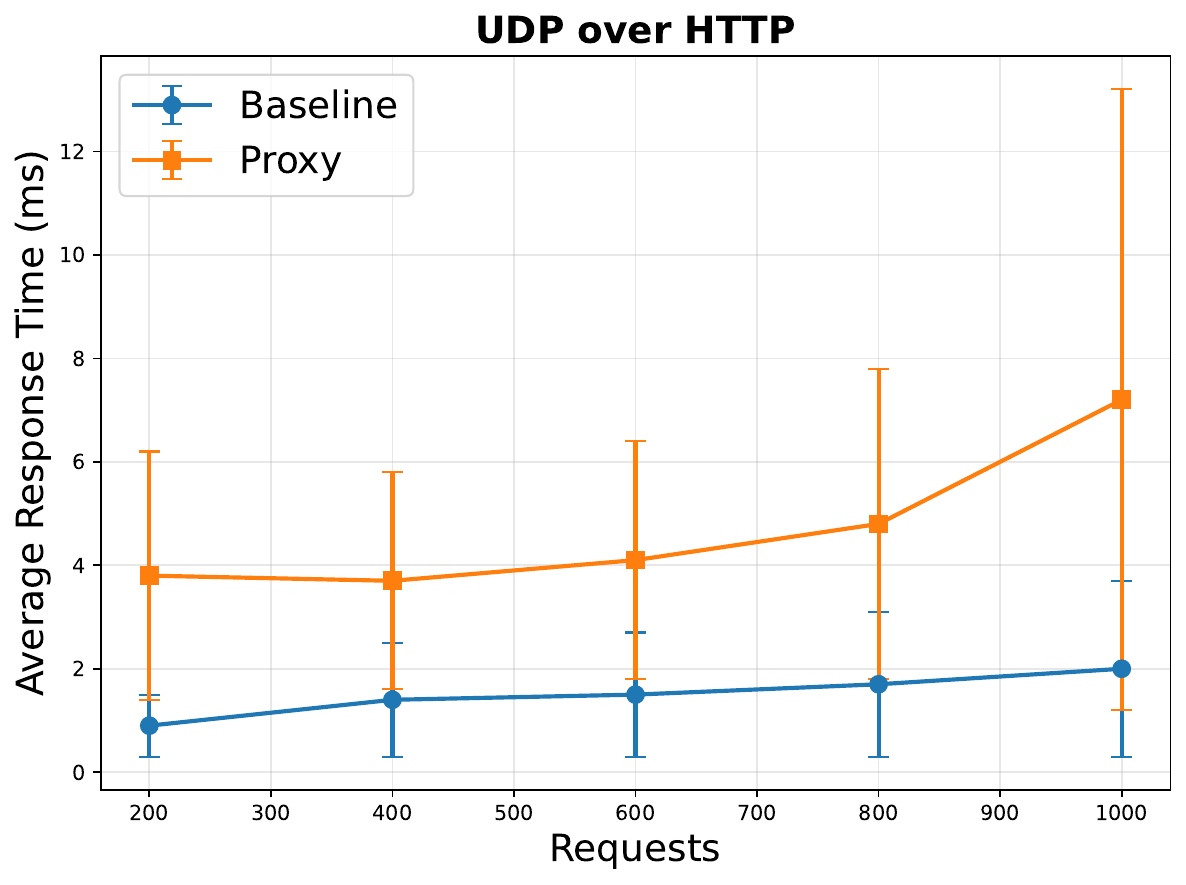}
        \caption{UDP over HTTP}
        \label{fig:udp}
    \end{subfigure}
    
    \caption{Performance comparison between Baseline and Proxy across different scenarios. Standard deviation is shown as error bars.}
    \label{fig:performance_comparison}
\end{figure*}

In addition to single-request analysis, we evaluated parallel requests scaled from 200 to 1000. Each request represents a unique client, and thus we emulated a scenario in which requests from up to 1000 clients arrived at the same time. Results are shown in Figure~\ref{fig:performance_comparison}. In all scenarios, end-to-end latency increases as concurrency rises, and no request failures were observed. With 1000 requests, the baseline approaches are 18.4\% and 110.5\% slower than the proxy approach for HTTP proxying and TCP tunneling, respectively. In contrast, for CONNECT and UDP, the baseline is 23.2\% and 72.2\% faster than the proxy approach.

Baseline UDP has neither reliability nor connection setup, while tunneling through HTTP introduces both reliability semantics and connection establishment on the proxy path. Even if we used HTTP/3 with QUIC datagrams \cite{rfc_9297, pauly_unreliable_2022}, this would remove reliability semantics, but the QUIC connection still requires setup and adds TLS 1.3 cryptographic overhead. In the proxied CONNECT setup, the client sends an HTTP/1.1 CONNECT to the client proxy, which forwards it to the server proxy. The server proxy opens a TCP connection to the backend and returns HTTP 200. After that, the client performs the TLS handshake directly with the backend through the tunnel. This process has high per-flow overhead, whereas the baseline simply establishes a direct client-to-server TLS connection. Despite this overhead, TCP tunneling over HTTP outperforms direct TCP because the tunnel uses HTTP/2: each new tunneled request becomes a new HTTP/2 stream between the proxies instead of establishing a fresh end-to-end TCP connection for every request. HTTP proxying also outperforms direct HTTP due to Envoy connection pooling, which keeps upstream connections open and reuses them, thereby reducing per-request handshake cost.

Note that Table~\ref{tab:perf} reports single-connection overhead. HTTP/2 connection establishment adds fixed setup cost that raises single-request latency, but tunneling over HTTP/2 benefits workloads with many requests. Therefore, although proxies increase single-connection overhead, they can reduce average end-to-end latency at scale by amortizing connection setup and leveraging HTTP/2 stream multiplexing and connection pooling. All of these advantages or disadvantages depend on proxy implementation, the processing they perform, and the number of proxies in the path.

\section{Conclusion and Future Work}\label{conclusion}
This paper presented Hermes, a general-purpose, portable, and adaptable networking architecture that addresses fundamental challenges in efficient service delivery over the Internet. Hermes extends an overlay of reconfigurable proxies from end-user devices to service endpoints and decouples networking logic from application logic, enabling adaptation to changing service and network conditions without modifying applications. \rev{From an architectural viewpoint, Hermes provides unified solutions to four critical challenges:} (i) end-to-end traffic management by extending service-provider control to end-user devices; (ii) backward compatibility through protocol translation and tunneling; (iii) consistent security and privacy models via unified control across network layers and segments; and (iv) an adaptable communication layer that complements standard protocols to meet service requirements.

\rev{Through our prototype implementation and experiments, we showed that} Hermes enables backward compatibility by tunneling traffic over HTTP, achieves a 100\% delivery success rate in intermittent connectivity scenarios where direct TCP fails by moving retry control into proxies, supports Layer 3 policy-based traffic routing, and serves as an efficient substrate for experimental architectures such as NDN. Performance measurements indicate that while Hermes improves various performance measures, it can add end-to-end latency. Across multiple use cases, we demonstrated that the benefits Hermes delivers outweigh this processing overhead. Additionally, the isolated HTTP tunneling and proxying overhead, as a core part of the overlay, is modest, typically under 2 ms per proxy pair traversal. Finally, we showed although proxies increase one-hop single-connection overhead, they can amortize this cost at scale through connection pooling and HTTP/2 multiplexing, thereby lowering end-to-end latency relative to direct no-proxy baselines.

\rev{Because Hermes enables unified management across network layers and network segments, future work will include studying control-plane algorithms for dynamic overlay optimization that leverage Hermes' unified capabilities. More mature adoption of Hermes will enable large-scale studies of its performance impact and a better understanding of how control-plane algorithms behave under different service requirements and business goals. We will also investigate AI-driven network management to automate operations and improve reliability. Finally, we consider extending Hermes into multi-operator overlays as a possible future direction.}

\section*{Acknowledgment}
\noindent
This work was supported by NSF CNS Award 2213672.

\clearpage

\bibliographystyle{IEEEtran}
\bibliography{reference}

\clearpage
\appendix
\section {Appendix}

\subsection {Research reproducibility}
All prototype implementations and use cases we presented in this paper are available at address https://github.com/Bfarkiani/Hermes.
\subsection {Additional information and configuration details}
This section includes additional information. 

\begin{figure}[h] 
    \centering 
    \begin{minipage}{0.5\textwidth}
        \centering
        \includegraphics[width=\linewidth, height=0.25\paperheight, keepaspectratio]{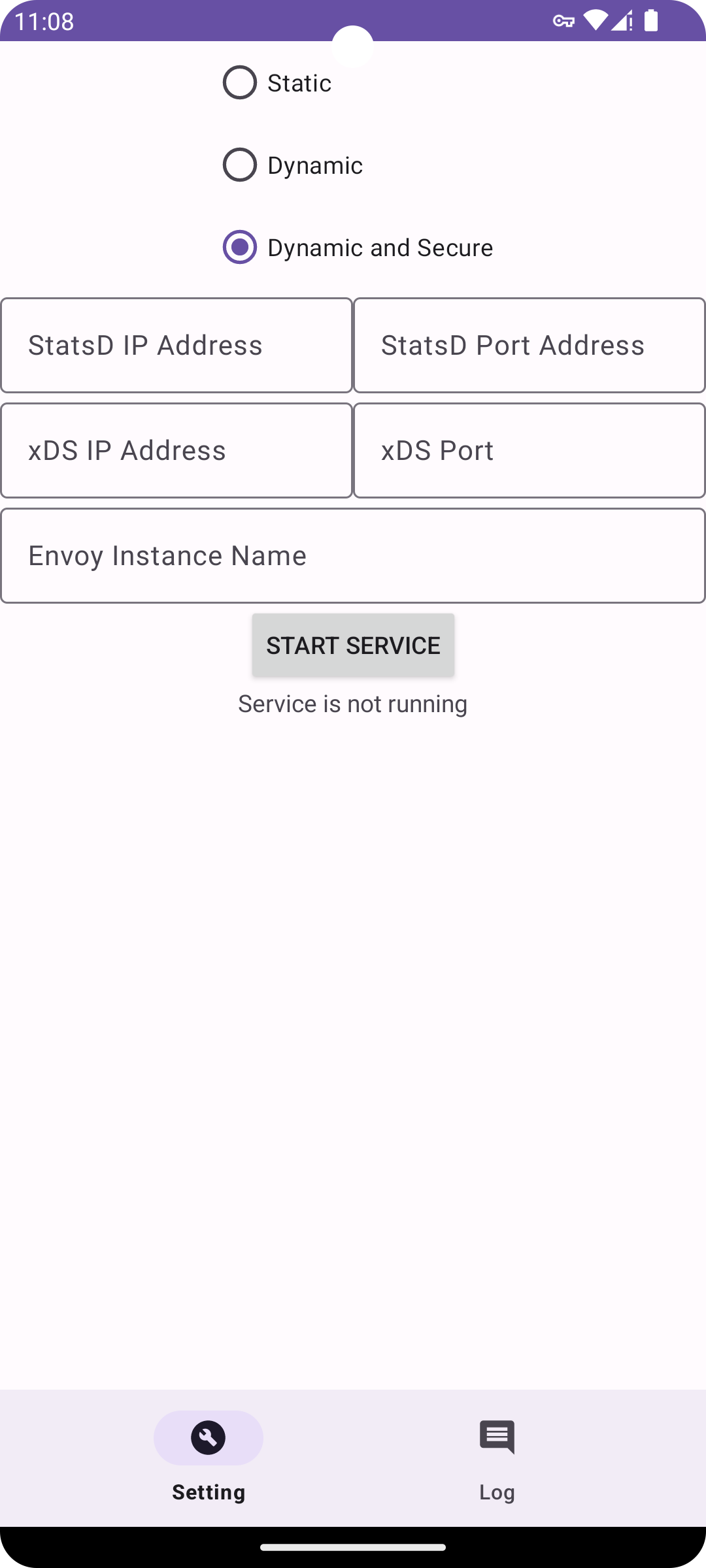} 
        \caption{The Hermes Android mobile client.}
        \label{fig:mobileAndroid}
    \end{minipage}\hfill
    \begin{minipage}{0.5\textwidth}
        \centering
        \includegraphics[width=\linewidth, height=0.25\paperheight, keepaspectratio]{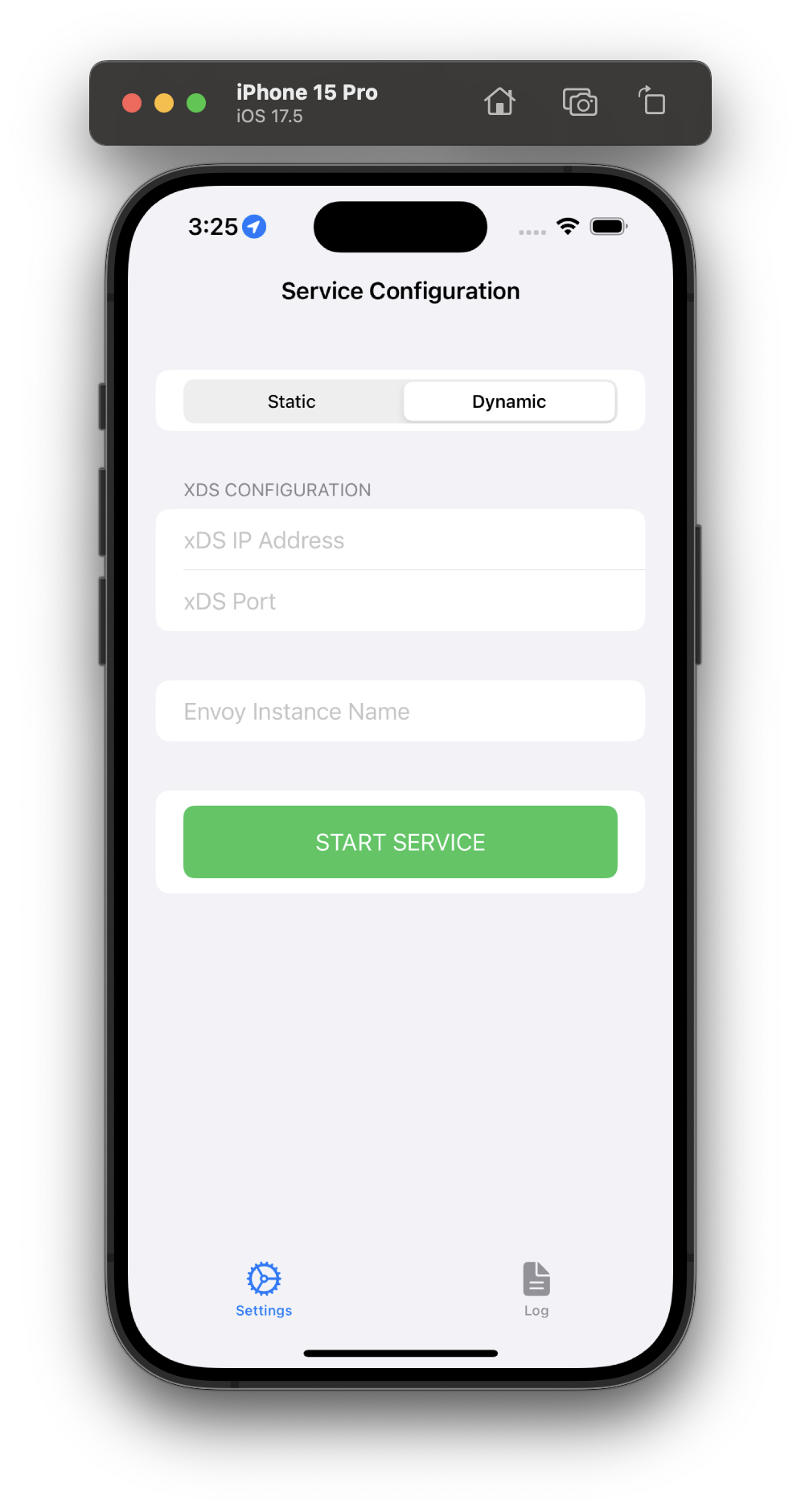}
        \caption{The Hermes iOS mobile client.}
        \label{fig:mobileIOS}
    \end{minipage}
\end{figure}

\begin{table}[htbp]
\caption{File sizes used in the experiments}
\label{tab:filesize}
\begin{tabular}{|l|l|l|}
\hline
\textbf{Use case} & \textbf{File name} & \begin{tabular}[c]{@{}l@{}}\textbf{Size of the downloaded}\\ \textbf{files on disk (bytes)}\end{tabular} \\ \hline
\multirow{2}{*}{Video} & 1 minute  \cite{nasa} & 5736233 \\ \cline{2-3} 
 & 2 minutes \cite{nasa} & 11688214 \\ \hline
\multirow{2}{*}{Intermittent} & 1 MB  \cite{gutenberg} & 1090352 \\ \cline{2-3} 
 & 5 MB \cite{gutenberg} & 5497060 \\ \hline
\multirow{4}{*}{NDN} & 0KB & 0 \\ \cline{2-3} 
 & 1MB \cite{gutenberg} & 1090352 \\ \cline{2-3} 
 & 26MB \cite{gutenberg} & 26236224 \\ \cline{2-3} 
 & 49MB & 50000000 \\ \hline
\end{tabular}
\end{table}

\begin{table}[htbp]
\caption{Nginx Configuration }
\label{tab:nginxConfig}
\begin{tabular}{|l|l|}
\hline
\textbf{Configuration} & \textbf{Value} \\ \hline
Cache-control & Public, max-age=300 \\ \hline
Sendfile & On \\ \hline
Send\_file\_max\_chunk & 1k \\ \hline
Keepalive\_timeout & 0 \\ \hline
If\_modified\_since & Off \\ \hline
\end{tabular}
\end{table}

\begin{table}[htbp]
\caption{Video use case node configuration}
\label{tab:Videoconfig}
\begin{tabular}{|l|l|}
\hline
\textbf{Node} & \textbf{Configuration} \\ \hline
End-user & \begin{tabular}[c]{@{}l@{}}Linux Ubuntu v20.04.6\\ ffmpeg v4.2.7 \cite{ffmpeg}\\ Envoy dev  \cite{envoy_dev}\\ curl v7.81.0 \cite{curl}\end{tabular} \\ \hline
Ingress & \begin{tabular}[c]{@{}l@{}}Envoy dev  \cite{envoy_dev}\end{tabular} \\ \hline
Dist 1 and 2 & \begin{tabular}[c]{@{}l@{}}Envoy dev \cite{envoy_dev}\\ OPA\cite{OPA}\end{tabular} \\ \hline
Server & \begin{tabular}[c]{@{}l@{}}Envoy dev \cite{envoy_dev} \\ Video server (Python) \\ ffmpeg \cite{ffmpeg} \end{tabular} \\ \hline
\end{tabular}
\end{table}

\begin{table}[htbp]
\caption{Video use case ffmpeg configuration}
\label{tab:ffmpeg}
\begin{tabular}{|ll|}
\hline
\multicolumn{2}{|l|}{\textbf{Video server}} \\ \hline
\multicolumn{1}{|l|}{read rate} & 30fps \\ \hline
\multicolumn{1}{|l|}{codec} & libx264 \\ \hline
\multicolumn{1}{|l|}{format} & \text{rtp\_mpegts} \\ \hline
\multicolumn{1}{|l|}{bufsize} & 3968k \\ \hline
\multicolumn{1}{|l|}{protocol} & RTP \\ \hline
\multicolumn{2}{|l|}{\textbf{End-user}} \\ \hline
\multicolumn{1}{|l|}{time out} & 90s (1min video), 150s (2mins video) \\ \hline
\multicolumn{1}{|l|}{protocol\_whitelist} & udp, rtp \\ \hline
\multicolumn{1}{|l|}{buffer\_size} & 26214400 \\ \hline
\multicolumn{1}{|l|}{fifo\_size} & 500000 \\ \hline
\end{tabular}
\end{table}

\begin{table}[htbp]
\caption{IP use case proxy configuration}
\label{IPconfig}
\begin{tabular}{|l|l|}
\hline
\multicolumn{1}{|c|}{\textbf{Parameter}} & \multicolumn{1}{c|}{\textbf{Value}} \\ \hline
drain-time-s & 5 \\ \hline
parent-shutdown-time-s & 10 \\ \hline
drain-strategy & immediate \\ \hline
timeout & 3s \\ \hline
idle\_timeout & 3s \\ \hline
connect\_timeout & 5s \\ \hline
close\_connections\_on\_host\_set\_change & TRUE \\ \hline
\end{tabular}
\end{table}

\begin{table}[htbp]
\caption{Intermittent use case node configuration}
\label{tab:Intermittentconfig}
\begin{tabular}{|l|l|}
\hline
\textbf{Node} & \textbf{Configuration} \\ \hline
End-user & \begin{tabular}[c]{@{}l@{}}Linux Ubuntu v20.04.6\\ Envoy dev  \cite{envoy_dev} \\ wget v1.21.2\cite{wget} \end{tabular}\\ \hline
Ingress, Proxies 2-4 & \begin{tabular}[c]{@{}l@{}}Envoy dev  \cite{envoy_dev}\end{tabular} \\ \hline
Server, Proxy 5 & \begin{tabular}[c]{@{}l@{}}Envoy dev \cite{envoy_dev}\\ Nginx \cite{nginx_http3}\end{tabular} \\ \hline
\end{tabular}
\end{table}

\begin{table}[htbp]
\caption{Intermittent use case retry parameters}
\label{tab:intermittentRetry}
\begin{tabular}{|l|l|}
\hline
\multicolumn{1}{|l|}{\textbf{Software}} & \multicolumn{1}{l|}{\textbf{Parameter}} \\ \hline
Wget & \begin{tabular}[c]{@{}l@{}}tries=10\\ waitretry=20\\ timeout=60\\continue\\ retry-connrefused\\ retry-on-http-error=503\end{tabular} \\ \hline
Envoy HTTP listener& \begin{tabular}[c]{@{}l@{}}retry\_on: 5xx, \\ refused-stream,\\ connect-failure, \\ gateway-error, reset\\ per\_try\_timeout =25s\\ num\_retries =2\end{tabular} \\ \hline
\end{tabular}
\end{table}

\subsection{{Security assumptions, threat model, and mitigations}}
\label{app:security}

Hermes, as an architecture, supports multiple operational models depending on how it is implemented and deployed. Here we structure the security discussion around two endpoints of the deployment design space and note that practical deployments can lie between them:
\begin{itemize}
\item \textit{\textbf{Mode A} (service-provider-operated, fully owned):} the service provider operates the overlay control plane and deploys and manages all non-end-user data-plane components. End-users' dependent proxies run on end-user devices (outside the service provider's administrative domain) and are remotely configurable in accordance with service-provider policy intent via authenticated bootstrap and update channels.
\item \textit{\textbf{Mode B} (third-party-operated, fully managed):} a single third-party overlay operator operates the overlay control plane and all non-end-user data-plane components as a managed service. The third-party operator exposes an interface through which the service provider supplies its own data-plane policy configurations. End-users' dependent proxies run on end-user devices outside the administrative domains of both the service provider and the third-party overlay operator, but are remotely configurable in accordance with service-provider policy intent via the third-party's authenticated bootstrap and update channels.
\end{itemize}

Hermes is a service-oriented communication substrate and is agnostic to whether the application payload is encrypted. As described in the main body of the paper, dependent proxies ensure that traffic entering the Hermes overlay is either HTTP or encapsulated in HTTP, so that intermediate proxies can apply routing and policy using Hermes metadata carried in HTTP headers and transport-layer metadata. Whether Hermes overlay-carried HTTP traffic is protected for confidentiality and integrity is a deployment choice that depends on service requirements and the threat model.

In deployments where confidentiality of the carried payload is not required, or where the application payload is already end-to-end encrypted, a service provider may carry traffic over cleartext HTTP, even when using CONNECT or CONNECT-UDP as the carriage mechanism (as in our NDN and IP use cases). Otherwise, as shown in the Intermittent use case, Hermes can (1) use hop-by-hop TLS encryption, or (2) tunnel traffic across one or more overlay segments using CONNECT or CONNECT-UDP, where payload confidentiality is provided by the inner protocol (e.g., HTTPS or SSH). More generally, Hermes can support confidentiality for overlay-carried traffic in several ways:
\begin{itemize} 
\item \textit{End-to-end encrypted from application:} if the application traffic is already end-to-end protected (e.g., HTTPS or SSH) and is proxy-aware, the application can tunnel its traffic across the overlay using CONNECT or CONNECT-UDP. The overlay does not decrypt the application payload; intermediate proxies only observe connect-time headers and transport-layer metadata to select routes and enforce policy. 
\item \textit{End-to-end encrypted from dependent proxy:} the dependent proxy can establish a TLS-protected tunnel by originating CONNECT or CONNECT-UDP and carrying the received traffic through that tunnel. In this mode, the application does not initiate CONNECT or CONNECT-UDP; it either sends traffic directly to the local dependent proxy's port, or the dependent proxy captures traffic via a tunnel interface. The application traffic may already be end-to-end encrypted (e.g., SSH), in which case the dependent proxy tunnels it without decrypting it. However, if the application sends cleartext, the dependent proxy observes data and becomes part of the trusted computing base for that traffic class. 
\item \textit{Hop-by-hop encryption:} overlay proxies can establish hop-by-hop TLS connections. If the payload is not end-to-end encrypted separately, any overlay component that terminates TLS has visibility into the payload on that segment. \end{itemize}

These options can be mixed across the overlay. For example, a deployment can tunnel traffic using CONNECT across one segment and use hop-by-hop TLS across another. All of these choices are deployment-dependent. These distinctions are especially important in Mode B, where the service provider may want to prevent a third-party overlay operator from observing payload while still relying on that operator for routing and traffic management.

As discussed in the main body of the paper, end-user devices are unmanaged devices with centrally defined and remotely delivered policy: the end-user has full control on their device, can start and stop proxy and assisting components. However, these components still authenticate to the overlay and receive service-provider policy and configuration via the control plane configuration channel. Throughout this appendix, we assume all non-end-user data-plane components (standalone proxies, service-side dependent proxies, and any assisting components deployed outside end-user devices) operate within an operator-controlled administrative domain and can be operationally managed, even when hosted on rented infrastructure (e.g., cloud VMs, CDN points of presence). In Mode A this domain is operated by the service provider; in Mode B it is operated by the third-party overlay operator. For simplicity, we treat an end-user's dependent proxy together with any assisting components on the end-user's device as a single on-device component; throughout this appendix, we use "endpoint" interchangeably to refer to this combined end-user's dependent proxy and assisting components.

We use two NIST references as baseline models for threats that are not unique to Hermes. First, NIST SP 800-233 provides proxy-model threat profiles for service meshes (including both sidecar and shared proxy models), and covers threats including compromised proxies, denial-of-service and resource consumption, and bypassing proxy enforcement. It also highlights how proxy placement and sharing affect blast radius~\cite{nist_sp800_233}. Second, NIST SP 800-144 provides guidelines on security and privacy in public cloud computing, and it helps make trust assumptions explicit when working with third-party overlays~\cite{nist_sp800_144}. Threats and mitigations for non-end-user proxies and the overlay control plane follow established proxy and cloud deployment threat models as treated by the baseline references above and are not restated here. In contrast to these baseline models, Hermes adds two architectural elements that are central to this appendix: (1) dependent proxies and assisting components on end-user devices with user-consented traffic capture capabilities, and (2) a bootstrap and update path whose compromise can directly affect endpoint operation. Mode B additionally introduces tenant scoping and multi-tenant isolation concerns that are specific to a third-party managed overlay.

This appendix makes assumptions explicit, defines adversary models, enumerates attack surfaces, and summarizes mitigation strategies at an architectural level, following common guidelines for security considerations~\cite{rfc3552}.  We consider the following assets and security goals:
\begin{itemize}
\item Payload confidentiality (when required): prevent unauthorized disclosure of the content carried through the overlay (e.g., HTTP message bodies or tunneled bytes via CONNECT/CONNECT-UDP).
\item Traffic integrity (for protected traffic classes): prevent unauthorized modification, injection, or replay of traffic.
\item Component authenticity: ensure only legitimate Hermes components can participate, and prevent unauthorized actors from impersonating endpoints, proxies, or the control plane.
\item Availability: preserve service continuity under benign failures and resist denial-of-service where feasible.
\item Policy correctness: enforce the intended treatment of end-user traffic within the overlay and prevent unauthorized policy changes.
\end{itemize}

In the remainder of this section, we first discuss the shared threat model between Mode A and Mode B. Then we discuss how these modes affect security models.

\subsubsection{Shared threat model}

Threats and mitigations in this subsection apply to both Mode A and Mode B and capture what Hermes adds beyond what \cite{nist_sp800_233} and \cite{ nist_sp800_144} provide.

\paragraph{Security assumptions}
We make the following assumptions explicit:

\begin{itemize}
\item End-user consent for capture: capture and proxying on end-user devices is opt-in and enabled only with explicit end-user permission. The intended captured traffic class is defined by service-provider policy and configuration delivered to the endpoint component.
\item End-user devices are higher-risk: end-user devices are not assumed to be physically secure; endpoint compromise is plausible and must be treated as part of the threat model.
\item Bootstrap and update authenticity: regardless of who operates the overlay control plane (service provider in Mode A, third-party overlay operator in Mode B), the content of end-user bootstrap and updates originates from service-provider policy intent and may be targeted by external attackers. Therefore, authenticity, integrity, and freshness of delivered endpoint updates must be enforced by a Hermes-compliant implementation.
%\item Payload protection is deployment dependent: Hermes can carry cleartext payload, hop-by-hop protected payload (with termination at selected proxies), or end-to-end tunneled payload (CONNECT or CONNECT-UDP) where intermediate proxies do not terminate the inner security association. In Mode B, if the third-party overlay operator is not trusted with payload, deployments should prefer end-to-end tunneling for sensitive traffic classes.
\end{itemize}

\paragraph{Adversary model}
We focus on the following adversaries:

\begin{itemize}
\item End-user application adversary (unprivileged): a malicious or compromised application on the end-user device that can generate arbitrary traffic and can abuse overlay resources by driving excessive or adversarial traffic through the dependent proxy.
\item Local device adversary (privileged): malware or an attacker with elevated privileges on the end-user device that can tamper with the endpoint process, the tunnel-device capture path, local configuration state, or endpoint-stored credentials.
\item Access-network adversary (on-path): an attacker on the path between the end-user device and the first Hermes hop that can eavesdrop, drop, delay, replay, or modify packets on that segment.
\item Endpoint update-channel adversary: an attacker who can compromise the endpoint update path by (1) impersonating the entity that endpoints treat as authoritative for updates, or (2) tampering with, injecting, replaying, or modifying policy and configuration.
\end{itemize}

%In end-to-end encryption mode, intermediaries that do not terminate TLS generally cannot read or modify application payload, but can enforce policy at tunnel establishment and apply connection-level controls; on-path adversaries can still cause denial-of-service by disrupting connectivity. In hop-by-hop mode, any proxy that terminates TLS is in the trusted computing base for payload confidentiality and integrity on that segment, and compromise of such a proxy can expose or modify payload.

\paragraph{Attack surfaces}
The Hermes architecture introduces the following attack surfaces:

\begin{itemize}
\item Dependent proxy and assisting components: end-user traffic can enter the dependent proxy either via explicit application or OS proxy configuration, or via tunnel-based capture. Compromise or misuse of these components can (1) divert user-selected traffic away from Hermes (bypassing intended mediation), (2) inject, substitute, or tamper with traffic and overlay metadata, (3) disrupt availability by dropping or delaying flows, or (4) unintentionally downgrade required protections for protected traffic classes, exposing sensitive traffic as cleartext.
\item Local credential and configuration state: the dependent proxy may store endpoint-side identity and authorization state required to participate in the overlay (e.g., tokens or certificates that identify the end-user), as well as cached policy and local configuration needed for operation during intermittent connectivity.
\item Bootstrap and update channel: endpoint enrolment, proxy configuration, and policy delivery could be abused to subvert the service provider's policy intent.
\item Telemetry and reporting channel: endpoint-to-control-plane observability signals that could be abused for decision poisoning or telemetry-plane denial-of-service.
\end{itemize}

\paragraph{Threats and mitigation strategies}
Table~\ref{tab:endpoint_threats} summarizes endpoint-specific threats and architecture-level mitigation strategies. Mitigations are stated as requirements or recommended properties of a Hermes-compliant implementation, independent of any particular proxy software or OS mechanism.

\begin{table*}[t]
\centering
\caption{Shared threats and architectural mitigations}
\label{tab:endpoint_threats}
\renewcommand{\arraystretch}{1.1}
\begin{tabular}{p{0.12\linewidth} p{0.22\linewidth} p{0.60\linewidth}}
\hline
Threat & Impact & Architectural mitigations and notes \\
\hline
T1. Endpoint compromise  &
Tampering with Hermes-related HTTP headers on mediated traffic; availability disruption &
Treat end-user devices as higher-risk nodes than non-end-user overlay components. Support automated quarantine and re-enrollment workflows for suspected compromised endpoints. Where available, deployments may additionally gate endpoint enrollment on integrity signals (e.g., binary integrity and attestation mechanisms) and continuously re-evaluate endpoint posture and allowed functionality. In upstream proxies, bind policy enforcement to the authenticated endpoint identity, and only accept Hermes-related headers that are permitted for that identity, verifying their values against identity-scoped allowlists. \\
\hline
T2. Bypass of intended Hermes mediation for a traffic class  &
Loss of Hermes enforcement and delivery semantics for the traffic class; bypass of the configured capture policy; potential exposure to the default network path  &
Hermes capture is opt-in at the device: if capture is not enabled, the traffic class is out of scope and does not use Hermes. The relevant threat is bypass of the configured capture policy for a traffic class that the service provider intended to send through Hermes (e.g., by avoiding the configured proxy path or escaping tunnel capture coverage). Mitigate by treating capture coverage as a correctness property of the endpoint configuration, reducing bypass opportunities in the local network stack where feasible, and clearly reporting tunnel/proxy status and effective capture coverage details to the user. \\
\hline
T3. Abuse of overlay resources  &
Availability impact on overlay and downstream services; resource exhaustion; operational risk  & 
Availability issues can arise from both intended traffic and non-intended traffic directed toward the dependent proxies. Treat endpoints as untrusted traffic sources for availability. Apply rate limits per endpoint identity and per destination at upstream proxies. Provide anomaly signals for sudden spikes (e.g., connection attempts, rejected traffic, or filtered traffic), with automated throttling or quarantine. For non-intended or policy-disallowed traffic classes, the primary mitigation is early filtering, together with automated throttling or quarantine. \\
\hline
T4. Endpoint credential theft and impersonation  &
Unauthorized overlay access using stolen endpoint credentials; misuse of the endpoint's authorized privileges; endpoint impersonation from another device or process &
Bind credentials to endpoint identity and role or scope. Prefer short-lived credentials with frequent rotation. Minimize the authorization scope of endpoint identities (least privilege), and support enrollment revocation and credential re-issuance workflows. Additionally, monitor for anomalous patterns (e.g., unusual traffic volume) and trigger throttling or quarantine when suspicious behavior is detected.\\
\hline
T5. Endpoint update-channel compromise  &
Unintended change of policy or security posture; policy bypass; inability to bootstrap or keep configuration aligned with current intent; endpoint isolation  &
Require authenticity and integrity for endpoint updates (e.g., mutual authentication and integrity-protected update bundles). Ensure endpoints can authenticate the authority they rely on for updates. Require monotonic versioning and freshness to prevent replay or rollback to prior policy versions, and enforce authorization so only permitted principals can update endpoints. Use stability guards (e.g., minimum hold times, cooldown periods, and thresholds) to reduce induced churn, and define fallback behavior under update failure. \\
\hline
T6. Endpoint telemetry abuse &
Misleading observability; incorrect control-plane decisions that can impact other endpoints; induced churn; resource exhaustion or unavailability of telemetry ingestion services  &
Treat endpoint-reported signals as potentially adversarial. Validate with infrastructure-side observations where feasible. If telemetry is used to trigger reconfiguration, use stability guards (e.g., smoothing, cooldown periods, and thresholds), restrict which signals can trigger changes, and prefer decisions driven by aggregated evidence rather than single endpoints. Protect telemetry infrastructure with authentication, per-identity rate limits, and sampling to prevent telemetry-plane flooding, if possible. \\
\hline
T7. Access-network on-path interference (between endpoint and first hop data-plane)  &
Eavesdropping or tampering for cleartext traffic classes; denial-of-service via dropping or delaying packets; disruption of tunnel establishment  &
For protected traffic classes, use authenticated, encrypted, and integrity-protected connections between the endpoint and the first upstream proxy (e.g., TLS) so on-path adversaries cannot eavesdrop, modify, or replay protected payload. Treat availability on the access segment as a residual risk; mitigate with bounded retries and backoff, and alternative upstream selection where available. \\
\hline
\end{tabular}
\end{table*}

\subsubsection{Mode A: service-provider operated overlay (fully owned)}

Mode A keeps the overlay control plane and non-end-user overlay components within the service provider's administrative domain. Shared threats and mitigations above apply directly. Risks and mitigations for non-end-user proxies and the control plane depend on placement and sharing choices and can be derived by consulting~\cite{nist_sp800_233}; this appendix does not repeat those proxy-model threats and requirements.

\subsubsection{Mode B: third-party operated overlay (fully managed)}

Mode B adds managed multi-tenant risks that are not present in the same form in Mode A. Baseline managed-service risks in shared cloud environments, including denial-of-service and availability impacts due to shared resources, are covered by \cite{nist_sp800_144}, and we do not restate those here. Instead, we focus on concerns that directly affect endpoints due to the multi-tenant environment.

\paragraph{Mode B trust assumptions}
We consider the following assumptions in Mode B.
\begin{itemize}
\item Tenant isolation: the third-party overlay operator enforces isolation between tenants for configuration and policy namespaces, runtime state, and shared proxy resources, and limits noisy-neighbor effects through resource governance.
\item Configuration and identity integrity: the service provider supplies policy intent through the third-party overlay operator interface. Either the service provider explicitly trusts the third-party overlay operator not to introduce unauthorized modifications, or the deployment reduces this trust requirement by using end-to-end authenticity and integrity mechanisms for tenant policy and identity artifacts, so unauthorized modification is detectable.
\item Co-tenant adversary: other tenants are treated as potentially adversarial. Co-tenants may attempt to exhaust shared resources or exploit isolation weaknesses.
\item Payload confidentiality boundary (Mode B default): overlay-operated proxies, including endpoint proxy software, may be untrusted with cleartext payload. Therefore, by default we assume sensitive traffic classes are end-to-end protected by the application. Encrypted traffic may reach the endpoint either because the application is proxy-aware and tunnels through the endpoint (e.g., via CONNECT/CONNECT-UDP), or because the endpoint captures already-encrypted traffic from a non-proxy-aware application via a tunnel interface. If the application sends cleartext, then the endpoint and any overlay component that can observe that cleartext become part of the trusted computing base for payload confidentiality on the segments they handle.
\end{itemize}
%Even in end-to-end tunneling mode, the overlay operator necessarily observes tunnel setup information and flow-level metadata needed for routing and traffic management; we treat this metadata exposure as an inherent deployment tradeoff rather than a Hermes-specific threat. %do not need to discuss this

\paragraph{Mode B attack surfaces.}
Mode B introduces additional attack surfaces that can affect end-users by changing what endpoints receive through enrollment, bootstrap, and policy or update distribution:
\begin{itemize}
\item Operator-provided tenant control interface and scoping: the third-party overlay operator's API and associated workflows for tenant onboarding and management, policy publishing, and endpoint enrollment determine which identities and policies are issued and delivered to end-users' endpoints.
\item Operator-managed endpoint update and distribution pipeline: the overlay operator's distribution of endpoint proxy binaries and update bundles, along with associated signing keys and rollout mechanisms, can directly compromise endpoints if subverted.
\item Multi-tenant endpoint participation and attribution: when a single endpoint proxy instance participates in multiple tenants, incorrect tenant scoping or state isolation (e.g., for enrollment, telemetry, updates, and policy state) can cause cross-tenant leakage, misattribution, or conflicting configuration.
\end{itemize}

\paragraph{Mode B additional threats and mitigations}
Table~\ref{tab:modeb_threats} summarizes additional threats that arise specifically in Mode B. These items complement the threats shared across Mode A and Mode B in Table~\ref{tab:endpoint_threats}.

\begin{table*}[t]
\centering
\caption{Additional threats and architectural mitigations for Mode B third-party operated overlays.}
\label{tab:modeb_threats}
\renewcommand{\arraystretch}{1.1}
\begin{tabular}{p{0.12\linewidth} p{0.22\linewidth} p{0.60\linewidth}}
\hline
Threat & Impact & Architectural mitigations and notes \\
\hline
B1. Unintended tenant-scoped artifact issuance or delivery &
Misdelivery of tenant-scoped policy and identity artifacts to endpoints; unintended policy enforcement; cross-tenant impact &
Explicitly bind each tenant's artifacts (policies and endpoint identities) to tenant identifiers, and require endpoints to validate tenant identifiers, signatures, and freshness before applying artifacts. Protect artifact publication and distribution with authenticated access controls, so only authorized principals can issue or update tenant-scoped artifacts. Enforce strict per-tenant namespace isolation for policy configuration and runtime state. Provide tenant-visible audit logs and change transparency to detect anomalies. \\
\hline
B2. Operator-managed endpoint update or binary supply chain compromise &
Backdoored endpoint proxies; targeted tenant compromise; credential theft; large-scale endpoint compromise across tenants &
Require signed binaries and signed updates with verifiable provenance, monotonic versions, and freshness. Provide tenant-visible transparency for deployed versions and rollout state. Support revocation and quarantine for affected endpoint identities, and optionally gate endpoint participation on integrity signals (e.g., binary integrity and attestation) where available. \\
\hline 
B3. Multi-tenant endpoint attribution and separation failure & Cross-tenant telemetry leakage or misattribution; incorrect control-plane decisions; induced churn due to conflicting tenant configurations & When a single endpoint participates in multiple tenants, require explicit tenant context binding for endpoint-to-control-plane interactions (e.g., enrollment, telemetry, and updates), and require the control plane to validate tenant identifiers on each request. Maintain per-tenant keys, credentials, caches, and policy state at the endpoint, and avoid cross-tenant state reuse. Where multiple tenants impose conflicting endpoint configurations, define deterministic conflict-resolution rules (or require explicit user selection) to prevent oscillation and unintended behavior. \\
\hline
B4. Confidentiality boundary violation (Mode B) &
Unintended exposure of sensitive traffic as cleartext to overlay-operated components (e.g., due to misconfiguration, downgrade of required protection, or unintended TLS termination) &
Where feasible, require application-originated end-to-end secure sessions. For cases where the endpoint must protect plaintext applications, require the endpoint to originate TLS to intended egress points using certificate validation and, where appropriate, certificate or identity pinning. Additionally, enforce downgrade resistance by policy (reject disallowed protection modes), and provide tenant-visible auditing of effective protection and termination points. \\
\hline
\end{tabular}
\end{table*}

\vfill
\end{document}